\documentclass[arxiv]{imsart}

\usepackage{graphicx}
\usepackage[space]{grffile}
\usepackage{latexsym}
\usepackage{textcomp}
\usepackage{tabulary}
\usepackage{booktabs,array,multirow}
\usepackage{amsfonts,amsmath}
\usepackage{mathrsfs}  
\usepackage{url}
\usepackage[colorlinks,citecolor=blue,urlcolor=blue]{hyperref}
\usepackage{etoolbox}
\usepackage[utf8]{inputenc}
\usepackage[english]{babel}
\usepackage[page,titletoc,title]{appendix}

\usepackage{siunitx}
\usepackage{lscape}
\usepackage{verbatim}
\usepackage{enumitem}
\usepackage[caption=false]{subfig}
\usepackage{xcolor}
\usepackage{mathtools}
\usepackage[numbers]{natbib}

\startlocaldefs

\newif\iflatexml\latexmlfalse

\AtBeginDocument{\DeclareGraphicsExtensions{.pdf,.PDF,.eps,.EPS,.png,.PNG,.tif,.TIF,.jpg,.JPG,.jpeg,.JPEG}}

\DeclareMathOperator\arctanh{arctanh}

\theoremstyle{plain}
\theoremstyle{definition}

\newtheorem{defn}{Definition}
\newtheorem{prop}{Proposition}
\newtheorem{theo}{Theorem}
\newtheorem{coro}{Corollary}
\newtheorem{lemm}{Lemma}
\newtheorem{rmk}{Remark}

\newcommand{\MReps}{{\rm MReps}}
\newcommand{\INVI}[1]{}
\newcommand{\IID}{IID}
\newcommand{\trp}{^{\tt T}}

\newcommand{\trace}{\mbox{\rm\bf tr}}

\newcommand{\E}{\mathrm{E}}
\newcommand{\Var}{\mathrm{Var}}
\newcommand{\AVar}{\mathrm{AsyVar}}
\newcommand{\Cov}{\mathrm{Cov}}
\newcommand{\Cor}{\mathrm{Cor}}
\newcommand{\CCC}{\mathrm{CCC}}
\newcommand{\PCC}{\mathrm{PCC}}
\newcommand{\MSE}{\mathrm{MSE}}
\newcommand{\MSPE}{\mathrm{MSPE}}
\newcommand{\bs}{\boldsymbol}
\let\txt\texttt

\newcommand{\RED}[1]{{\color{red} #1}}
\newcommand{\BLUE}[1]{{\color{blue} #1}}
\newcommand{\VEC}[1]{{\mbox{VEC}\left(#1\right)}}
\newcommand{\TRACE}[1]{{\mbox{tr}(#1)}}

\newcommand{\Sgn}{\mbox{\rm Sgn}}
\newcommand{\norm}[1]{\left\lVert #1 \right\rVert}

\newcommand{\conas}{\stackrel{{a.s.}}{\longrightarrow}}

\allowdisplaybreaks
\vfuzz \hfuzz
\makeatletter
\renewcommand*\env@matrix[1][\arraystretch]{%
  \edef\arraystretch{#1}%
  \hskip -\arraycolsep
  \let\@ifnextchar\new@ifnextchar
  \array{*\c@MaxMatrixCols c}}
\makeatother

\endlocaldefs

\begin{document}
\begin{frontmatter}
\title{Maximum Agreement Linear Predictors}
\runtitle{/ Maximum Agreement Linear Predictor //}

\begin{aug}
\author[A]{Taeho Kim},
\author[B]{Pierre Chauss\'e},
\author[C]{Matteo Bottai},
\author[D]{Gheorghe Doros},
\author[E]{Mihai Giurcanu},
\author[F]{George Luta},
\author[G]{Edsel A.\ Pe\~na}
\address[A]{Department of Mathematics, Lehigh University.}
\address[B]{Department of Economics, University of Waterloo.}
\address[C]{Division of Biostatistics, Institute of Environmental Medicine, Karolinska Institutet.}
\address[D]{Department of Biostatistics, School of Public Health, Boston University.}
\address[E]{Department of Public Health Sciences, University of Chicago.}
\address[F]{Department of Biostatistics, Bioinformatics \& Biomathematics, Georgetown University.}
\address[G]{Department of Statistics, University of South Carolina.}


\end{aug}


\begin{abstract}
This paper studies predictor functions motivated by maximizing a measure of agreement with the predictand. Specifically, it examines distributional properties and predictive performance of the estimated maximum agreement linear predictor (MALP), the linear predictor maximizing Lin's concordance correlation coefficient (CCC) between the predictor and the predictand. It is compared and contrasted, theoretically and through computer experiments, with the estimated least-squares linear predictor (LSLP), with respect to some performance measures. Finite-sample and asymptotic properties are obtained, and confidence intervals and prediction intervals are also presented. Predictors are illustrated using two real data sets: an eye data set and a body fat data set. Results indicate that the estimated MALP is a viable alternative to the estimated LSLP if one desires a predictor whose predicted values possesses higher agreement with the predictand values, as measured by the CCC.

\smallskip


\smallskip

\end{abstract}%

\tableofcontents

\end{frontmatter}

\section{Introduction}
\label{sec-intro}

The invention, construction, or development of prediction functions, or more colorfully, `prediction machines', is one of the most important and consequential endeavors of statisticians, mathematicians, machine learners, data scientists, and artificial intelligence researchers, with deep implications and utilities in many facets of science, engineering, medicine, health, economics, business, politics, and in general, society. The most fundamental of all such prediction problems is that there is a response random variable, $Y$, called a {\em predictand}, and a set of concomitant random variables, represented by a random vector, $X$, called a {\em predictor}, and the goal is to predict the value of $Y$ when one observes or sets the value $X$ to $x_0$. Regression models, many machine learning algorithms, artificial neural nets, deep learners, and large language models, are about prediction machines! This paper concerns a prediction machine developed to maximize `agreement' between the predictand and predictor machine or function. But, how should we measure agreement?

The classical correlation coefficient introduced by \citet{Galt:1889} and mathematically formulated by \citet{Pear:1896}, nowadays called Pearson's correlation coefficient (PCC), a manifestation of Stigler's Law of Eponymy, continues to be one of the most-used statistical tools as it provides a quantitative measure of the degree of linear association between two random variables, see \citet{LN:1988,Stig:1989} for its historical account.  With its simple mathematical formulation and low computational cost, its importance still persists in the era of Big Data where researchers have to deal with a large number of variables and/or a large number of observations. While {there is no question about} the pervasive utility of the PCC, there are certain situations where researchers are more interested in the degree of agreement from the vantage point of the $45^{\circ}$ line through the origin, such as in calibration, imputation, and linear equating.
In such scenarios, the PCC is not the most appropriate measure of agreement since {it views} the degree of linear association from a vantage point that could be any line. 

\citet{Lin:1989} introduced a measure of agreement called the concordance correlation coefficient ($\CCC$). For a bivariate random vector $(Y,X)$ with mean vector $(\mu_\txt{Y},\mu_\txt{X})$ and covariance matrix $\begin{bmatrix}[0.5]\sigma^2_\txt{Y}&\sigma_\txt{YX}\\ \sigma_\txt{YX}&\sigma_\txt{X}^2  \end{bmatrix}$, so has PCC $\rho= \frac{\sigma_{\txt{YX}}}{\sigma_\txt{Y}\sigma_\txt{X}}$, the $\CCC$ between $Y$ and $X$ is defined below. 
\begin{defn}
For a bivariate random vector $(Y,X)$, its concordance correlation coefficient is
\begin{eqnarray}
		\CCC[Y,X]  \ =	\ \rho^c & := &  1-\frac{\E[(Y-X)^2]}{\Var[Y]+\Var[X]+(\E[Y]-\E[X])^2} 
		\nonumber \\
	& = & \frac{2\rho\sigma_\txt{Y}\sigma_\txt{X}}{\sigma_\txt{Y}^2+\sigma_\txt{X}^2+(\mu_\txt{Y}-\mu_\txt{X})^2}. \label{DEF_CCC}
\end{eqnarray}
\end{defn}
The $\CCC$\ definition clearly requires that $Y$ and $X$ possess the same units or unitless. In addition, this entails that $Y$ and $X$ be measurements of the same characteristic, trait, or phenomenon. For brevity, we use $\rho$ and $\rho^c$ instead of $\rho_\txt{YX}$ and $\rho^c_\txt{YX}$ whenever the correlations are between the variables $Y$ and $X$; however, the subscripts are specified in other situations.
Observe that $\E[(Y-X)^2]$ is twice the expected perpendicular squared Euclidean distance from the $45^\circ$ line of the random point $(Y,X)$ {and its denominator represents the same expected squared distance when $Y$ and $X$ are independent}. Hence, it is a measure of the degree of concordance between $Y$ and $X$, with a small absolute value indicative of high concordance. Thus, geometrically speaking, $\rho^c$ measures a particular linear association of $Y$ and $X$ from the vantage point of the $45^\circ$ line through the origin. Properties of the $\CCC$ (see \citet{Lin:1989}) in relation to the $\PCC$, are (i) $-1 \ \le \ \rho^c \ \le \ 1$; (ii) $\rho \ = \ 0$ if and only if $\rho^c \ = \ 0$; and (iii) $|\rho^c| \ \le \ |\rho|$ with  equality holding if and only if $\mu_\txt{Y} \ = \ \mu_\txt{X}\text{ and }\sigma^2_\txt{Y} \ = \ \sigma^2_\txt{X}$. 
%

\begin{figure}[ht]
 \centering
 \includegraphics[height=150pt]{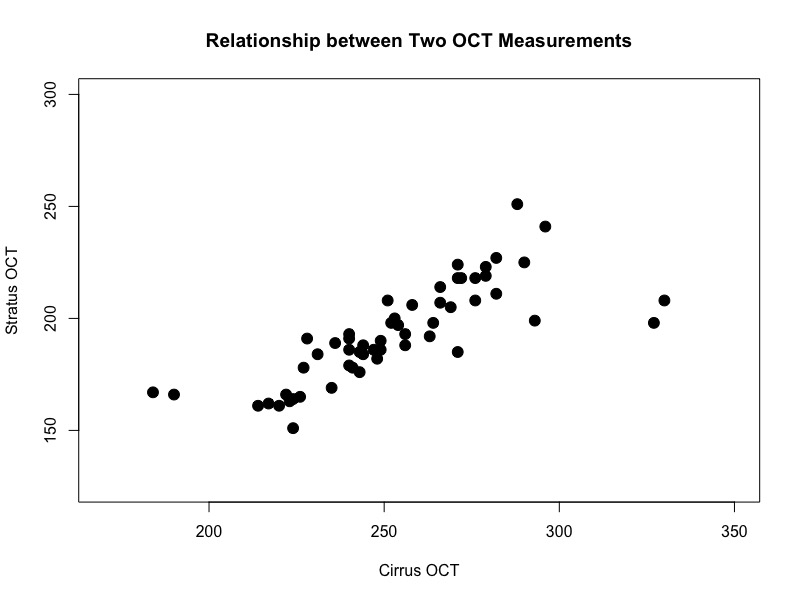}
 \caption{ The scatter plot between Stratus and Cirrus OCTs based on 56 observations in the data set in \citet{APDS:2011}.}
 \vspace{-10pt}
 \label{OCT_DATA}
 \end{figure}

As a concrete example of the use of the $\CCC$,  \citet{APDS:2011}, in an opthalmology study, developed a conversion formula between {the measurements obtained from} two optical coherence tomography (OCT) approaches: Stratus OCT (a conventional approach) and Cirrus OCT (an advanced approach). Figure \ref{OCT_DATA} shows a scatter plot of the paired OCT sample data that was analyzed.  
 Their formula aimed to maximize the $\CCC$ between the two OCT measurements. The idea of using the CCC to develop a predictor (or, a conversion formula, in the case of \citet{APDS:2011}'s setting) was studied in \citet{BKLLP:2022} for the case where the joint probability distribution of $(Y,X)$ is completely known, that is, the functional form of the distribution is known and there are no unknown parameters. Thus, given a class of predictors $\mathcal{H}=\{\tilde{Y}(X):E[\tilde{Y}^2(X)]<\infty\}$, an optimal predictor $\tilde{Y}^\star\in\mathcal{H}$ was developed to maximize the $\CCC$ between $Y$ and $\tilde{Y}$. Since $\CCC$ is a measure of agreement, the resulting predictor was called a maximum agreement predictor in the class $\mathcal{H}$, called the MAP$_\mathcal{H}$, or simply MAP.

 In the current paper, we consider the situation where the joint probability distribution of $(Y,X)$ is known only to belong to the class of joint distributions with finite fourth moments, or its functional form is known but there are unknown finite-dimensional parameters. We aim to examine the statistical properties of the estimator of the maximum agreement predictor, specifically the maximum agreement {\em linear} predictor (MALP). Thus, this paper extends the results in \citet{BKLLP:2022} to the realistic setting where the joint distribution of $(Y,X)$ is not known, or if its functional form is known, there still are  unknown parameters. {However, it still does {\em not} fully extends \citet{BKLLP:2022} since this paper deals only with {\em linear} predictors.} 

When $X$ is one-dimensional, i.e., $p=1$, the class of linear predictors is $\mathcal{H}_\txt{LP} = \{\tilde{Y}:\  \tilde{Y}(x) = \alpha + x\beta\}$. The resulting MALP, derived in Section \ref{sec: Predictors}, is comparable to the classical least-squares linear predictor (LSLP), $\tilde{Y}^\dagger$, which is the linear predictor $\tilde{Y}(x) = \alpha + \beta x \in \mathcal{H}_\txt{LP}$  which minimizes the mean-squared prediction error (MSPE): $\tilde{Y} \mapsto E[(Y-\tilde{Y}(X))^2]$, where expectation is with respect to the joint distribution of $(Y,X)$. When the parameters are known, these two predictors are given by
\begin{equation}\label{SIMPLE_CASE}
\begin{array}{c}
 	\tilde{Y}^\star(x_\txt{0}) \ = \ \mu_\txt{Y}+\Sgn(\rho)\left(\frac{\sigma_\txt{Y}}{\sigma_\txt{X}}\right)(x_\txt{0}-\mu_\txt{X})\;\text{ and }\; 
  \tilde{Y}^{\dagger}(x_\txt{0}) \ = \ \mu_\txt{Y}+\rho\left(\frac{\sigma_\txt{Y}}{\sigma_\txt{X}}\right)(x_\txt{0}-\mu_\txt{X}),
  \end{array}
 \end{equation}
 where $\Sgn(a) = \{-1, 0, +1\}$ according to whether $a \{<,=,>\} 0$, respectively.

We now outline the main contents of this paper. In Section \ref{sec: Review}, we provide a brief literature review regarding different correlation and agreement measures. In Section \ref{sec: Predictors}, we formally define the MALP with known/unknown parameters.  Section \ref{sec: Asymptotics} investigates the consistency and asymptotic normality of the estimated MALP and LSLP, with some proofs relegated to Appendices. Section \ref{sec: Simulations} presents simulation studies assessing adequacy of asymptotic approximations and comparing the estimated MALP with the estimate LSLP. Section \ref{sec: Intervals} provides confidence and prediction interval procedures for the MALP. Illustrative data analyses using two data sets, one being the eye data set, and the other a well-used body fat data set, are presented in Section \ref{sec: Illustrations}. Section \ref{sec: Conclusions} closes with some concluding remarks.  

\section{Literature Review}
\label{sec: Review}
 
As previously mentioned, the historical account for the development of the $\PCC$ can be found in \citet{Stig:1989}. In addition, \citet{LN:1988} provided the history of $\PCC$ as well as its conceptual derivation and interpretation from geometric and algebraic perspectives. 
In prediction settings, the coefficient of determination is the squared $\PCC$ between the predictand and the predictor, hence extending the $\PCC$ to more general settings. Regarding coefficients of determination, \citet{Kvaa:1985} discussed several different versions and their non-equivalence causes misleading interpretations when used in data analysis.
The $\PCC$ is also one of the most widely-used procedures in other disciplines, with several papers describing its discipline-specific uses and statistical interpretations. Some of the well-cited works are \citet{SBS:2018} and \citet{Muka:2012} in medical research; \citet{BP:1980} in engineering; \citet{Chap:2012} in chemistry; and \citet{Ozer:1985} in psychology. 

There are also works that generalize the $\PCC$ in other directions. With $\PCC$, a zero correlation does not necessarily imply independence, except for the jointly normally-distributed case. \citet{SRB:2007} introduced the concept of a distance correlation coefficient which has the property that if it equals zero then the variables are independent. \citet{Rayw:2007} developed a procedure that can handle different types of data -- nominal, ordinal, or interval type -- for data mining applications. There have been efforts to measure non-linear associations, see for example, \citet{DS:2009,LHS:2013,RRFG:2011}. The paper by \citet{STNF:2013} provided a comparative study regarding the $\PCC$ and its extensions, particularly for gene expression data in biology. Lastly, a survey paper by \citet{TOS:2022} examined shortcomings of the $\PCC$ and described the development of alternative procedures during the past two decades. 

The concept of agreement concerns the concordance between two numerical or categorical variables. As discussed previously, the usual Pearson correlation would not be appropriate to measure agreement. For continuous variables, the intra-class correlation coefficient and the Bland-Altman plot are classical methods for assessing  agreement \citep{RPA:2017,WP:2010}; {whereas, for categorical variables, Cohen's kappa ($\kappa$) and its extensions are the typical measures of agreement \citep{BCMS:1999}.} The \emph{intra-class} correlation coefficient (ICC) was originally proposed in the classic book by \citet{Fish:1925}
(see also \citet{Fish:2021}).
The ICC's essential difference from the \emph{inter-class} correlation, which is the PCC, is in the use of the pooled data for computing the sample mean and the sample variance. To expand on this, let $\{(Y_i,X_i), i=1,\ldots,n\}$ be a paired sample data. Denote the basic sample statistics by 
\begin{eqnarray*}
& \bar{Y} = \frac{1}{n} \sum_{i=1}^n Y_i;\; \bar{X} = \frac{1}{n} \sum_{i=1}^n X_i;\; & \\
&  S_\txt{Y}^2 = S_\txt{YY} = \frac{1}{n} \sum_{i=1}^n (Y_i-\bar{Y})^2;\; S_\txt{X}^2 = S_\txt{XX} = \frac{1}{n} \sum_{i=1}^n (X_i-\bar{X})^2;\; & \\
& S_\txt{YX} = S_\txt{XY} = \frac{1}{n} \sum_{i=1}^n (Y_i-\bar{Y}) (X_i-\bar{X}). &
\end{eqnarray*}
Whereas the sample PCC is given by
%
$r = {S_\txt{YX}}/{(S_\txt{Y} S_\txt{X})}$,
%
the ICC is defined as
\begin{displaymath}
r^\txt{I} = \frac{1}{S_P^2} \frac{1}{n} \sum_{i=1}^n (Y_i - \bar{Z})(X_i - \bar{Z})
\end{displaymath}
where $\bar{Z} = \frac{1}{2}(\bar{Y} + \bar{X})$ is the pooled mean and $S_P^2 = \frac{1}{2n}\sum_{i=1}^n [(Y_i-\bar{Z})^2 +  (X_i - \bar{Z})^2]$ is the pooled variance. Therefore, the $Y$ and $X$ observations are getting standardized by the  pooled mean and the pooled standard deviation in the computation of the ICC. As such this becomes a reasonable measure of the concordance between the two variables. Later, various ICCs were developed in parametric and nonparametric settings, see, \citet{MB:1994} for their comparisons and limitations. Named after its inventors, the Bland-Altman (BA) plot, also known as the Tukey mean-difference plot, is a scatterplot that can be utilized to evaluate agreement; see \citet{BA:1983,BA:1986,BA:1999}. In particular, \citet{BA:1986} criticized the use of correlation approaches to measure agreement due to their sensitivity to sample heterogeneity. {With the $x$-axis and the $y$-axis displaying the difference and the average between the two variables, respectively, the BA plot describes the bias between two measurements with 95\% limits of agreement.}

Since introduced by \citet{Lin:1989,Lin:1992}, the concordance correlation coefficient has been used to measure  agreement in reproducibility, assay validation, and other types of agreement studies. The results were generalized to both continuous and categorical variables as well as with random and fixed covariates in \citet{LHWY:2002} and \citet{LHW:2012}. \citet{CJ:1994} compared the ICCs and $\CCC$ and proposed a method to estimate the $\CCC$ through variance components of a mixed effects model. \citet{AN:1997} described some shortcomings of the $\CCC$ as an agreement measure. 
In addition, there have been recent developments based on the $\CCC$ in emotion recognition (\citet{WRMS:2016,TTNS:2017}), in canonical analysis (\citet{Lipo:2022}), in image data analysis within the context of spatial statistics (\citet{VPER:2020}), and in quantile functional regression for clinical imaging application (\citet{YBRM:2020}).

The previously-mentioned paper by \citet{BKLLP:2022} proposed a new prediction procedure that maximizes the CCC between predictor and predictand, called the maximum agreement predictor. Under the bivariate normal distribution, the form of the resulting predictor coincides with the form of $\tilde{Y}^\star$ in (\ref{SIMPLE_CASE}). In fact, the particular form of $\tilde{Y}^\star$ with a single covariate has appeared in several previous studies and was developed via different motivations or rationales, three of which are mentioned below.
\begin{itemize}
 \item[(i)] {\em Geometric mean regression:} While the least-squares regression typically only assumes the randomness of $Y$, the geometric mean regression also assumes the randomness of $X$. Then,  in the $xy$-plane, the slope of the regression of $Y$ on $X$ and the slope of the regression of $X$ on $Y$ are $S_\txt{YX}/S_\txt{XX}$ and $S_\txt{YY}/S_\txt{YX}$, respectively. When these two slopes are combined by taking their geometric mean, the resulting combined slope is $\Sgn\{S_\txt{YX}\}(S_\txt{YY}/S_\txt{XX})$ (see \citet{DY:1997}), which coincides with the slope of $\tilde{Y}^\star$ in (\ref{SIMPLE_CASE}). 

 \item[(ii)] {\em Least-triangle regression:} Another way to obtain the same slope is via the so-called least-triangle regression (see \citet{BSE:1988}), which could be described as follows. Given $n$ points $(Y_i,X_i), i=1,\ldots,n$, and a line $L(x;c,d) = c + dx$, then for the $i$th point there is a unique right-triangle whose sides are parallel to the $x$-axis and the $y$-axis, respectively, and with hypotenuse on the line. Denoting by $A_i(c,d)$ the area of the $i$th right-triangle, and letting $Q(c,d) = \sum_{i=1}^n A_i(c,d)$ be the total area of the $n$ triangles, the least-triangle regression line $L(x;c^*,d^*)$ is the line  associated with the $(c,d)$ that minimizes $Q(c,d)$ for all $(c,d) \in \Re_2$. It turns out that for this optimal line the slope is $d^* = \Sgn\{S_\txt{YX}\}(S_\txt{YY}/S_\txt{XX})$ and $Q(c^*,d^*) = \frac{n}{2} \sqrt{S_\txt{YY} S_\txt{XX}}$. Note that the $X_i$s and $Y_i$s are being treated symmetrically in this approach.
 
 \item[(iii)] {\em Reduced major-axis regression:} As introduced in \cite{KH:1950}, this approach solves the problem of the symmetric regression procedure that minimizes the sum of squared perpendicular distances. See, \citet{IFAB:1990} and \citet{WWFW:2006} for comparisons to other asymmetric and symmetric approaches. Also, see \citet{Smit:2009} for its usage. For the general accounts of symmetric regression,  refer to chapter 12 in \citet{AS:1998} and also to chapter 12 in \citet{Taag:2008}.

\end{itemize}
Lastly, in \citet{FH:2014}, the form of MALP in (\ref{SIMPLE_CASE}) was utilized in {a health science setting} as one of the mappings from the {space of} values of a source instrument to the equivalent {space of} values of a target instrument.  This method, called the linear equating approach (see \citet{KB:2014}), was designed to alleviate the variance shrinkage issue of the least-squares regression-based predictor relative to the predictand, by matching the means and variances of the {respective measurements obtained from the two instruments}.

\section{Maximum Agreement and Least-Squares Linear Predictors}
\label{sec: Predictors}

\subsection{MALP and LSLP: Known Distribution}
\label{SECTION_KNOWN_PRED}


Let $(\Psi,\mathscr{F},\textbf{P})$ be the probability space on which all random entities are defined. Let $(Y, X\trp)\trp:(\Psi,\mathscr{F})\;\to\;(\Re \times \Re_p,\mathscr{B}\times\mathscr{B}_p)$ where $\mathscr{B}$ and $\mathscr{B}_{p}$ are the Borel $\sigma$-fields of $\Re$ and $\Re_{p}$, respectively. Thus, $(Y, X\trp)\trp$ is a $(p+1) \times 1$ random vector with range space  $\mathcal{Y} \times \mathcal{X} \subseteq \Re \times \Re_p$. Let $F = F_{(Y,X)}$ be the induced joint distribution function of $(Y,X\trp)\trp$: $F(y,x) = \textbf{P}\{Y \le y; X \le x\}$. We assume that $\E[Y^4] < \infty$ and $\E[X_i^4] < \infty$ for each $i = 1, \ldots, p$.  Denote its mean vector by $(\mu_\txt{Y},\mu_\txt{X}\trp)\trp$ and the components of the covariance matrix by $\Var[Y] = \sigma_\txt{Y}^2$, $\Cov[X,X] = \Sigma_\txt{XX}$, and $\Cov[Y,X] = \Sigma_\txt{YX}$, where $\Sigma_\txt{XX}$ is assumed non-singular with the inverse $\Sigma_\txt{XX}^{-1}$. Let $(Y_i,X_i\trp)\trp, i=1,2,\ldots,n,$ be independent and identically distributed (\IID) copies, that is, a random sample, of $(Y,X\trp)\trp$.
In this section, we consider linear predictor functions for $Y$, given $X = x_0$. 
For our purposes, we will only call a function a {\em predictor} if its value is computable given $X=x_0$, so it does not depend on {\em unknown} quantities.
%
Given $\tilde{Y}\in\mathcal{H}$, where $\mathcal{H}$  is a class of predictors of $Y$ based on $X$, we define the corresponding moments, if they exist, via $\mu_{\tilde{\txt{Y}}}   =  \E[\tilde{Y}(X)]$, $\sigma^2_{\tilde{\txt{Y}}}  =  \Var[\tilde{Y}(X)]$, and $\sigma_{\txt{Y}\tilde{\txt{Y}}}  =  \Cov[Y,\tilde{Y}(X)]$.	
Then, the $\CCC$ between $Y$ and $\tilde{Y}$ from (\ref{DEF_CCC}) is
\begin{equation}\label{DEF_CCC_AGREE}
\CCC[Y,\tilde{Y}] \ = \  \rho^c_{\txt{Y}\tilde{\txt{Y}}} \	= \ \frac{2\sigma_{\txt{Y}\tilde{\txt{Y}}}}{\sigma_\txt{Y}^2+\sigma^2_{\tilde{\txt{Y}}}+(\mu_\txt{Y}-\mu_{\tilde{\txt{Y}}})^2}.
\end{equation}
A CCC-based optimal predictor in the class $\mathcal{H}$, in the sense of maximizing agreement, is the element of  $\mathcal{H}$ maximizing $\CCC[Y,\tilde{Y}]$. 
Two classes of predictor functions that are of main interest are the class of {\em linear} predictors given by
{\begin{equation}\label{SPACE_LINEAR_PREDICTOR}
	\mathcal{H}_\txt{LP} \ = \ \left\{\tilde{Y}: \Re_p \rightarrow \Re \ | \ \tilde{Y}(x)=\alpha+\beta x:\;(\alpha,\beta)\in\mathbb{R}\times\mathbb{R}_p\right\},
\end{equation}
where $\beta = \Sigma_\txt{YX} \Sigma_\txt{XX}^{-1}$ is a row vector,} and the more general class of predictors given by
\begin{equation*}
	\mathcal{H}_\txt{P} \ = \ \left\{\tilde{Y}: \Re_p \rightarrow \Re \ | \ \E[\tilde{Y}^2(X)]<\infty \right\}.
\end{equation*}
Note that $\mathcal{H}_\txt{LP} \subset \mathcal{H}_\txt{P}$. In the developments below, we first assume that the joint distribution $F$ is completely known, so that there are no unknown parameters.

\begin{defn}\label{MAX_AGREEMENT_LIN_PREDICTOR}
	The maximum agreement linear predictor, denoted by $\tilde{Y}^\star(x_0)=\tilde{Y}^\star(x_0;\alpha,\beta) = \alpha + \beta x_0 \in  \mathcal{H}_\txt{LP}$, satisfies
$\rho^c_{\txt{Y}\tilde{\txt{Y}}^\star}  =  \sup_{\tilde{Y}\in\mathcal{H}_\txt{LP}}\rho^c_{\txt{Y}\tilde{\txt{Y}}}$;
whereas, the maximum agreement predictor, $\tilde{Y}^{\star\star}(x_0) \in \mathcal{H}_\txt{P}$, satisfies $\rho^c_{\txt{Y}\tilde{\txt{Y}}^{\star\star}}  =  \sup_{\tilde{Y}\in\mathcal{H}_\txt{P}}\rho^c_{\txt{Y}\tilde{\txt{Y}}}$.
\end{defn}

\citet{BKLLP:2022} established that, {provided $\Var[\E[Y|X]] > 0$,} the {\em unique} MAP is
\begin{equation}
    \label{MAP}
    \tilde{Y}^{\star\star}(x_0) = \E[Y] + \sqrt{\frac{\Var[Y]}{\Var[\E[Y|X]]}} (\E[Y|X=x_0] - \E[Y]).
\end{equation}
Thus, if the joint distribution $F$ satisfies the linearity property on the conditional mean given by
\begin{equation}
    \label{linearity conditional mean}
    \E[Y|X=x_0] =  \alpha + \beta x_0
\end{equation}
{with $\beta \ne 0$ and $X$ is not degenerate to guarantee that $\Var[\E[Y|X]] > 0$,} then the MAP becomes
$$\tilde{Y}^{\star\star}(x_0) = (\alpha + \beta \mu_\txt{X}) + \sqrt{\frac{\sigma_\txt{Y}^2}{\beta \Sigma_\txt{XX} \beta^{\trp}}} \beta (x_0 - \mu_\txt{X}).$$
The $(\alpha,\beta)$ minimizing the mapping $(\alpha,\beta) \mapsto \E[(Y - (\alpha + \beta X))^2]$, the least-squares optimization, are
$$\beta = \Sigma_\txt{YX} \Sigma_\txt{XX}^{-1} ;\  \alpha = \E[Y] -  \beta \E[X]; \ \mbox{and}\ 
\Var[\E[Y|X]] = \Sigma_\txt{YX} \Sigma_\txt{XX}^{-1} \Sigma_\txt{XY}.$$
{Note that condition (\ref{linearity conditional mean}) is not sufficient to always obtain $\beta = \Sigma_\txt{YX} \Sigma_\txt{XX}^{-1}$, {\em even} if we impose the homoscedasticity condition that $\Var[Y|X=x_0]$ does not depend on $x_0$. The least-squares optimization is one way to uniquely determine the values of $\alpha$ and $\beta$, but there could be other approaches, such as minimizing a weighted least-squares criterion or the expected absolute deviation criterion.}
The specific expressions for $\alpha$ and $\beta$ arising from the least-squares criterion lead to the least-squares linear predictor (LSLP) given by
\begin{equation}
    \label{LSLP}
\tilde{Y}^\dagger(x_0) = \mu_\txt{Y} +  \Sigma_\txt{YX} \Sigma_\txt{XX}^\txt{-1}(x_0 - \mu_\txt{X}),
\end{equation}
Now, if we use these $\alpha$ and $\beta$ in the expression for the MAP under condition (\ref{linearity conditional mean}), we obtain
\begin{equation}
    \label{MALP}
\tilde{Y}^\star(x_0) = \mu_\txt{Y} + \sqrt{\frac{\sigma_\txt{Y}^2}{\Sigma_\txt{YX} \Sigma_\txt{XX}^{-1} \Sigma_\txt{XY}}}  \Sigma_\txt{YX} \Sigma_\txt{XX}^{-1}  (x_0 - \mu_\txt{X}) =
\mu_\txt{Y} + \frac{1}{\gamma}  \Sigma_\txt{YX} \Sigma_\txt{XX}^{-1}  (x_0 - \mu_\txt{X}),
\end{equation}
where $\gamma^2 = {\Sigma_\txt{YX} \Sigma_\txt{XX}^{-1} \Sigma_\txt{XY}}/{\sigma_\txt{Y}^2}$ and  $\gamma = + \sqrt{\gamma^2}.$
Note that $\gamma^2$ is the coefficient of multiple determination associated with the multiple linear regression model that is related to the LSLP.
Since the predictor in (\ref{MALP}) belongs to the class $\mathcal{H}_\txt{LP}$, then this is also the MALP.

However, in general, the joint distribution $F$ need not satisfy the linearity property (\ref{linearity conditional mean}), but we may still opt to utilize linear predictors, that is, those predictors in the class $\mathcal{H}_\txt{LP}$ instead of in the more general class $\mathcal{H}_\txt{P}$.
There is then the question of whether, in the more restrictive class $\mathcal{H}_\txt{LP}$ and when condition (\ref{linearity conditional mean}) may not valid, if the MALP is as in (\ref{MALP}) and if it is unique. It should be pointed out that the question of uniqueness is {\em not} a trivial one since when one tries to find a predictor which maximizes its PCC with the predictand, then there exists an infinite number of maximizers, hence non-uniqueness ensues \citep{BKLLP:2022}. Theorem \ref{THM_MALP} below provides an affirmative answer to this uniqueness question and establishes that this MALP is indeed the predictor in (\ref{MALP}). 
\begin{theo}
\label{THM_MALP}
The MALP at $X=x_0$ in Definition \ref{MAX_AGREEMENT_LIN_PREDICTOR} is unique and given by (\ref{MALP}). In addition, the achieved maximum $\CCC$ has the value $\rho^\txt{c}_{\txt{Y}\tilde{\txt{Y}}^\star} = \gamma$.
\end{theo}

\begin{proof}
The optimization is implemented via Lagrange multipliers. For details regarding the optimization and the proof of uniqueness, {refer to Section A.1 in the Appendices.}
\end{proof}

We further point out that Theorems 1 and 2 in \citet{BKLLP:2022} proved the uniqueness of the MAP in $\mathcal{H}_\txt{P}$ and that both the MAP and the least-squares predictor $x_0 \mapsto \E[Y|X=x_0]$ also maximize the PCC with the predictand among the predictors in $\mathcal{H}_\txt{P}$. Additional observations are as follows (see also \citet{BKLLP:2022} and \citet{Chri:2022}).
%
%
From (\ref{MALP}) and (\ref{LSLP}), observe that $\sigma_{\tilde{\txt{Y}}^\dagger}^2  =  \Var[\tilde{Y}^\dagger(X)] = \Sigma_{\txt{YX}} \Sigma_{\txt{XX}}^{-1} \Sigma_{\txt{XY}}$ and $\sigma_{\tilde{\txt{Y}}^\star}^2  =  \Var[\tilde{Y}^\star(X)] = \sigma_\txt{Y}^2$,
so that $$\Var[\tilde{Y}^\star(X)] \ge \Var[\tilde{Y}^\dagger(
X)].$$ 
Thus, with $\rho_{\txt{Y}\tilde{\txt{Y}}^\dagger}=\sigma_{\tilde{\txt{Y}}^\dagger}/\sigma_\txt{Y} = \gamma$, note that the MALP can be expressed in terms of the LSLP via $\tilde{Y}^\star(x_0)  \ = 
  \  \left(1-\frac{1}{\gamma}\right)\mu_\txt{Y}  +  \frac{1}{\gamma}\tilde{Y}^\dagger(x_0)$. 
This amounts to a re-centering and a re-scaling of the LSLP so its mean is equal to $E(Y)$ and its variance is equal to $\Var[Y]$. This is referred to as the ``calibration approach'' to obtaining the MALP from the LSLP, and this provides a computational advantage in practice since $\tilde{Y}^\dagger(x_0)$ can be obtained conveniently from existing statistical packages.

{In fact, under condition (\ref{linearity conditional mean}), the MALP, aside from maximizing the $\CCC[Y,\hat{Y}]$ among all $\hat{Y} \in \mathcal{H}_\txt{P}$, can also be obtained by minimizing the mean-squared prediction error between the predictor and the predictand, {\em under} the conditions that $\E[\hat{Y}(X)] = \E[Y]$ and $\Var[\hat{Y}(X)] = \Var[Y]$, among all predictors $\hat{Y}$ in $\mathcal{H}_\txt{P}$; see \citet{Chri:2022}. A similar constrained optimization idea is also found in the constrained kriging approach in spatial statistics; see \citet{Cres:2015}. 
In \citet{Ghos:1992}, the MALP has also appeared in a simultaneous Bayesian estimation context, where it is called a constrained Bayes predictor. Additional related works can be found in the measurement error literature, specifically the moment reconstruction method \cite{FreEtAl2004} and the moment-adjusted imputation method \cite{ThoSteDav2011}.
}

\subsection{Estimated MALP and LSLP: Unknown Distribution}
\label{SECTION_INFER}

In the preceding subsection, we assumed that the joint distribution $F$ is completely known, hence there are no unknown parameters. Clearly that setting is unrealistic, hence we consider in this subsection the setting where the joint distribution is not known, or if its functional form is known, that there are still unknown parameters. In this case, the MALP in (\ref{MALP}) and the LSLP in (\ref{LSLP}) are not legitimate predictors in the sense that their values are not computable..

To obtain the associated predictors, we replace the unknown parameters in the MALP and the LSLP by their empirical counterparts or their estimators based on the sample data. 
For brevity, denote this random sample by $(\mathbf{Y},\mathbf{X})$, and let the associated sample mean vectors and covariance matrices  be
\begin{eqnarray*}
& \bar{Y} = \frac{1}{n} \sum_{i=1}^n Y_i;\; \bar{X} = \frac{1}{n} \sum_{i=1}^n X_i;\;S_\txt{Y}^2 = \frac{1}{n} \sum_{i=1}^n (Y_i - \bar{Y})^2;\; & \\
&  S_\txt{XX} = \frac{1}{n} \sum_{i=1}^n (X_i - \bar{X})(X_i - \bar{X})\trp;\;
S_\txt{XY} = S_\txt{YX}\trp = \frac{1}{n} \sum_{i=1}^n (Y_i - \bar{Y})(X_i - \bar{X})\trp.  &  
\end{eqnarray*}
Respectively, $(\bar{Y},\bar{X},S_\txt{Y}^2,S_\txt{XX},S_\txt{YX})$ are the method-of-moments estimators (MMEs) of $(\mu_\txt{Y}, \mu_\txt{X}, \sigma_\txt{Y}^2, \Sigma_\txt{XX}, \Sigma_\txt{YX})$. Substituting these estimators for their counterparts in the MALP and LSLP, the {\em estimated} maximum agreement linear predictor (EMALP) of $Y$ at $X = x_0$, based on $(\mathbf{Y},\mathbf{X})$, is then
\begin{equation}
\label{EMALP}
\hat{Y}^\star(x_0;(\mathbf{Y},\mathbf{X})) \equiv \hat{Y}^\star(x_0) = \bar{Y} + \sqrt{\frac{S_\txt{Y}^2}{S_\txt{YX} S_\txt{XX}^{-1} S_\txt{XY}}} S_\txt{YX} S_\txt{XX}^{-1} (x_0 - \bar{X})
= \bar{Y} + \frac{1}{\hat{\gamma}} S_\txt{YX} S_\txt{XX}^{-1} (x_0 - \bar{X})
\end{equation}
where
$\hat{\gamma}^2 = {S_\txt{YX} S_\txt{XX}^{-1} S_\txt{XY}}/{S_\txt{Y}^2}$ and $\hat{\gamma} =  \sqrt{\hat{\gamma}^2}.$
Likewise, the {\em estimated} least-squares linear predictor (ELSLP) of $Y$ at $X = x_0$, based on $(\mathbf{Y},\mathbf{X})$, is
\begin{equation}
\label{ELSLP}
\hat{Y}^\dagger(x_0;(\mathbf{Y},\mathbf{X})) \equiv \hat{Y}^\dagger(x_0) = \bar{Y} + S_\txt{YX} S_\txt{XX}^{-1} (x_0 - \bar{X}).
\end{equation}
Observe that the EMALP is a linear combination of $\bar{Y}$ and the ELSLP via the equation
$$\hat{Y}^\star(x_0;(\mathbf{Y},\mathbf{X}))  = \left(1 - \frac{1}{\hat{\gamma}}\right) \bar{Y} + \left(\frac{1}{\hat{\gamma}}\right) \hat{Y}^\dagger(x_0;(\mathbf{Y},\mathbf{X})).$$
It is certainly possible that we could improve statistical properties of these estimated predictors by using estimators of the relevant parameters that are more efficient than the MMEs, such as the maximum likelihood estimators (MLEs). On the other hand, when $(Y,X)$ has a multivariate normal distribution (MVN), then the MMEs above coincide with the MLEs. In addition, since under the MVN joint distribution, condition (\ref{linearity conditional mean}), regarding the linearity of the conditional mean of $Y$ given $X=x_0$, holds, then the MALP is in fact also the MAP, and the resulting EMALP will also be the {\em estimated} MAP.

In the next sections, we examine the asymptotic and finite-sample properties of the EMALP $\hat{Y}^\star(x_0)$ as the sample size $n \rightarrow \infty$. Finite-sample properties will be studied through computer simulation studies. Since asymptotic and finite-sample  properties of the ELSLP $\hat{Y}^\dagger(x_0)$ are well-studied, we will not focus much, nor provide many details, in obtaining properties of the ELSLP. However, we will still make comparisons between the EMALP and the ELSLP, since the ELSLP is the most well-known predictor, hence serves as a baseline predictor for which other predictors can be compared.

\section{Asymptotic Properties of EMALP and ELSLP}
\label{sec: Asymptotics}

\subsection{Consistency and Asymptotic Normality}
\label{subsec: asy normality of EMALP and ELSLP}

By invoking Kolmogorov's Strong Law of Large Numbers (SLLN) and under the condition that fourth moments are finite, $\Sigma_\txt{XX}$ is nonsingular, and $\Sigma_\txt{YX} \ne 0$ (cf., \citep{Res14})), it is immediate that as $n \rightarrow \infty$,
$$\hat{Y}^\star(x_0) \conas \tilde{Y}^\star(x_0) \equiv \mu_\txt{Y} + \sqrt{\frac{\sigma_\txt{Y}^2}{\Sigma_\txt{YX} \Sigma_\txt{XX}^{-1} \Sigma_\txt{XY}}} \Sigma_\txt{YX} \Sigma_\txt{XX}^{-1} (x_0 - \mu_\txt{X}).$$
As such, $\hat{Y}^\star(x_0)$ is strongly, hence also weakly (or in probability), consistent for $\tilde{Y}^\star(x_0)$, the population-level MALP. The other asymptotic property of major interest is the asymptotic distribution of $\hat{Y}^\star(x_0)$. For the purpose of obtaining its asymptotic distribution, we present a result below that will provide simplifications in the subsequent derivations. 
%

\begin{prop} 
\label{prop-reduction}
Let $(\mathbf{Y},\mathbf{X})$ be an \IID\ sample $\{(Y_i,X_i\trp)\trp, i=1,2,\ldots,n\}$ from a distribution $F_\txt{(Y,X)}(y,x)$ with mean vector $\mu_\txt{(Y,X)} = \begin{bmatrix}[0.5] \mu_\txt{Y} \\ \mu_\txt{X} \end{bmatrix}$, covariance matrix $\Sigma_\txt{(Y,X)} = \begin{bmatrix}[0.5] \sigma_\txt{Y}^2 & \Sigma_\txt{YX} \\ \Sigma_\txt{XY} & \Sigma_\txt{XX}\end{bmatrix}$, where $\Sigma_\txt{XX}$ is nonsingular with inverse $\Sigma_\txt{XX}^{-1}$ and $\Sigma_\txt{YX} \ne 0$. For $x_0 \in \mathcal{X} \subseteq \Re_p$ and defining $w_0 = \Sigma_\txt{XX}^{-1/2} (x_0 - \mu_\txt{X})$, we have
$$\hat{Y}^\star(x_0;(\mathbf{Y},\mathbf{X})) = \mu_\txt{Y} + \sigma_\txt{Y} \hat{Y}^\star(w_0;(\mathbf{V},\mathbf{W})),$$
where $(\mathbf{V},\mathbf{W})$ represents $\{(V_i,W_i\trp)\trp, i=1,2,\ldots,n\},$ with
$V_i = (Y_i - \mu_\txt{Y})/\sigma_\txt{Y} \quad \mbox{and} \quad W_i = \Sigma_\txt{XX}^{-1/2} (X_i - \mu_\txt{X}),$
which are \IID\ from the distribution 
%
$F_\txt{(V,W)}(v,w;\mu_\txt{(Y,X)},\Sigma_\txt{(Y,X)}) = F_\txt{(Y,X)}(\mu_\txt{Y} + \sigma_\txt{Y} v, \mu_\txt{X} + \Sigma_\txt{XX}^{1/2} w),$
%
 whose mean vector is $\mu_\txt{(V,W)} = \begin{bmatrix}[0.5] 0 \\ 0 \end{bmatrix}$ and covariance matrix $\Sigma_\txt{(V,W)} = \begin{bmatrix}[0.5] 1 & \Omega\trp \\ \Omega & I\end{bmatrix}$ with $$\Omega = \Sigma_\txt{XX}^{-1/2} \Sigma_\txt{XY} \sigma_\txt{Y}^{-1}.$$
\end{prop}

\begin{proof}
This follows immediately from the transformations $V_i = (Y_i - \mu_\txt{Y})/\sigma_\txt{Y}$ and  $W_i = \Sigma_\txt{XX}^{-1/2} (X_i - \mu_\txt{X})$, the location-scale equivariance properties of the sample mean vector and the sample covariance matrix, and the expressions for $\hat{Y}^\star(x_0; (\mathbf{Y},\mathbf{X}))$ and $\hat{Y}^\star(w_0; (\mathbf{V},\mathbf{W}))$.
\end{proof} 

Because of Proposition \ref{prop-reduction}, it suffices to establish the asymptotic distribution under the distribution of $(V,W)$, that is, it suffices to obtain the asymptotic distribution of
$$\hat{Y}^\star(w_0;(\mathbf{V},\mathbf{W})) = \bar{V} + \sqrt{\frac{S_\txt{V}^2}{S_\txt{VW} S_\txt{WW}^{-1} S_\txt{WV}}} S_\txt{VW} S_\txt{WW}^{-1} (w_0 - \bar{W})$$
when $(\mathbf{V},\mathbf{W}) \equiv \{(V_i,W_i\trp)\trp, i=1,2,\ldots,n\},$ are \IID\ from the joint distribution $F_\txt{(V,W)}(\cdot,\cdot; \mu_\txt{(Y,X)},\Sigma_\txt{(Y,X)})$ in Proposition \ref{prop-reduction}. Moments of orders three and four will depend on $(\mu_\txt{YX},\Sigma_\txt{YX})$ and the specific form of the distribution of $F_\txt{(Y,X)}$.

Let $(V,W\trp)\trp$ as in Proposition \ref{prop-reduction} and denote by $\mathcal{V} \subseteq \Re$ the range space of $V$ and $\mathcal{W} \subseteq \Re_p$ the range space of $W$. Define the object (which could be viewed as a tensor or a list structure)
$$T = \{V,W,V^2,W^{\otimes 2},VW\}.$$
We shall let $\mbox{VEC()}$ be the vectorization operator, so for an $r \times c$ matrix $M =\begin{bmatrix}[0.5] m_1&m_2&\cdots&m_c\end{bmatrix}$, where $m_j$, $j=1,2,\ldots,c,$ are the column vectors, 
$\VEC{M} \equiv \begin{bmatrix}[0.5]m_1\trp\;&m_2\trp\;&\cdots\;&m_c\trp\end{bmatrix}\trp,$
an $rc$-dimensional column vector. Define the column vector 
$$T^* = \VEC{T} = \begin{bmatrix}V\;&W\trp\;&V^2\;&\VEC{W^{\otimes 2}}\trp\;&W\trp V\end{bmatrix}\trp.$$
%
The mean vector and covariance matrix of $T^*$ will be denoted, respectively, by
\begin{align*}\nu \equiv \E[T^*] = \begin{bmatrix}0\;&0\trp\;&1\;&\VEC{I}\trp\;&\Omega\trp\end{bmatrix}\trp
\;\text{ and }\;
\Gamma  \equiv  \Cov[T^*,T^*]  =  \begin{bsmallmatrix}
1 & \Omega\trp & \Gamma_{13} & \Gamma_{14} & \Gamma_{15} \\
\Omega & I & \Gamma_{23} & \Gamma_{24} & \Gamma_{25} \\
\Gamma_{31} & \Gamma_{32} & \Gamma_{33} & \Gamma_{34} & \Gamma_{35} \\
\Gamma_{41} & \Gamma_{42} & \Gamma_{43} & \Gamma_{44} & \Gamma_{45} \\
\Gamma_{51} & \Gamma_{52} & \Gamma_{53} & \Gamma_{54} & \Gamma_{55} 
\end{bsmallmatrix},
\end{align*}
where $\Gamma_{rc} = \Cov[T_r^*,T_c^*]$ for $r,c=1,2,\ldots,5$ are symmetric covariance matrices that depend on the specific joint distribution.
In terms of $V$ and $W$, we have these elements as follows:
\begin{align*}\Gamma_{13} =& \Cov\left[V,V^2\right];\; \Gamma_{14} = \Cov\left[V,\VEC{W^{\otimes 2}}\right];\; \Gamma_{15} = \Cov[V,WV];\\
\Gamma_{23} =&  \Cov\left[W,V^2\right];\; \Gamma_{24} = \Cov\left[W,\VEC{W^{\otimes 2}}\right];\; 
\Gamma_{25} = \Cov[W,WV]; \\
\Gamma_{33} =& \Cov\left[V^2,V^2\right];\;
\Gamma_{34} = \Cov\left[V^2,\VEC{W^{\otimes 2}}\right];\;
\Gamma_{35} = \Cov\left[V^2,WV\right]; \\
\Gamma_{44} =&
\Cov\left[\VEC{W^{\otimes 2}},\VEC{W^{\otimes 2}}\right];\;
\Gamma_{45} = \Cov\left[\VEC{W^{\otimes 2}},WV\right];\;
\Gamma_{55} =  \Cov[WV,WV].
\end{align*}
%
%
As such the elements of $\Gamma$ are functions of the joint moments of $V$ and $W$ up to the fourth order, that is, in terms of the joint moments of form
$$\E\left[V^a W_j^b W_k^c\right]\  \mbox{for}\ a, b, c \in \{0,1,2,3,4\}; a + b + c \le 4; j,k = 1,2,\ldots,p.$$ 
In particular, measures of skewness and kurtosis of $V$ and $W$ play roles in these expressions. But, note that each $W_j, j=1,2,\ldots,p,$ is not just a function of $X_j$, but rather, it is a linear function of {\em all} the $X_k, k=1,2,\ldots,p$.
The computations of the elements of $\Gamma$ may be accomplished in terms of the untransformed $(Y,X)$ if the joint distribution $F$ does not belong to a location-scale family of distributions. In terms of $(Y,X)$, we have that (using the notation $\Cov[Q] \equiv \Cov[Q,Q]$ for a column vector $Q$)
\begin{displaymath}
    \Gamma = 
    \Cov\left[
    \begin{array}{c}
    \sigma_\txt{Y}^{-1}(Y-\mu_\txt{Y}) \\
    \Sigma_\txt{XX}^{-1/2}(X - \mu_\txt{X}) \\
    \sigma_\txt{Y}^{-2} (Y - \mu_\txt{Y})^2 \\
    \VEC{\Sigma_\txt{XX}^{-1/2}(X - \mu_\txt{X})^{\otimes 2} \Sigma_\txt{XX}^{-1/2}} \\
     \Sigma_\txt{XX}^{-1/2}(X - \mu_\txt{X}) (Y-\mu_\txt{Y}) \sigma_\txt{Y}^{-1}
    \end{array}
    \right].
\end{displaymath}
For example, in terms of $Y$ and $X$, we have
\begin{eqnarray*}
\Gamma_{14} & = & \Cov\left[V,\VEC{W^{\otimes 2}}\right] =
\Cov\left[ \sigma_\txt{Y}^{-1}(Y-\mu_\txt{Y}),\VEC{\Sigma_\txt{XX}^{-1/2}(X - \mu_\txt{X})^{\otimes 2} \Sigma_\txt{XX}^{-1/2}} \right] \\
& = &  \left(\frac{1}{\sigma_\txt{Y}}
\VEC{
\Sigma_\txt{XX}^{-1/2} \left\{\Cov\left[Y,X^{\otimes 2}\right] - 2 \mu_\txt{X} \Sigma_\txt{YX}   \right\} \Sigma_\txt{XX}^{-1/2} }\right)\trp.   
\end{eqnarray*}
{Before proceeding further, we comment that an alternative approach to deriving the asymptotic results is to directly deal with the $(Y,X)$ instead of first making the transformation into $(V,W)$. Both approaches have their distinct advantages and disadvantages, but they lead to the same asymptotic results for the EMALP and ELSLP.}

Let $(V_i,W_i), i=1,2,\ldots,n$, be \IID\ copies of $(V,W)$, and let $T_i, i=1,2,\ldots,n$, be the associated objects, and define the object
$$A = \{A_1,A_2,A_3,A_4,A_5\} = \frac{1}{n} \sum_{i=1}^n T_i.$$
Thus,
$$A_1 = \bar{V},\; A_2 = \bar{W},\; A_3 = \frac{1}{n} \sum_{i=1}^n V_i^2,\; A_4 = \frac{1}{n} \sum_{i=1}^n W_i^{\otimes 2},\; A_5 = \frac{1}{n} \sum_{i=1}^n W_iV_i.$$
Recall that a sequence of random variables $\{R_n, n=1,2,\ldots\}$ is said to be asymptotically normal with asymptotic mean $\eta_n$ and asymptotic variance $\tau^2$, abbreviated $R_n \sim \mathcal{AN}\left(\eta_n,\frac{1}{n}\tau^2\right)$ if $\sqrt{n}(R_n - \eta_n)/\tau$ converges in distribution to a standard normal random variable. The following lemma now follows by the Multivariate Central Limit Theorem for \IID\ random vectors (cf., \citep{Res14}).
\begin{lemm} 
\label{lemmaA}
As $n\longrightarrow \infty$, $\VEC{A} \sim \mathcal{AN}\left(\nu,\frac{1}{n} \Gamma\right)$.
\end{lemm}

Next, define the object $B =\{B_1,B_2,B_3,B_4,B_5\}$
with components
\begin{eqnarray*}
B_1  =  A_1,\;
B_2  =  A_2,\; 
B_3 =  A_3 - A_1^2,\;
B_4 =  A_4 - A_2^{\otimes 2},\;
B_5  =  A_5 - A_2 A_1.
\end{eqnarray*}
Thus, note that $B_1 = \bar{V}$, $B_2 = \bar{W}$, $B_3 = S_\txt{V}^2$, $B_4 = S_\txt{WW}$, and $B_5 = S_\txt{WV}$.

\begin{lemm}
\label{lemmaB}
As $n \longrightarrow \infty$, $\VEC{B} \sim \mathcal{AN}\left(\nu,\frac{1}{n}\Gamma\right)$.
\end{lemm}

\begin{proof}
This follows immediately from Lemma \ref{lemmaA}, the Delta Method \citep{Res14}, and since the transformation from $\VEC{A}$ to $\VEC{B}$ has Jacobian matrix equal to the Identity matrix.
\end{proof}

Next, for $w_0 \in \mathcal{W}$, define the mapping $g(w_0;B)$ via
$$C(w_0) = g(w_0;B) = B_1 + \sqrt{\frac{B_3}{B_5\trp B_4^{-1} B_5}} B_5\trp B_4^{-1} (w_0 - B_2).$$
Note that $\hat{Y}^\star(w_0;(\mathbf{V},\mathbf{W})) = g(w_0;B) = C(w_0)$.

\begin{lemm}
\label{Jacobian}
With $\gamma^2 = \Omega\trp\Omega = ||\Omega||^2$ and $\gamma = +\sqrt{\gamma^2}$, and under the condition that $\Omega \ne 0$ or $\gamma \ne 0$, the Jacobian (row) vector of the transformation from $B$ to $C(w_0)$ is
\begin{eqnarray*}
J(w_0) & = & \left.\frac{\partial g(w_0;B)}{\partial\VEC{B}\trp}\right\vert_{B=\nu} \\ & = &  \left[ \begin{matrix}
1\;&\; -\frac{\Omega\trp}{\gamma}\;&\;\frac{1}{2} \frac{(\Omega\trp w_0)}{\gamma}\;&\; -\frac{(\Omega\trp \otimes w_0\trp)}{\gamma} + \frac{1}{2} \frac{(\Omega\trp w_0)}{\gamma^3} (\Omega\trp \otimes \Omega\trp)\;&\;
\frac{w_0\trp}{\gamma} - \frac{(\Omega\trp w_0)}{\gamma^3} \Omega\trp\end{matrix} \right].
\end{eqnarray*}
\end{lemm}

\begin{proof}
The expression is obtained by using the chain rule of differentiation and matrix differentiation results in \citet[pp. 204--205]{MN:2019}. Key results used in obtaining this Jacobian vector are
\begin{align*}
\left.\tfrac{\partial}{\partial (\VEC{B_4})\trp}(B_5\trp B_4^{-1} (w_0 - B_2))\right\vert_ {B = \nu} =&  -(\Omega\trp \otimes w_0\trp);\\ 
\left.\tfrac{\partial}{\partial (\VEC{B_4})\trp}(B_5\trp B_4^{-1} B_5)\right\vert_{ B = \nu} =&  -(\Omega\trp \otimes \Omega).
\end{align*}
\end{proof}

\begin{theo}[Asymptotic Normality]
\label{prop: asy normality 1}
If $\Omega \ne 0$, then as $n \rightarrow \infty$, $$\hat{Y}^\star(w_0;(\mathbf{V},\mathbf{W})) = C(w_0) \sim \mathcal{AN}\left(\frac{\Omega\trp w_0}{\gamma},\frac{1}{n} \Xi(w_0)\right),$$
where
$\Xi(w_0) = J(w_0) \Gamma J(w_0)\trp.$
\end{theo}

\begin{proof}
This follows immediately from the Delta Method and Lemma \ref{Jacobian}.
\end{proof}

\begin{coro}
\label{coro: asy normality 2}
If $\Sigma_\txt{XX}$ is nonsingular and $\Sigma_\txt{YX} \ne 0$, then for $x_0 \in \mathcal{X}$, as $n \longrightarrow \infty$, $$\hat{Y}^\star(x_0;(\mathbf{Y},\mathbf{X})) \sim \mathcal{AN}\left(\mu_\txt{Y} + \sigma_\txt{Y} \frac{\Omega\trp w_0}{\gamma}, \frac{1}{n} \sigma_\txt{Y}^2 \Xi(w_0)\right)$$
where $w_0 = \Sigma_\txt{XX}^{-1/2} (x_0 - \mu_\txt{X})$ and $\Omega = \Sigma_\txt{XX}^{-1/2} \Sigma_\txt{XY} \sigma_\txt{Y}^{-1}$.
\end{coro}

\begin{proof}
    This follows immediately from Proposition \ref{prop-reduction} and Theorem \ref{prop: asy normality 1}.
\end{proof}

Also, as mentioned earlier, we will not dwell much on the properties of the ELSLP which are well-known (e.g., see \citet{Rao:1973}). We simply state the following results. Strong consistency again follows from Kolmogorov's SLLN so we have
    $$\hat{Y}^\dagger(x_0;(\mathbf{Y},\mathbf{X})) \  \mbox{converges strongly to}\  \tilde{Y}^\dagger(x_0) = \mu_\txt{Y} + \Sigma_\txt{YX} \Sigma_\txt{XX}^{-1} (x_0 - \mu_\txt{X}).$$
\begin{theo}[Asymptotics of ELSLP]
    \label{theo: asymptotics for ELSLP}
    Under the conditions of Proposition \ref{prop-reduction}, as $n \rightarrow \infty$,
    %
    $$\hat{Y}^\dagger(w_0;(\mathbf{V},\mathbf{W})) \equiv \bar{V} + S_\txt{VW} S_\txt{WW}^{-1} (w_0 - \bar{W}) \sim \mathcal{AN}\left(\Omega\trp w_0,\frac{1}{n} \Xi_\txt{LS}(w_0)\right)$$
    where $\Xi_\txt{LS}(w_0) = J_\txt{LS}(w_0) \Gamma (J_\txt{LS}(w_0))\trp$ with the Jacobian vector equal to
    $$J_\txt{LS}(w_0) = \left[ 1\quad -\Omega\trp\quad 0 \quad -(\Omega\trp \otimes w_0\trp)\quad  w_0\trp \right].$$
    As such, the ELSLP has the asymptotic distribution
    $$\hat{Y}^\dagger(x_0;(\mathbf{Y},\mathbf{X})) \sim \mathcal{AN}\left(\tilde{Y}^\dagger(x_0) = \mu_\txt{Y} + \Sigma_\txt{YX} \Sigma_\txt{XX}^{-1} (x_0 - \mu_\txt{X}),\frac{1}{n} \sigma_\txt{Y}^2 \Xi_\txt{LS}(w_0)\right).$$
\end{theo}

{Earlier we have seen that $\Var[\tilde{Y}^\star(x_0)] \ge \Var[\tilde{Y}^\dagger(x_0)]$. We conjecture that this inequality also holds true for the asymptotic variances of the EMALP and ELSLP under a general joint distribution $F$ for $(Y,X)$, that is, $\AVar[\hat{Y}^\star(x_0)] \ge \AVar[\hat{Y}^\dagger(x_0)]$. Later, we actually establish this inequality for the asymptotic variances under the special case where $(Y,X)$ has a multivariate normal distribution, by applying the Cauchy-Schwartz Inequality on the closed-form asymptotic variances of the EMALP and ELSLP (see the comment immediately after Corollary \ref{coro: asymptotics for ELSLP under MVN} on page \pageref{coro: asymptotics for ELSLP under MVN}). The intuition behind this general conjecture is that when $\hat{\gamma}$ is close to 1, the two predictors are almost equal. When $\hat{\gamma}$ is not close to 1, then the EMALP has the additional component $1/\hat{\gamma}$ relative to the ELSLP which will introduce more variability into the EMALP relative to the ELSLP. As yet another reason why the general conjecture may hold is because the variance of the ELSLP will tend to be smaller than the variance of $Y$, whereas the variance of the EMALP is forced to be close to the variance of $Y$. However, we have not {\em yet} found a proof for this general conjecture, which is tantamount to showing that $\Xi(w_0) \ge \Xi_\txt{LS}(w_0)$.}

{Throughout the manuscript, our setting for prediction problems is that both $X$ and $Y$ are random. Another framework, which is more appropriate with designed experiments, is to view the $X$ as fixed, rather than random. This is actually an easier problem to deal with, though when considering asymptotic results, certain restrictions on the sequence of $X$ matrices as $n \rightarrow \infty$  are required, such as $\bar{X}$, $S_{\txt{X}\txt{X}}$, and $S_\txt{YX}$  converging in probability to some vectors or matrices $\mu_\txt{X}$, $\Sigma_{\txt{X}\txt{X}}$, and $\Sigma_\txt{YX}$, respectively.}

\subsection{Consistent Estimators of the Asymptotic Variances}
\label{subsec: estimating asymptotic variance}

To perform other statistical procedures, such as constructing prediction intervals, there is a need to estimate the asymptotic variances of the EMALP and the ELSLP. Consistent estimators of these variances are obtained by consistently estimating the Jacobian vectors $J(w_0)$ and $J_\txt{LS}(w_0)$ and also $\Gamma$. To do so given the sample data $(\mathbf{Y},\mathbf{X}) = \{(Y_i,X_i): i=1,2,\ldots,n\}$, we first transform to $(\hat{\mathbf{V}},\hat{\mathbf{W}}) = \{(\hat{V}_i,\hat{W}_i): i=1,2,\ldots,n\}$ via the transformations
$$\hat{V}_i = \frac{Y_i - \bar{Y}}{S_\txt{Y}} \quad \mbox{and} \quad \hat{W}_i = S_\txt{XX}^{-1/2}(X_i - \bar{X}).$$
We then form for $i=1,2,\ldots,n,$ the vectors
$$\hat{T}_i^* = \left(\hat{V}_i,\hat{W}_i\trp,\hat{V}_i^2,\VEC{(\hat{W}_i)^{\otimes 2})}\trp,(\hat{W}_i\hat{V}_i)\trp\right).$$
With 
$\bar{{\hat{T}}}^* = \frac{1}{n} \sum_{i=1}^n \hat{T}_i^*,
$
the sample mean of the $\hat{T}_i^*$, a consistent estimator of $\Gamma$ is the sample covariance matrix associated with $\{\hat{T}_i^*, i=1,2,\ldots,n\}$, given by
\begin{equation}
\label{estimator of Gamma}
\hat{\Gamma} = \frac{1}{n} \sum_{i=1}^n \left(\hat{T}_i^* - \bar{\hat{T}}^*\right)^{\otimes 2} = \frac{1}{n} \sum_{i=1}^n \left(\hat{T}_i^*\right)^{\otimes 2} - \left(\bar{\hat{T}}^*\right)^{\otimes 2}.
\end{equation}
The $(2,1)$th element of $\hat{\Gamma}$ is $\hat{\Omega}$, a consistent estimator of $\Omega$. With $\hat{w}_0 = S_\txt{XX}^{-1/2} (x_0 - \bar{X})$, consistent estimators $\hat{J}(\hat{w}_0)$ and $\hat{J}_\txt{LS}(\hat{w}_0)$ of $J(w_0)$ and $J_\txt{LS}(w_0)$, respectively, are obtained by plugging in $\hat{\Omega}$ for $\Omega$ and $\hat{w}_0$ for $w_0$ in the respective expressions of these Jacobian vectors. Consistent estimators of the asymptotic variances of the EMALP and ELSLP are then, respectively,
$$\hat{\sigma}_\txt{MA}^2(x_0) = S_\txt{Y}^2 \hat{J}(\hat{w}_0) \hat{\Gamma} (\hat{J}(\hat{w}_0))\trp \quad \mbox{and} \quad \hat{\sigma}_\txt{LS}^2(x_0) = S_\txt{Y}^2 \hat{J}_\txt{LS}(\hat{w}_0) \hat{\Gamma} (\hat{J}_\txt{LS}(\hat{w}_0))\trp$$
where $S_\txt{Y}^2$ is the sample variance of $Y_i, i=1,2,\ldots,n$.

\subsection{Special Cases}
\label{subsec: special cases}

In this section we provide explicit expressions of $\Xi(w_0)$ for a variety of situations. Recall that if we know the $\Xi(w_0)$, then we have the complete characterization of the asymptotic distribution of $\hat{Y}^\star(w_0;(\mathbf{V},\mathbf{W}))$, as well as $\hat{Y}^\star(x_0;(\mathbf{Y},\mathbf{X}))$.


\subsubsection{Case of \texorpdfstring{$p=1$}{p=1}}

For notation, for $a,b \in \{0,1,2,3,4\}$, let
$$\eta_{a,b} = \E\left[V^a W^b\right] = \frac{1}{\sigma_\txt{Y}^a \sigma_\txt{X}^b} \E\left[(Y-\mu_\txt{Y})^a (X-\mu_\txt{X})^b\right].$$
Thus, we have
$\eta_{1,0} = \eta_{0,1} = 0;\; \eta_{2,0} = 1 = \eta_{0,2} = 1;\; \eta_{1,1} = \rho = \Cor[V,W].$
Note that under certain distributions, such as the location-scale family, the computations of these elements of $\Gamma$ will be simplified. If the distribution is not of the location-scale family type, then the calculations may not be so simple. For the case $p=1$, the symmetric matrix $\Gamma = \Cov[T^*,T^*]$ becomes equal to
\begin{equation}
\label{Gamma p 1}
\Gamma=\begin{bmatrix}[0.5]
\;\;1\;\; & \;\;\rho\;\;\;\; & \eta_{3,0}\phantom{-1} & \eta_{1,2}\phantom{-1} & \eta_{2,1}\phantom{- \rho^2} \\
& \;\;1\;\;\;\; & \eta_{2,1}\phantom{-1} & \eta_{0,3}\phantom{-1} & \eta_{1,2}\phantom{- \rho^2} \\
&& \eta_{4,0}-1 & \eta_{2,2} - 1 & \eta_{3,1} - \rho\phantom{^2} \\
&&& \eta_{0,4} - 1 & \eta_{1,3} - \rho\phantom{^2} \\
&&&& \eta_{2,2} - \rho^2
\end{bmatrix}.
\end{equation}
The Jacobian vector in Lemma \ref{Jacobian} becomes, after simplifying,
\begin{displaymath}
J(w_0) = \begin{bmatrix}1\;&\; -\Sgn(\rho)\;&\; \frac{1}{2}\Sgn(\rho)w_0\;&\; -\frac{1}{2} \Sgn(\rho) w_0\;&\; 0\end{bmatrix}.
\end{displaymath}

\begin{prop}[Under a General $F$]
\label{prop: Xi when p=1, general distribution}
If $p=1$ and for any distribution of $(V,W)$ such that fourth moments are finite, $\E[V] = \E[W] = 0$, $\Var[V] = \Var[W] = 1$, $\Cor[V,W] = \rho \ne 0$, and $\gamma = |\rho|$, we have that
$$\Xi(w_0) = 2(1-\gamma) + \frac{w_0^2}{4} [\eta_{4,0} + \eta_{0,4} - 2\eta_{2,2}] + \Sgn(\rho) w_0 [\eta_{3,0} - \eta_{1,2}] + w_0 [\eta_{0,3} - \eta_{2,1}].$$
\end{prop}

\begin{proof}
Plugging in $J(w_0)$ and $\Gamma$ in $\Xi(w_0) = J(w_0) \Gamma J(w_0)\trp$, then simplifying, yields the expression for $\Xi(w_0)$ in the statement of the Proposition.
\end{proof}

When the distribution of $(V,W)$ is bivariate normal, we obtain:

\begin{prop}[Under $F = BVN$]
\label{prop: Xi when p=1, normal distribution}
If, in Proposition \ref{prop: Xi when p=1, general distribution}, the distribution of $(V,W)$ is bivariate normal, then
$$\Xi(w_0) = (1 - \gamma^2) \left(\frac{2}{1+\gamma} + w_0^2\right) = (1-\rho^2) \left(\frac{2}{1+|\rho|} + w_0^2\right).$$
\end{prop}

\begin{proof}
Under the condition that
$(V,W) \sim \mathcal{BVN}\left([0,0],\begin{bmatrix}[0.5]1 & \rho \\ \rho & 1\end{bmatrix}\right)$
we easily obtain, or recalling well-known results, that
$\eta_{3,0} = \eta_{0,3} = 0;\; \eta_{1,2} = \eta_{2,1} = 0;\; \eta_{2,2} = 1 + 2\rho^2;\; \eta_{4,0} = \eta_{0,4} = 3.$
Substituting these values into the expression for $\Xi(w_0)$ in Proposition \ref{prop: Xi when p=1, general distribution}, then simplifying, yields the expression for $\Xi(w_0)$ in the statement of the Proposition.
\end{proof}

\begin{coro}
\label{coro: asy normality p=1 general}
    Under $(Y,X) \sim \mathcal{BVN}\left([\mu_\txt{Y},\mu_\txt{X}],\begin{bmatrix}[0.5] \sigma_\txt{Y}^2 & \rho \sigma_\txt{Y} \sigma_\txt{X} \\ \rho \sigma_\txt{Y} \sigma_\txt{X} & \sigma_\txt{X}^2 \end{bmatrix}\right)$, 
    then, with $w_0 = (x_0 - \mu_\txt{X})/\sigma_\txt{X}$, we have
    $$\hat{Y}^\star(x_0;(\mathbf{Y},\mathbf{X})) \sim \mathcal{AN}\left(\mu_\txt{Y} + \Sgn(\rho) \frac{\sigma_\txt{Y}}{\sigma_\txt{X}} (x_0 - \mu_\txt{X}), \frac{1}{n} \sigma_\txt{Y}^2 \Xi(w_0)\right).$$
\end{coro}

\begin{proof}
    Follows immediately from Proposition \ref{prop: Xi when p=1, normal distribution}.
\end{proof}

\subsubsection{Case of \texorpdfstring{$p\ge 1$}{p>= 1}}

Unless more specific structure is placed on the joint distribution of $(V,W)$, or equivalently on the joint distribution of $(Y,X)$, it may not be possible to obtain a more explicit or closed-form expression for $\Xi(w_0)$ other than the general expression given in Theorem \ref{prop: asy normality 1}.
However, under multivariate normality we are able to obtain an explicit and closed-form expression for $\Xi(w_0)$. 

\begin{prop}[MVN case, $p \ge 1$]
\label{prop: under MVN, p >= 1}
If in Theorem \ref{prop: asy normality 1} we have
$(V,W) \sim \mathcal{MVN}\left([0,\mathbf{0}],\begin{bmatrix}[0.5] 1 & \Omega\trp \\ \Omega & \mathbf{I}\end{bmatrix}\right),$
with $\Omega \ne 0$ and with $\gamma^2 = \lVert\Omega\rVert^2$ and $\gamma = + \sqrt{\gamma^2}$, then
$$\Xi(w_0) = (1-\gamma^2) \left\{\frac{2}{1+\gamma} + \frac{\lVert w_0 \rVert^2}{\gamma^2} - \left(\frac{1-\gamma^2}{\gamma^4}\right) (\Omega\trp w_0)^2\right\}.$$
\end{prop}

\begin{rmk}

In an earlier version of this work we had obtained the expression for $\Xi(w_0)$ given in Proposition \ref{prop: under MVN, p >= 1} via a proof that assumes at the {\em outset} the MVN condition. That proof, which was quite interesting, but lengthy and tedious, exploited sampling distributional properties of the MVN, such as the $t$, $F$, and Wishart distributions, and the independence of the sample mean vector and sample covariance matrix. However, that approach did {\em not} generalize to enable us to obtain asymptotic normality when the joint distribution of $(V,W)$ or $(Y,X)$ is not multivariate normal. Our Theorem \ref{prop: asy normality 1} in the current manuscript, in contrast, established asymptotic normality for the general case. The proof presented below shows that we recover that earlier result for $\Xi(w_0)$, given in the statement of Proposition \ref{prop: under MVN, p >= 1}, from the general expression of $\Xi(w_0)$ in Theorem \ref{prop: asy normality 1}.
\end{rmk}

\begin{proof}
Below, to indicate dimensionality of the identity matrix, $I_k$ will be the $k \times k$ Identity matrix. Also, if $M$ is a $p \times p$ matrix, the $p^2 \times p^2$ commutation matrix $K_{p^2}$ (cf., \citet[pp. 54--55]{MN:2019}) is such that
$$K_{p^2} \VEC{M} = \VEC{M\trp}.$$
Recall also the notation $\Omega^{\otimes 2} \equiv \Omega \Omega\trp$.
Under the given MVN joint distribution for $(V,W)$, we have the $\Gamma$ matrix to equal
\begin{equation}
\label{gamma matrix under MVN}
\Gamma^\txt{MVN} =
\begin{bmatrix}[0.5]
1 & \Omega\trp & 0 & 0 & 0 \\
\Omega & I_p & 0 & 0 & 0 \\
0 & 0 & 2 & 2 \left(\VEC{\Omega^{\otimes 2}}\right)\trp & 2 \Omega\trp \\ 
0 & 0 & 2 \VEC{\Omega^{\otimes 2}} & I_{p^2} + K_{p^2} & \left(\Omega\trp \otimes I_p\right) + \left(I_p \otimes \Omega\trp\right) \\
0 & 0 & 2\Omega & \left(\Omega \otimes I_p\right) + \left(I_p \otimes \Omega\right) & I_p + \Omega^{\otimes 2}
\end{bmatrix}.
\end{equation}
From Lemma \ref{Jacobian}, the Jacobian vector is
\begin{eqnarray*}
J(w_0) & = &
\begin{bmatrix}
1\;&\; -\frac{\Omega\trp}{\gamma}\;&\; \frac{1}{2} \frac{\Omega\trp w_0}{\gamma}\;&\; -\frac{(\Omega\trp \otimes w_0\trp)}{\gamma} + \frac{1}{2} \frac{(\Omega\trp w_0)}{\gamma^3} (\Omega\trp \otimes \Omega\trp)\;&\;
\frac{w_0\trp}{\gamma} - \frac{(\Omega\trp w_0)}{\gamma^3} \Omega\trp\end{bmatrix}.
\end{eqnarray*}
The asymptotic variance of $\hat{Y}^\star(w_0;(\mathbf{V},\mathbf{W}))$ is then by Proposition \ref{prop: asy normality 1}
$$\Xi^\txt{MVN}(w_0) = J(w_0) \Gamma^\txt{MVN} J(w_0)\trp = \TRACE{\Gamma^\txt{MVN} J(w_0)\trp J(w_0)}.$$
where $\TRACE{M}$ is the trace of matrix $M$. As the simplifications that follow demonstrate, this $\Xi^\txt{MVN}(w_0)$ turns out to have a very nice closed-form expression given by
$$\Xi^\txt{MVN}(w_0) = (1-\gamma^2) \left\{\frac{2}{1+\gamma} + \frac{\lVert w_0 \rVert^2}{\gamma^2} - \left(\frac{1-\gamma^2}{\gamma^4}\right) (\Omega\trp w_0)^2\right\}.$$
We now provide the details of the simplification leading to the above closed-form expression. For this purpose, we first recall some identities (see \citet[chapter 2]{MN:2019}) pertaining to the Kronecker product and the $\VEC{}$ operator. Below, $A, B, C, D$ are matrices, $a, b$ are vectors, $\alpha$ is a scalar, and $I$ is the identity matrix. 
\begin{align*}
&(A \otimes B)\trp = (A\trp \otimes B\trp);\;\;
(A \otimes B)(C \otimes D) = (AC) \otimes (BD);\\
&(\alpha \otimes A) = \alpha A = A\alpha = A \otimes \alpha;\;\;  
(a\trp \otimes b) = ba\trp = (b \otimes a\trp);\\
&\VEC{AB} = (I \otimes A) \VEC{B} = (B\trp \otimes I) \VEC{A} =  (B\trp \otimes A) \VEC{I}.
\end{align*}

Expanding the quadratic form $\Xi^\txt{MVN}(w_0) = J(w_0) \Gamma^{MVN} J(w_0)\trp$, we obtain
\begin{eqnarray*}
&&\Xi^\txt{MVN}(w_0)   =   \left\{(1)(1)(1)\right\} + \left\{\frac{\Omega\trp}{\gamma} I_{p} \frac{\Omega}{\gamma}\right\} + 
\left\{\frac{1}{2} \frac{(\Omega\trp w_0)}{\gamma} (2) \frac{1}{2} \frac{(\Omega\trp w_0)}{\gamma}\right\} + \\
&& \left\{ 
\left(
-\frac{( \Omega\trp  \otimes w_0\trp)}{\gamma} + \frac{1}{2} \frac{(\Omega\trp w_0)}{\gamma^3} (\Omega\trp \otimes \Omega\trp)\right)
\left(I_{p^2} + K_{p^2}\right) \right. \\
&& \left. \left(
-\frac{(\Omega \otimes w_0)}{\gamma} + \frac{1}{2} \frac{(\Omega\trp w_0)}{\gamma^3} (\Omega \otimes \Omega)\right)
\right\} + \\
&& \left\{
\left(\frac{w_0\trp}{\gamma} -  \frac{(\Omega\trp w_0)}{\gamma^3}  \Omega\trp\right) 
\left(I_{p} + \Omega^{\otimes 2}\right) 
\left(\frac{w_0}{\gamma} -  \frac{(\Omega\trp w_0)}{\gamma^3}  \Omega\right)
\right\} +  
(2) (1)(\Omega\trp)\left(-\frac{\Omega}{\gamma}\right) +  \\
&&  (2) \frac{1}{2} \frac{(\Omega\trp w_0)}{\gamma} 2 (\VEC{\Omega^{\otimes 2}})\trp \left(-\frac{(\Omega \otimes w_0)}{\gamma} + 
\frac{1}{2} \frac{(\Omega\trp w_0)}{\gamma^3} (\Omega \otimes \Omega)\right) +  \\
&& (2) \frac{1}{2} \frac{(\Omega\trp w_0)}{\gamma} (2 \Omega\trp) \left(\frac{w_0}{\gamma} -  \frac{(\Omega\trp w_0)}{\gamma^3}  \Omega\right) +  \\
&& (2) \left(
-\frac{( \Omega\trp  \otimes w_0\trp)}{\gamma} + \frac{1}{2} \frac{(\Omega\trp w_0)}{\gamma^3} (\Omega\trp \otimes \Omega\trp)\right) 
\left((\Omega \otimes I_p) + (I_p \otimes \Omega)\right)  \left(\frac{w_0}{\gamma} -  \frac{(\Omega\trp w_0)}{\gamma^3}  \Omega\right).
\end{eqnarray*}
To present in an organized manner the simplifications of the more complicated terms above, we present them as lemmas.

\begin{lemm}
\label{term 1}
\begin{eqnarray*}
&&\left(
-\frac{( \Omega\trp  \otimes w_0\trp)}{\gamma} + \frac{1}{2} \frac{(\Omega\trp w_0)}{\gamma^3} (\Omega\trp \otimes \Omega\trp)\right)
\left(I_{p^2} + K_{p^2}\right) 
\left(
-\frac{(\Omega \otimes w_0)}{\gamma} + \frac{1}{2} \frac{(\Omega\trp w_0)}{\gamma^3} (\Omega \otimes \Omega)\right)\\
&&= \lVert w_0 \rVert^2 - \frac{1}{2} \frac{(\Omega\trp w_0)^2}{\gamma^2}.\end{eqnarray*}
\end{lemm}

\begin{proof}
First observe that
$K_{p^2}(\Omega \otimes w_0) = (w_0 \otimes \Omega).$
This is so since 
\begin{eqnarray*}
 K_{p^2}(\Omega \otimes w_0)   =  K_{p^2} \VEC{w_0 \otimes \Omega\trp} 
 = \VEC{(w_0 \otimes \Omega\trp)\trp} = \VEC{w_0\trp \otimes \Omega} = w_0 \otimes \Omega. 
\end{eqnarray*}
Therefore we have that
\begin{eqnarray*}
K_{p^2} \left(-\frac{(\Omega \otimes w_0)}{\gamma} + \frac{1}{2} \frac{(\Omega\trp w_0)}{\gamma^3} (\Omega \otimes \Omega)\right)  
& = & \left(
-\frac{(w_0 \otimes \Omega)}{\gamma} +  \frac{1}{2} \frac{(\Omega\trp w_0)}{\gamma^3} (\Omega \otimes \Omega)\right).
\end{eqnarray*}
Hence, the term involving $K_{p^2}$ in the lemma equals, applying the identity $(A \otimes B) (C \otimes D) = (AC) \otimes (BD)$,
\begin{eqnarray*}
\lefteqn{
\left(
-\frac{( \Omega\trp  \otimes w_0\trp)}{\gamma} + \frac{1}{2} \frac{(\Omega\trp w_0)}{\gamma^3} (\Omega\trp \otimes \Omega\trp)\right)
K_{p^2} 
\left(
-\frac{(\Omega \otimes w_0)}{\gamma} + \frac{1}{2} \frac{(\Omega\trp w_0)}{\gamma^3} (\Omega \otimes \Omega)\right)  } \\
& = & \left(
-\frac{(\Omega\trp \otimes w_0\trp)}{\gamma} + \frac{1}{2} \frac{(\Omega\trp w_0)}{\gamma^3} (\Omega\trp \otimes \Omega\trp)\right) 
\left(
-\frac{(w_0 \otimes \Omega)}{\gamma} + \frac{1}{2} \frac{(\Omega\trp w_0)}{\gamma^3} (\Omega \otimes \Omega)\right) \\
& = & \left\{ \frac{(\Omega\trp w_0)^2}{\gamma^2} -
2\left(\frac{1}{2}\right) \frac{(\Omega\trp w_0)}{\gamma^4} ((w_0\trp \Omega) \otimes (\Omega\trp \Omega)) + 
\frac{1}{4} \frac{(\Omega\trp w_0)^2}{\gamma^6} ((\Omega\trp \Omega) \otimes (\Omega\trp \Omega))  \right\} \\
& = & \left\{
 \frac{(\Omega\trp w_0)^2}{\gamma^2} -  \frac{(\Omega\trp w_0)^2}{\gamma^4} \gamma^2 + \frac{1}{4}  \frac{(\Omega\trp w_0)^2}{\gamma^6} \gamma^4
 \right\} 
  =  \frac{1}{4} \frac{(\Omega\trp w_0)^2}{\gamma^2}.
\end{eqnarray*}
On the other hand, the term involving $I_{p^2}$ becomes
\begin{eqnarray*}
\lefteqn{
\left(
-\frac{( \Omega\trp  \otimes w_0\trp)}{\gamma} + \frac{1}{2} \frac{(\Omega\trp w_0)}{\gamma^3} (\Omega\trp \otimes \Omega\trp)\right)
I_{p^2} 
\left(
-\frac{(\Omega \otimes w_0)}{\gamma} + \frac{1}{2} \frac{(\Omega\trp w_0)}{\gamma^3} (\Omega \otimes \Omega)\right)  } \\
& = & \left(
-\frac{(\Omega\trp \otimes w_0\trp)}{\gamma} + \frac{1}{2} \frac{(\Omega\trp w_0)}{\gamma^3} (\Omega\trp \otimes \Omega\trp)\right) 
\left(
-\frac{(\Omega \otimes w_0)}{\gamma} + \frac{1}{2} \frac{(\Omega\trp w_0)}{\gamma^3} (\Omega \otimes \Omega)\right) \\
& = & \left\{ \frac{(\Omega\trp \Omega) (w_0\trp w_0)}{\gamma^2} -
2\left(\frac{1}{2}\right) \frac{(\Omega\trp w_0)}{\gamma^4} ((w_0\trp \Omega) \otimes (\Omega\trp \Omega)) + 
\frac{1}{4} \frac{(\Omega\trp w_0)^2}{\gamma^6} ((\Omega\trp \Omega) \otimes (\Omega\trp \Omega))  \right\} \\
& = & \left\{
 \frac{(\gamma^2 \lVert w_0 \rVert^2)}{\gamma^2} -  \frac{(\Omega\trp w_0)^2}{\gamma^4} \gamma^2 + \frac{1}{4}  \frac{(\Omega\trp w_0)^2}{\gamma^6} \gamma^4
 \right\} 
  =  \lVert w_0 \rVert^2 -  \frac{3}{4} \frac{(\Omega\trp w_0)^2}{\gamma^2}.
\end{eqnarray*}
Hence, the term in this lemma equals
$$\lVert w_0 \rVert^2 -  \frac{3}{4} \frac{(\Omega\trp w_0)^2}{\gamma^2} + \frac{1}{4} \frac{(\Omega\trp w_0)^2}{\gamma^2} = \lVert w_0 \rVert^2 - \frac{1}{2} \frac{(\Omega\trp w_0)^2}{\gamma^2}.$$
\end{proof}

\begin{lemm}
\label{term 2}
$$\left(\frac{w_0\trp}{\gamma} -  \frac{(\Omega\trp w_0)}{\gamma^3}  \Omega\trp\right) 
\left(I_{p} + \Omega^{\otimes 2}\right) 
\left(\frac{w_0}{\gamma} -  \frac{(\Omega\trp w_0)}{\gamma^3}  \Omega\right) = \left( \frac{\lVert w_0 \rVert^2}{\gamma^2} - \frac{(\Omega\trp w_0)^2}{\gamma^4} \right).$$
\end{lemm}

\begin{proof}
Since 
$\Omega\trp \left(\frac{w_0}{\gamma} - \frac{(\Omega\trp w_0)}{\gamma^3} \Omega\right) = 0$,
then this term simplifies to
\begin{eqnarray*}
 \left\Vert \frac{w_0}{\gamma} -  \frac{(\Omega\trp w_0)}{\gamma^3}  \Omega \right\Vert^2 
& = & \frac{\lVert w_0\rVert^2}{\gamma^2} - 2 \frac{w_0\trp}{\gamma} \frac{(\Omega\trp w_0)}{\gamma^3} \Omega + \frac{(\Omega\trp w_0)^2}{\gamma^6} (\Omega\trp\Omega) 
 =  \frac{\lVert w_0\rVert^2}{\gamma^2}  - \frac{(\Omega\trp w_0)^2}{\gamma^4}.
\end{eqnarray*}
\end{proof}

\begin{lemm}
\label{term 3}
$$(2) \frac{1}{2} \frac{(\Omega\trp w_0)}{\gamma} 2 (\VEC{\Omega^{\otimes 2}})\trp \left(-\frac{(\Omega \otimes w_0)}{\gamma} + 
\frac{1}{2} \frac{(\Omega\trp w_0)}{\gamma^3} (\Omega \otimes \Omega)\right) = -(\Omega\trp w_0)^2.$$
\end{lemm}

\begin{proof}
First observe that
$$\VEC{\Omega^{\otimes 2}} = \VEC{\Omega\Omega\trp} = (I \otimes \Omega) \VEC{\Omega\trp} = (I \otimes \Omega) \Omega = \Omega \otimes \Omega$$
so that
\begin{align*}
&\left(\VEC{\Omega^{\otimes 2}}\right)\trp (\Omega \otimes w_0) = (\Omega\trp \otimes \Omega\trp)(\Omega \otimes w_0) = (\Omega\trp\Omega) \otimes (\Omega\trp w_0) = \gamma^2 (\Omega\trp w_0);\\
&\left(\VEC{\Omega^{\otimes 2}}\right)\trp (\Omega \otimes \Omega) = (\Omega\trp \otimes \Omega\trp)(\Omega \otimes \Omega) = (\Omega\trp\Omega) \otimes (\Omega\trp \Omega) = \gamma^4.
\end{align*}
The term therefore becomes
\begin{eqnarray*}
 2 \left\{
-\frac{(\Omega\trp w_0)}{\gamma^2} \gamma^2 (\Omega\trp w_0) + \frac{1}{2} \frac{(\Omega\trp w_0)^2}{\gamma^4} \gamma^4 \right\}
& = & 2 \left\{ -(\Omega\trp w_0)^2 + \frac{1}{2} (\Omega\trp w_0)^2 \right\} = -(\Omega\trp w_0)^2.
\end{eqnarray*}
\end{proof}

\begin{lemm}
\label{term 4}
$$(2) \frac{1}{2} \frac{(\Omega\trp w_0)}{\gamma} (2 \Omega\trp) \left(\frac{w_0}{\gamma} -  \frac{(\Omega\trp w_0)}{\gamma^3}  \Omega\right) = 0.$$
\end{lemm}

\begin{proof} 
Result is immediate from $\Omega\trp \left(\frac{w_0}{\gamma} - \frac{(\Omega\trp w_0)}{\gamma^3} \Omega\right) = 0$.
\end{proof}

\begin{lemm}
\label{term 5}
\begin{eqnarray*}
&& (2) \left(
-\frac{(\Omega\trp \otimes w_0\trp )}{\gamma} + \frac{1}{2} \frac{(\Omega\trp w_0)}{\gamma^3} (\Omega\trp \otimes \Omega\trp)\right) 
\left((\Omega \otimes I_p) + (I_p \otimes \Omega)\right) 
\left(\frac{w_0}{\gamma} -  \frac{(\Omega\trp w_0)}{\gamma^3}  \Omega\right)\\    
&&=  -2 \lVert w_0\rVert^2 + 2 \frac{(\Omega\trp w_0)^2}{\gamma^2}.
\end{eqnarray*}
\end{lemm}

\begin{proof}
The left-hand side of the equation in the statement of the lemma, using the identity $(A \otimes B) (C \otimes D) = (AC) \otimes (BD)$, becomes
\begin{eqnarray*}
&&2\left(  -\frac{(\Omega\trp \Omega \otimes w_0\trp I)}{\gamma} - \frac{(\Omega\trp I \otimes w_0\trp \Omega)}{\gamma} +  \right. \\
&& \left. \frac{1}{2} \frac{(\Omega\trp w_0)}{\gamma^3} (\Omega\trp\Omega \otimes \Omega\trp I) + \frac{1}{2} \frac{(\Omega\trp w_0)}{\gamma^3} (\Omega\trp I \otimes \Omega\trp \Omega) \right)
\left(\frac{w_0}{\gamma} - \frac{(\Omega\trp w_0)}{\gamma^3} \Omega\right)
 \\
& = & 
2 \left( -\frac{\gamma^2 w_0\trp}{\gamma}  -\frac{(\Omega\trp \otimes w_0\trp\Omega)}{\gamma}  + \frac{1}{2} \frac{(\Omega\trp w_0)}{\gamma^3} \gamma^2 \Omega\trp + \frac{1}{2} \frac{(\Omega\trp w_0)}{\gamma^3} \gamma^2 \Omega\trp \right)\left(\frac{w_0}{\gamma} - \frac{(\Omega\trp w_0)}{\gamma^3} \Omega\right) \\
& = & 2(-\gamma w_0\trp) \left(\frac{w_0}{\gamma} - \frac{(\Omega\trp w_0)}{\gamma^3} \Omega\right)  =  -2\lVert w_0 \rVert^2 + 2 \frac{(\Omega\trp w_0)^2}{\gamma^2}. 
\end{eqnarray*}
\end{proof}

Going back now to proving Proposition \ref{prop: under MVN, p >= 1}, we could put back these simplified expressions in $\Xi^\txt{MVN}(w_0)$ to obtain:
\begin{eqnarray*}
\Xi^\txt{MVN}(w_0) & = & 1 + 1 + \frac{1}{2} \frac{(\Omega\trp w_0)^2}{\gamma^2} + \left\{\lVert w_0\rVert^2 - \frac{1}{2} \frac{(\Omega\trp w_0)^2}{\gamma^2}\right\} + \\
& & \left\{ \frac{\lVert w_0 \rVert^2}{\gamma^2} - \frac{(\Omega\trp w_0)^2}{\gamma^4} \right\}+ 2(-\gamma) - (\Omega\trp w_0)^2 + 0 + \left\{ -{2} \lVert w_0 \rVert^2 + {2} \frac{(\Omega\trp w_0)^2}{\gamma^2}\right\} \\
& = & 2 - 2\gamma + \frac{\lVert w_0 \rVert^2}{\gamma^2} - \lVert w_0 \rVert^2  + (\Omega\trp w_0)^2 \left(-\frac{1}{\gamma^4} -1 + \frac{2}{\gamma^2}\right) \\
& = & 2\frac{(1-\gamma^2)}{1 + \gamma} + \frac{(1-\gamma^2)}{\gamma^2} \lVert w_0 \rVert^2 - \frac{(1-\gamma^2)^2}{\gamma^4} (\Omega\trp w_0)^2 \\
& = & (1 - \gamma^2) \left\{ \frac{2}{1+\gamma} + \frac{\lVert w_0 \rVert^2}{\gamma^2}  - \frac{(1 - \gamma^2)}{\gamma^4} (\Omega\trp w_0)^2\right\} .
\end{eqnarray*}
This completes the proof of the proposition.
\end{proof}
Note that the expression of $\Xi^\txt{MVN}(w_0)$ in Proposition \ref{prop: Xi when p=1, normal distribution} can be recovered from the result of Proposition \ref{prop: under MVN, p >= 1} by setting $p=1$. 

\begin{coro}
If $p=1$, then $\Xi^\txt{MVN}(w_0)$ in Proposition \ref{prop: under MVN, p >= 1} becomes $\Xi(w_0) = (1-\gamma^2)\{2/(1+\gamma) + w_0^2\}$. 
\end{coro}

\begin{proof}
Immediate by noting that, when $p = 1$,
$$\frac{\lVert w_0 \rVert^2}{\gamma^2} - \frac{(1-\gamma^2)}{\gamma^4}(\Omega\trp w_0)^2 = \frac{w_0^2}{\rho^2} - \frac{(1-\rho^2)}{\rho^4}(\rho w_0)^2 =  w_0^2.$$
\end{proof}

\begin{coro}[Asymptotic normality under MVN, $p \ge 1$]
\label{coro: under MVN, p >=1, predictor}
If $$(Y,X) \sim \mathcal{MVN}\left([\mu_\txt{Y},\mu_\txt{X}], \begin{bmatrix}[0.5] \sigma_\txt{Y}^2 & \Sigma_\txt{YX} \\ \Sigma_\txt{XY} & \Sigma_\txt{XX} \end{bmatrix}\right)$$ with $\Sigma_\txt{XX}$ nonsingular and $\Sigma_\txt{YX} \ne 0$, then 
$$\hat{Y}^\star(x_0;(\mathbf{Y},\mathbf{X})) \sim \mathcal{AN}\left(\tilde{Y}^\star(x_0) \equiv \mu_\txt{Y} + \frac{1}{\gamma} \Sigma_\txt{YX} \Sigma_\txt{XX}^{-1} (x_0 - \mu_\txt{X}), \frac{1}{n} \sigma_\txt{MA:MVN}^2(x_0)\right)$$
with
$\gamma^2 = {\Sigma_\txt{YX}\Sigma_\txt{XX}^{-1}\Sigma_\txt{XY}}/{\sigma_\txt{Y}^2}$
and
\begin{eqnarray*}
\sigma_\txt{MA:MVN}^2(x_0)  =  \sigma_\txt{Y}^2 (1-\gamma^2) \left\{\frac{2}{1+\gamma} + \frac{1}{\gamma^2} \lVert\Sigma_\txt{XX}^{-1/2} (x_0-\mu_\txt{X})\rVert^2 - \frac{(1-\gamma^2)}{\sigma_\txt{Y}^2 \gamma^4} \lVert\Sigma_\txt{YX}\Sigma_\txt{XX}^{-1}(x_0-\mu_\txt{X})\rVert^2\right\}.
\end{eqnarray*}
\end{coro}

\begin{proof}
This result immediately follows from Corollary \ref{coro: asy normality 2}, Proposition \ref{prop: under MVN, p >= 1}, and since $w_0 = \Sigma_\txt{XX}^{-1/2}(x_0 - \mu_\txt{X})$.
\end{proof}

We also present the result for the ELSLP under the MVN assumption.

\begin{coro}[Asymptotic normality of ELSLP under MVN, p $\ge$ 1]
    \label{coro: asymptotics for ELSLP under MVN}
    If $(Y,X)$ has an MVN distribution, then as $n \rightarrow \infty$,
    $$\hat{Y}^\dagger(x_0;(\mathbf{Y},\mathbf{X})) \sim \mathcal{AN}\left(\tilde{Y}^\dagger(x_0) \equiv \mu_\txt{Y} +  \Sigma_\txt{YX} \Sigma_\txt{XX}^{-1} (x_0 - \mu_\txt{X}),\frac{1}{n} \sigma_\txt{LS:MVN}^2(x_0)\right)$$
    where $$\sigma_\txt{LS:MVN}^2(x_0) = \sigma_Y^2 (1 - \gamma^2) \left\{1 + (x_0 - \mu_\txt{X})\trp \Sigma_\txt{XX}^{-1} (x_0 - \mu_\txt{X})\right\}.$$
    %
\end{coro}

\begin{proof}
    Follows immediately from Theorem \ref{theo: asymptotics for ELSLP} and the $\Gamma = \Gamma^\txt{MVN}$ given in (\ref{gamma matrix under MVN}).
\end{proof}

When $F$ is multivariate normal, we could compare the asymptotic variances of the EMALP and the ELSLP. 
By Cauchy-Schwartz Inequality, we find that $
   \left\{{\Sigma_{\txt{YX}}\Sigma_{\txt{XX}}^{-1}} (x_0 - \mu_\txt{X})\right\}^2  \le \{(x_0 - \mu_\txt{X}) \Sigma_\txt{XX}^{-1} (x_0 - \mu_\txt{X})\} (\sigma_\txt{Y}^2 \gamma^2)$. 
As a consequence, we have the inequality, for $p \ge 1$,
\begin{eqnarray*}
  \sigma_\txt{MA:MVN}^2(x_0)  & \ge &  \sigma_\txt{Y}^2 (1-\gamma^2) \left\{\frac{2}{1+\gamma} + (x_0 - \mu_\txt{X}) \Sigma_\txt{XX}^{-1} (x_0 - \mu_\txt{X})\trp \right\} \\
    & \stackrel{\gamma \le 1}{\ge} &  \sigma_\txt{Y}^2 (1-\gamma^2) \left\{1 + (x_0 - \mu_\txt{X}) \Sigma_\txt{XX}^{-1} (x_0 - \mu_\txt{X})\trp \right\}   =  \sigma_\txt{LS:MVN}^2(x_0).
\end{eqnarray*}
Thus, we have shown that, for $p \ge 1$, we have
$\sigma_\txt{MA:MVN}^2(x_0) \ge \sigma_\txt{LS:MVN}^2(x_0)$, that is, the ELSLP has a smaller asymptotic variance than the EMALP. This is {\em not} a surprising result though since the least-squares predictor is derived by minimizing a variance. In addition, note that the EMALP and ELSLP predictors have different asymptotic means, which are the population-level MALP and the LSLP, respectively. 



{

\subsubsection{Case of \texorpdfstring{$p\ge1$}{p>=1}, Non-Normal \texorpdfstring{$(Y,X)$}{(Y,X)}}
\label{NON_NORMAL_TWO_GROUP}

In this subsubsection we also provide an example where bivariate or multivariate normality do not hold true. This will demonstrate the generality of the setting that we have considered. For simplicity, we consider only the case with $p=1$.
We assume that covariate $X$ is a Bernoulli random variable with parameter $\theta$, while the response variable $Y$, given $X=x_0$, has a normal distribution with mean $\E[Y|X=x_0] = \alpha + x_0 \beta$ and variance $\Var[Y|X=x_0] = \sigma^2$. Thus, this is a two-group model, such as for instance having a control group ($x = 0$) and a treated group ($x = 1$). The model is equivalent to the structural model
\begin{align}\label{structural_model}
X \sim \text{Bernoulli}(\theta) \quad\text{and}\quad Y = \alpha + X\beta + \sigma \epsilon\;\;\text{with}\;\;\epsilon \sim \mathcal{N}(0,1),
\end{align}

where $X$ and $\epsilon$ are independent. In this model, the linearity in $x_0$  of the conditional mean of $Y$, given $X=x_0$, is thus {\em trivially (since $x \in \{0,1\}$)} satisfied. Iterated rules of expectation, variance, and covariance yield the following parameters: 
$$\mu_\txt{X} = \theta; \mu_\txt{Y} = \alpha +\theta \beta; \sigma_\txt{X}^2 = \theta (1-\theta); \sigma_\txt{Y}^2 = \sigma^2 + \theta(1-\theta) \beta^2; \sigma_\txt{YX} =  \theta(1-\theta) \beta,$$
so that
$$\Omega = \Sgn(\beta) \sqrt{\frac{\theta(1-\theta) \beta^2}{\sigma^2 + \theta(1-\theta) \beta^2}}; \quad \gamma = |\Omega| = +\sqrt{\frac{\theta(1-\theta) \beta^2}{\sigma^2 + \theta(1-\theta) \beta^2}}.$$
{When $\beta = 0$, provided that $\theta \in (0,1)$, then the covariance $\Omega = 0$. But this is equivalent to the situation where $\E[Y|X=0] = \E[Y|X=1]$, that is, the two group means coincide. In this case, the MALP will not be defined since $\gamma = 0$. See the discussion about the MAP and MALP in subsection \ref{SECTION_KNOWN_PRED} on page \pageref{SECTION_KNOWN_PRED}.}

Transform $(Y,X)$ into $(V,W)$ via
$$V = \frac{Y - (\alpha + \theta \beta)}{\sqrt{\sigma^2 + \theta(1-\theta) \beta^2}} \quad \mbox{and} \quad W = \frac{X - \theta}{\sqrt{\theta(1-\theta)}}.$$
With $T^* = (V, W, V^2, W^2, WV)\trp$, we then have the $5 \times 5$ covariance matrix
$\Gamma =  \Cov[T^*,T^*],$
whose elements are obtained using the iterated rules. We already know that $\Gamma_{11} = 1$, $\Gamma_{22} = 1$, and $\Gamma_{12} = \Gamma_{21} = \Omega$. The other elements of $\Gamma$ will involve values of 
$\zeta_{ab,cd} = \Cov[Y^a X^b, Y^c X^d]$
%
%
for  $a,b,c,d \in \{0,1,2\}$. 
%
%
Thus, for example, we have
\begin{eqnarray*}
\Gamma_{23} & = & \Cov\left[W,V^2\right] = \Cov\left[\frac{X-\mu_\txt{X}}{\sigma_\txt{X}},\left(\frac{Y-\mu_\txt{Y}}{\sigma_\txt{Y}}\right)^2\right] = \frac{1}{\sigma_\txt{X} \sigma_\txt{Y}^2} \left\{\Cov\left[X,Y^2\right] - 2\mu_\txt{Y} \Cov[X,Y]\right\} \\ & = & \frac{1}{\sigma_\txt{X} \sigma_\txt{Y}^2} \left(\zeta_{01,20} - 2\mu_\txt{Y} \zeta_{01,10}\right) = \gamma\left\{1 - 2(\alpha + \theta\beta)(2\alpha + \beta)\right\},
\end{eqnarray*}
where $\zeta_{01,10} = \sigma_\txt{YX} = \theta (1-\theta) \beta$ and $\zeta_{01,20} =  (2\alpha + \beta) \beta \theta (1-\theta)$, and the final expression obtained after some simplifications and using the intermediate step $\Cov[X,X^2] = \Cov[X,X] = \Var[X]$, since $X$ is Bernoulli. Other elements of $\Gamma$ are obtained similarly, though we do not present them since they will be consistently estimated anyhow in practice as described in the last portion of this subsubsection.

If we have a random sample $(\mathbf{Y},\mathbf{X}) = \{(Y_i,X_i), i=1,2,\ldots,n\}$, then we are able to obtain explicit forms of the EMALP and the ELSLP as follows. Let the statistics for the two groups be
\begin{eqnarray*}
& n_0 = \sum_{i=1}^n (1-X_i);\quad  n_1 = \sum_{i=1}^n X_i; \quad  \bar{Y}_0 = \frac{1}{n_0} \sum_{i=1}^n (1-X_i) Y_i; \quad \bar{Y}_1 = \frac{1}{n_1} \sum_{i=1}^n X_i Y_i; & \\
& \hat{\theta} = \frac{n_1}{n}; \quad S_0^2 = \frac{1}{n_0} \sum_{i=1}^n (1-X_i) (Y_i - \bar{Y}_0)^2; \quad S_1^2 = \frac{1}{n_1} \sum_{i=1}^n X_i (Y_i - \bar{Y}_1)^2. & 
\end{eqnarray*}
In terms of these group statistics, we have
\begin{eqnarray*}
  &  \bar{X} = \hat{\theta}; \quad \bar{Y} = \hat{\theta} \bar{Y}_1 + (1-\hat{\theta}) \bar{Y}_0; 
 \quad S_\txt{X}^2 = \hat{\theta} (1-\hat{\theta}); \quad S_\txt{YX} = \hat{\theta}(1-\hat{\theta}) (\bar{Y}_1 - \bar{Y}_0); & \\
  & S_\txt{Y}^2 = \hat{\theta} S_1^2 + (1-\hat{\theta}) S_0^2 +  \hat{\theta} (1 - \hat{\theta}) (\bar{Y}_1 - \bar{Y}_0)^2. &
\end{eqnarray*}
Then, we get that
$$\hat{\gamma} =  +\sqrt{\frac{\hat{\theta}(1-\hat{\theta}) (\bar{Y}_1 - \bar{Y}_0)^2}{\hat{\theta} S_1^2 + (1-\hat{\theta}) S_0^2 +  \hat{\theta}(1-\hat{\theta}) (\bar{Y}_1 - \bar{Y}_0)^2}}.$$
Note that when $\bar{Y}_1 = \bar{Y}_2$ and either $S_1^2 > 0$ or $S_0^2 > 0$, then $\hat{\gamma} = 0$, so when this occurs, the EMALP is not defined. However, under the condition that $\beta \ne 0$, then provided that $\theta \in (0,1)$, by the SLLN, $(\bar{Y}_1 - \bar{Y}_0)$ will converge with probability one to $\beta$, which will then be non-zero. The ELSLP under this two-group model becomes
$$\hat{Y}^\dagger(x_0) =
\begin{cases}
\begin{array}{ccc}
\bar{Y}_1 & \mbox{if} & x_0 = 1 \\
\bar{Y}_0 & \mbox{if} & x_0 = 0
\end{array}
\end{cases};
$$ 
whereas, provided $\hat{\gamma} \ne 0$, the EMALP is
$$\hat{Y}^\star(x_0) =
\begin{cases}
\begin{array}{ccc}
\frac{1}{\hat{\gamma}} \left\{\bar{Y}_1 - (1-\hat{\gamma}) \left[\bar{Y}_0 + \hat{\theta}(\bar{Y}_1 - \bar{Y}_0)\right]\right\} & \mbox{if} & x_0 = 1 \\
\frac{1}{\hat{\gamma}} \left\{\bar{Y}_0 - (1-\hat{\gamma}) {\left[\bar{Y}_0 + \hat{\theta}(\bar{Y}_1 - \bar{Y}_0)\right]}\right\} & \mbox{if} & x_0 = 0
\end{array}
\end{cases}.
$$ 
%
According to subsection \ref{subsec: estimating asymptotic variance}, we could then consistently estimate the associated covariance matrix $\Gamma$ by first doing the transformations, for $i=1,2,\ldots,n$,
$\hat{V}_i = {(Y_i - \bar{Y})}/{S_\txt{Y}}$ and $\hat{W}_i = {(X_i - \hat{\theta})}/{S_\txt{X}}$,
then computing the sample covariance matrix of $\{T_i^* = (V_i,W_i,V_i^2,W_i^2,W_iV_i), i=1,2,\ldots,n\}$  to yield $\hat{\Gamma}$.
However, we note that it is possible that there could be better estimators of $\Gamma$ than this $\hat{\Gamma}$ since the model has more structure. Under this non-normal model, the asymptotic results for the EMALP and ELSLP hold true using the $\Gamma$ described above and with the corresponding Jacobian vectors. 
}

{

\subsection{On Mis-Specified Linear Predictors}

As mentioned earlier, if the joint distribution $F$ of $(Y,X)$ satisfies (\ref{linearity conditional mean}), the condition of linearity of the conditional mean of $Y$, given $X=x_0$, then the MALP is also the MAP. What happens when this linearity condition is not satisfied, that is, when
$x_0 \mapsto \E[Y|X=x_0] \ne \alpha + \beta\trp x_0?$
The MAP should be as in (\ref{MAP}):
$$\tilde{Y}^{\star\star}(x_0) = \E[Y] + \sqrt{\frac{\Var(Y)}{\Var[\E[Y|X]]}} \{\E[Y|X=x_0] - \E[Y]\},$$
but the EMALP $\hat{Y}^\star(x_0)$ will converge strongly to the MALP
$$\tilde{Y}^\star(x_0) \equiv \mu_\txt{Y} + \sqrt{\frac{\sigma_\txt{Y}^2}{\Sigma_\txt{YX} \Sigma_\txt{XX}^{-1} \Sigma_\txt{XY}}} \Sigma_\txt{YX} \Sigma_\txt{XX}^{-1} (x_0 - \mu_\txt{X}),$$
which could be quite different from $\tilde{Y}^{\star\star}(x_0)$. As such, when condition (\ref{linearity conditional mean}) is not satisfied, caution must be exercised when using the EMALP (but, in the same token, this is also an issue with the ELSLP) for predicting the value of $Y$ at $X=x_0$, since the prediction could be quite biased, the `bias' term loosely used here. Nonetheless, the asymptotic results for the EMALP still hold true since the asymptotic mean specified in the result is the MALP, and the same comment holds for the ELSLP.
}

\section{Empirical Studies on Estimated MALP}
\label{sec: Simulations}

In this section, we present the results of computer experiments to i) assess the quality of the mean and variance approximations provided by the asymptotic mean and variance of the estimated MALP; ii) assess the adequacy of the approximate normality; and iii) assess the quality of EMALP and ELSLP in terms of CCC, PCC, and MSE, though this MSE is actually the `mean squared prediction error' (MSPE). For these computer experiments, we focus on the simple setting with $p =1$ where $F$, the distribution of $(Y,X)$, is a bivariate normal distribution. However, {Section A.2 in the Appendices} includes the results of the first experiment with $p=2$ for a trivariate normal distribution, as well as the corresponding results for the estimated LSLP. In addition, it also contains two additional experiments concerning the performance and quality of the normal approximations of the EMALP and ELSLP, respectively. 

\subsection{Computer Experiment 1}
\label{EXP_1}
In the first set of computer experiments, we assign the arbitrarily chosen parameter values $(\mu_\txt{X}=5,\;\mu_\txt{Y}=5,\;\sigma_\txt{X}=2,\;\sigma_\txt{Y}=4,\;\rho)$, where $\rho$ is replaced by a value from the set $\{0.05, 0.5,0.9\}$. Then, given a sample size $n \in \{30,50,200\}$, we obtain the vector $x_0 = [x_\txt{0j}]_{j=1}^9$ whose elements are the deciles of the marginal distribution $\mathcal{N}(\mu_\txt{X},\sigma_\txt{X}^2)$ of $X$. For each replication $l = 1, \ldots, \MReps$, a random sample of size $n$, given by $(Y_l,X_l) = \{(Y_{li},X_{li})\}_{i=1}^n$, is generated from $F$, the bivariate normal distribution with the assigned parameters. For this sample, $\hat{Y}_l^\star(\cdot;(Y_l,X_l))$ is obtained, and then the predictions $\hat{Y}_{\txt{0}jl}^\star = \hat{Y}_l^\star(x_{\txt{0}j};(Y_l,X_l))$ are computed. For each $j = 1,\ldots,9$, the means, variances, boxplots, and histograms of $\hat{Y}_{\txt{0}j}^\star = \{\hat{Y}_{\txt{0}jl}^\star\}_{l=1}^\MReps$ are obtained. 

We compare these empirical means and variances with the approximations provided by the asymptotic results in Corollary \ref{coro: asy normality p=1 general} for different values of $n$ and $\rho$. Observe that by the Weak Law of Large Numbers (WLLN), as $\MReps$ increases, the empirical means and variances will converge in probability to the true means and variances of the $\hat{Y}_{\txt{0}j}^\star$s. Thus, for this experiment, we want to ascertain how good the asymptotic approximations for the means and variances are for a given combination $(n,\rho)$. 
\begin{figure}[ht]
    \centering
    \includegraphics[height=230pt]{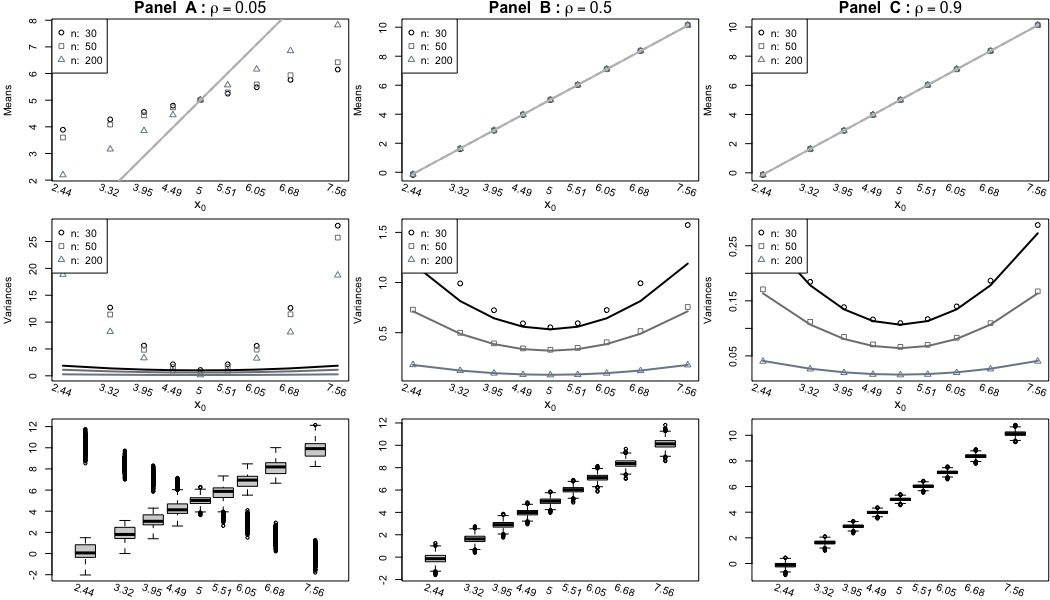}
    \caption{ The approximation quality for the EMALP for different $(\rho,n)$ at $x_0$'s: the empirical means and variances (dotted) of $\hat{Y}_0^\star$s with their asymptotic approximations (solid curves). Also depicted are the boxplots for the case where $n=200$.}
    \label{Simul_11_MAP}
\end{figure}
Figure \ref{Simul_11_MAP} depicts the results pictorially, with the first row being for the means, the second row for the variances, and the third row providing the boxplots but only for the case $n = 200$; while the three columns (panels A, B, C) are for the different values of $\rho$. 
For each of these plots, the abscissa shows the values of the $x_{\txt{0}j}$'s. In the first and second rows, the dotted points are the empirical results, whereas the solid curves are the asymptotic approximations. When $\rho=0.5$ and $\rho = 0.9$ in panel B and panel C, respectively, the asymptotic approximations are adequate, even for $n = 30$, but with a slight degradation for the variance approximation when $n = 30$. Note that in these cases, the boxplots do not anymore show mixture distributions at $n = 200$. When $\rho=0.05$ in panel A, there is a non-negligible discrepancy between the empirical means and variances with the asymptotic means and variances, even when $n$ is large. The boxplots in the third row also show that the distributions are far from normal. The reason for this is that when $\rho$ is small, the sample correlation coefficient $r$ has a high probability of having the opposite sign as that of $\rho$, and since for the MALP there is a term involving $\Sgn(r)$, a discontinuous function at $r = 0$, this creates a bifurcation making the sampling distribution of the $\hat{Y}_{\txt{0}j}^\star$ to be a mixture distribution. Of course, when we keep increasing $n$ and when $|\rho| \ne 0$, the mixture component distribution induced by the oppositely-signed $r$ becomes less likely, since $r$ converges to $\rho$ in probability as $n \rightarrow \infty$. As a general suggestion, a rule of thumb for the minimum recommended sample size is to have $n= \left\lceil\left(\tfrac{1.96}{\arctanh(\tilde{\rho})}\right)^2\right\rceil,$ derived using the Fisher $z$-transformation, when advance or prior information provides an approximate value of $\rho$ given by $\tilde{\rho}$. With this sample size, the MALP will be using an $r$-value whose sign is the same as the true $\rho$-value approximately 97.5\% of the time. If the $\tilde{\rho}$ value was not obtainable in advance, which probably will be the case in practice, then we substitute the Pearson's sample correlation coefficient $r$ for $\tilde{\rho}$ in the formula to {check if the use of MALP is reasonable for the observed sample size.}

\begin{figure}[ht]
    \centering
    \includegraphics[height=135pt]{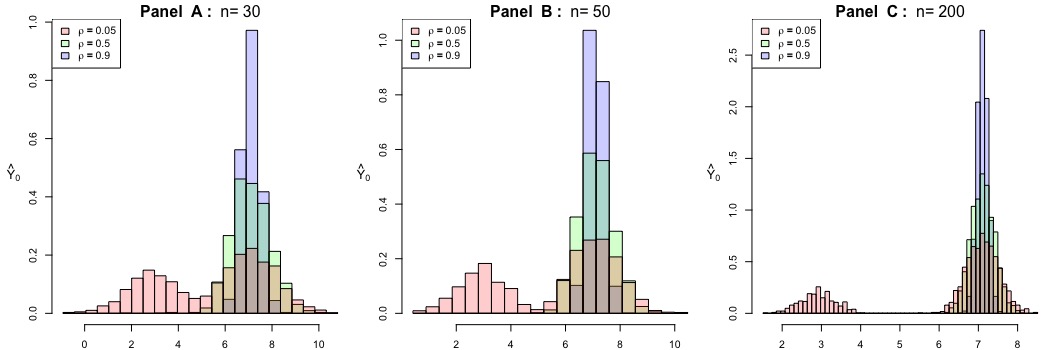}
    \caption{Empirical histograms of the EMALP $\hat{Y}_0^\star$s at $x_0=6.05$ for different values of $(\rho,n)$ based on 2000 replications.}
    \label{Simul_12_MAP}
\end{figure}

To examine further the distributional aspects, Figure \ref{Simul_12_MAP} provides the histograms of the $\hat{Y}_{\txt{0}j}^\star$ at $x_0=6.05$, the 70th percentile of the $X$ distribution, for the different values of $(n,\rho)$.  
We chose the 70th percentile, instead of the mean or median, in order to illustrate that the sampling distribution of $\hat{Y}_{\txt{0}j}^\star$ could be a mixture of two quite different distributions when $\rho$ is small. So, in each panel, the distribution of $\hat{Y}_{\txt{0}j}^\star$ for $\rho=0.05$ is bimodal, whereas two other distributions with $\rho=0.5$ and $0.9$ are bell-shaped. But, as $n$ increases, the left component of the mixture becomes less likely and the normal distribution approximation improves. 
It should be pointed out that a value of $\rho = .05$ is too low when trying to predict $Y$ from $X$ in practical situations. However, to give further insights regarding when the asymptotic expressions for the mean and variance become adequate for small $|\rho|$-values, we performed additional simulations at $\rho = .05$ and with sample sizes of $n \in \{1000, 3000, 5000\}$.  
\begin{figure}[ht]
\centering
\subfloat{\includegraphics[height=230pt]{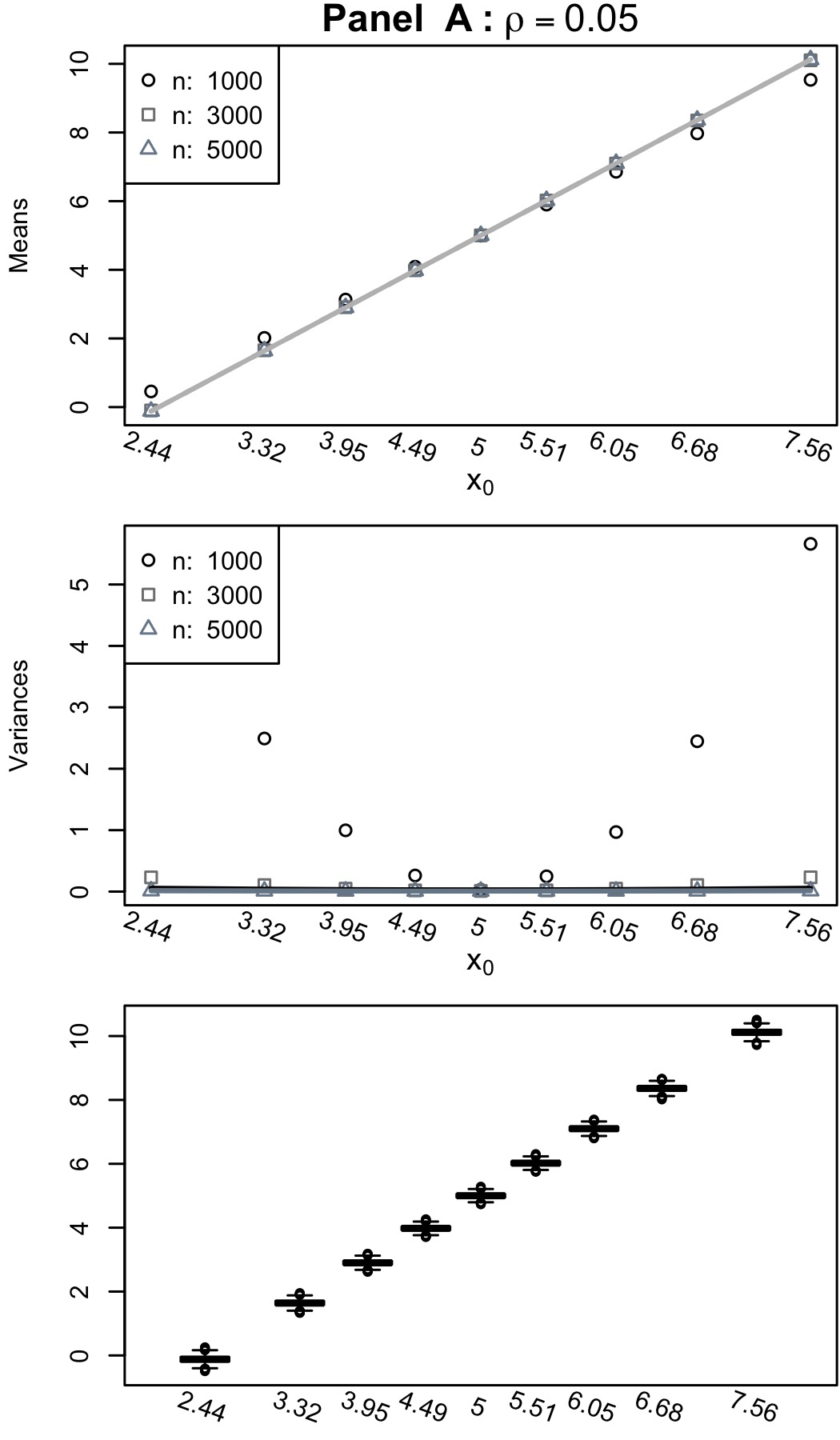}}
\subfloat{\includegraphics[height=230pt]{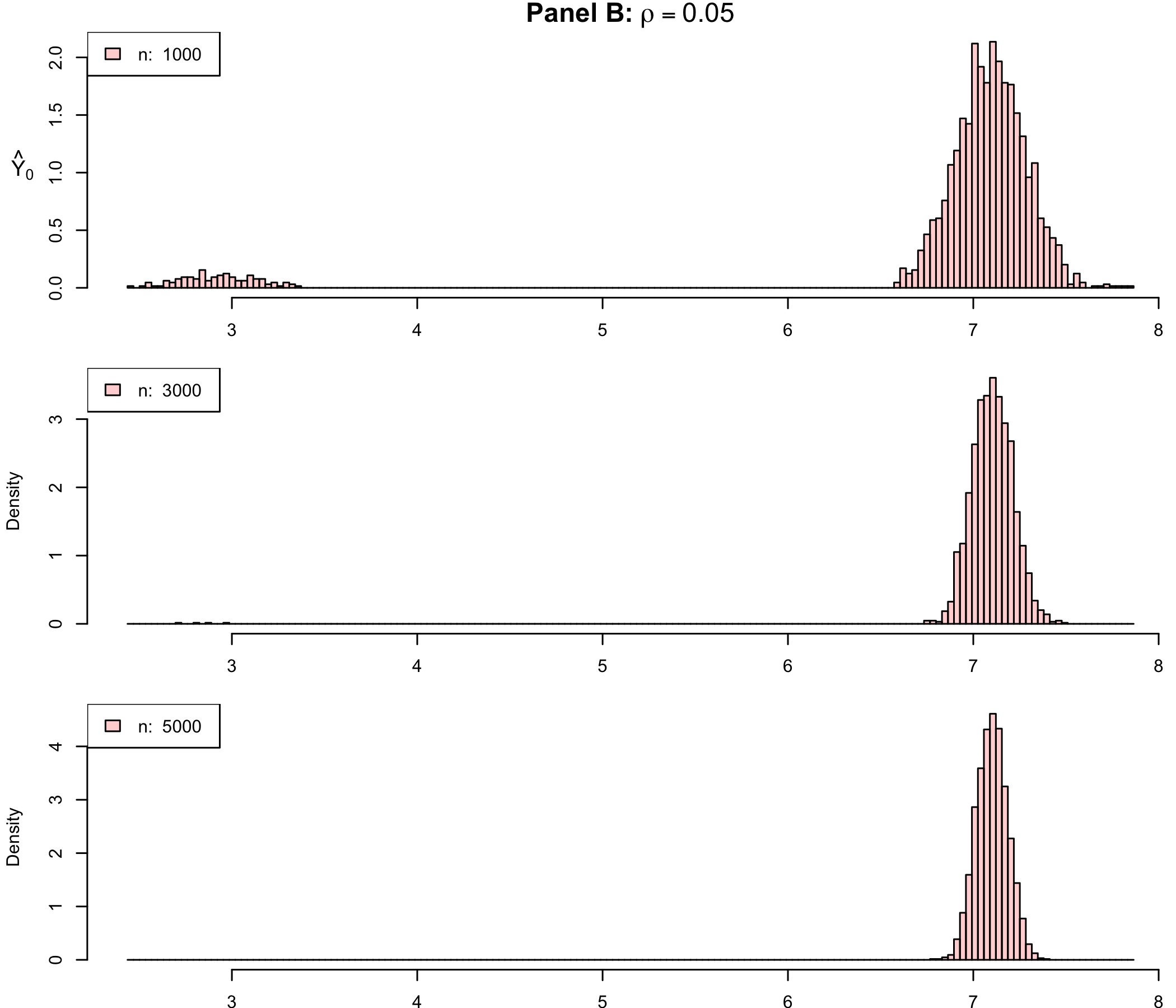}}
    \caption{{\small The approximation quality for the EMALP for $\rho=0.05$ and large $n=1000,3000,5000$: Panel A shows the empirical means and variances (dotted) of $\hat{Y}_0^\dagger$s with their asymptotic approximations (solid curves), along with the boxplots for the case when $n=5000$. Panel B presents the empirical histograms of the EMALP at $x_0=6.05$ for large $n$'s.}}
    \label{Simul_11_MALP_Large}
\end{figure}
The results of these additional simulation runs in Figure \ref{Simul_11_MALP_Large} showed that for large sample sizes $n$ and when $|\rho|$ is close to zero, the asymptotic expressions become adequate. When $\rho$ is close to zero and the sample size is not large, the asymptotic expressions should not be used, but instead computational approaches such as the bootstrap or jackknife should be employed instead to estimate standard errors. This is supported by the results of additional simulations we have performed.

We also performed simulations when $p=2$ and the distribution $F$  is trivariate normal with $\gamma=$0.05, 0.5, and 0.9. While the resulting behaviors are similar to the case with $p=1$,  the bi-modalities in the empirical sampling distribution histograms in Panel A (small $\gamma$) of {Figure 2 in Section B.2 of the Appendices} are less pronounced compared to the case with $p=1$. A possible explanation for this behavior is that the MALP does not contain the sign function when $p > 1$, compared to when $p=1$, so there is some smoothness with respect to $\gamma$. Observe, however, that the histograms for $n=200$ become closer to being bell-shaped when $|\gamma|$ increases, as can be seen in Panels B and C in Figure 2. For this experiment, the same procedures were done for the estimated LSLP $\hat{Y}_l^\dagger(\cdot;(Y_l,X_l))$ and the results regarding the predictions $\hat{Y}_{\txt{0}jl}^\dagger = \hat{Y}^\dagger(x_{\txt{0}j};(Y_l,X_l))$ are presented in {Figures 3 and 4 in Section B.2 of the Appendices}. Observe that since there is no $\Sgn(r)$ term in the LSLP, the asymptotic mean and variance approximations, including the normal approximations, are better compared to the case of the estimated MALP.

\subsection{Computer Experiment 2}
\label{EXP_2}
\begin{figure}[ht!]
\centering
	\includegraphics[height=135pt]{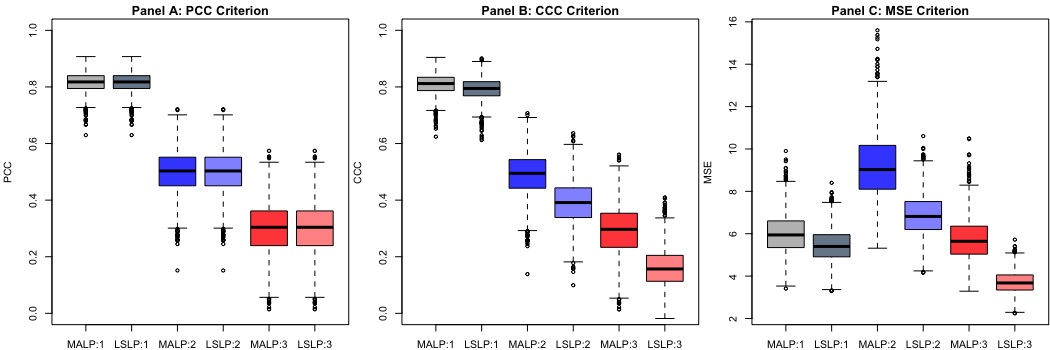}
		\caption{ Comparisons on prediction performance between EMALP and ELSLP: the side-by-side boxplots based on 2000 empirical PCC, CCC, and MSE values between $Y_0$ and $\hat{Y}_0$ for the different parameter sets.}
        \label{Simul_2_Comp}
\end{figure}

In this set of experiments, we investigated further the difference between EMALP and ELSLP, {with the performance evaluations done on a test sample data that is independent of the training data, as is usually done in machine learning settings}. For the $l$th replication with $l = 1,\ldots,\MReps$, a random sample $(Y_l,X_l) =  \{(Y_{li},X_{li})\}_{i=1}^n$ of size $n=100$, and another random sample $(Y_{0l},X_{0l}) =  \{(Y_{0li},X_{0li})\}_{i=1}^m$  of size $m=100$ are generated from the bivariate normal distribution for each parameter set in Table \ref{Parameter_Set}. 
\begin{table}[ht!]
	\caption{ Parameter sets for the bivariate normal distribution, together with contour plot of the associated distributions.}
\label{Parameter_Set}

\begin{minipage}{0.55\linewidth}
\centering

\begin{tabular}{c|ccccc|c}
	\hline\hline
&$\mu_\txt{X}$& $\mu_\txt{Y}$& $\sigma_\txt{X}$& $\sigma_\txt{Y}$& $\rho$&$\rho^\txt{c}_\txt{YX}$\\ \hline
	Set 1 & 5 &5 & 2 &4 &0.816&0.653\\
	{\color{blue}Set 2} & 8 & 4 & 3 & 3 & 0.5&0.265\\
	{\color{red}Set 3} &9   &1   &4   &2    &0.3   & 0.057\\   
	\hline\hline
\end{tabular}
\end{minipage}\hfill
\begin{minipage}{0.45\linewidth}
		\centering
	\includegraphics[height=125pt]{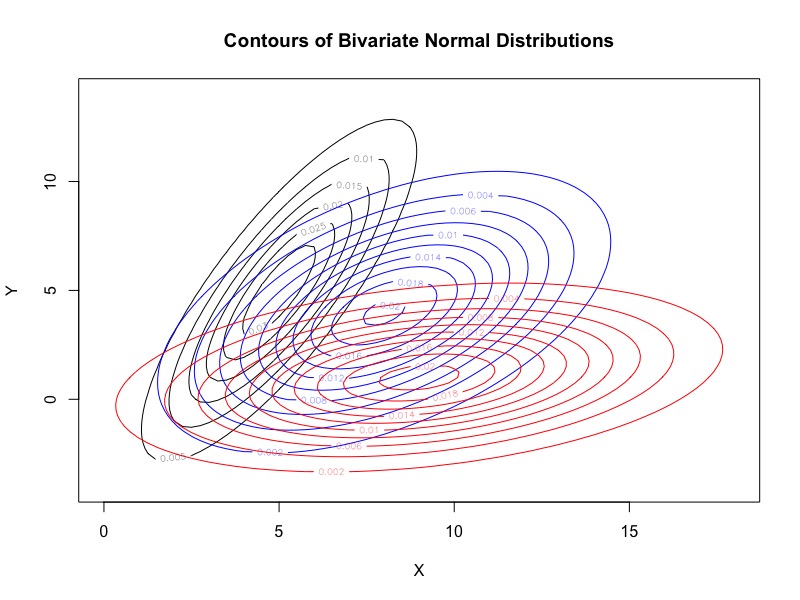}
\end{minipage}
\end{table}
Using the first sample to construct the estimated MALP and LSLP, the predicted values of $Y_{0l}$, given $X_{0l}$, for the second sample are obtained, denoted by $\hat{Y}_{0l}^\star$ and $\hat{Y}_{0l}^\dagger$, respectively. The performances of the estimated predictors are measured by the $\PCC$, $\CCC$, and $\MSE$ between $Y_0$ and $\hat{Y}_0$:
\begin{eqnarray}
	\hat{\rho}_{\txt{y}_\txt{0}\hat{\txt{y}}_\txt{0}}  &=&  \widehat{\PCC}\left(y_0,\hat{y}_0\right) \ = \ \frac{S_{\txt{y}_\txt{0}\hat{\txt{y}}_0}}{S_{\txt{y}_\txt{0}}S_{\hat{\txt{y}}_0}};   \\  \label{PERF_MEASURES}
	\hat{\rho}^c_{\txt{y}_\txt{0}\hat{\txt{y}}_\txt{0}}  &=&  \widehat{\CCC}\left(y_0,\hat{y}_0\right)  \ = \  \frac{2S_{\txt{y}_\txt{0}\hat{\txt{y}}_\txt{0}}}{S_{\txt{y}_\txt{0}}^2+S^2_{\hat{\txt{y}}_\txt{0}}+(\bar{y}_\txt{0}-\bar{\hat{y}}_\txt{0})^2};\nonumber \\  
	\hat{\mathcal{M}}_{\txt{y}_\txt{0}\hat{\txt{y}}_\txt{0}}  &=&  \widehat{\MSE}\left(y_0,\hat{y}_0\right) \ = \ \frac{1}{\txt{MReps}}\sum_{l=1}^\txt{MReps}\left(y_{0l}-\hat{y}_{0l}  \right)^2. \nonumber
\end{eqnarray}
Figure \ref{Simul_2_Comp} consists of the boxplots of the three performance measures based on $\{(Y_{0l},\hat{Y}_{0l}^\star)\}_{l=1}^\MReps$ and $\{(Y_{0l},\hat{Y}_{0l}^\dagger)\}_{l=1}^\MReps$. 
Panel A shows the prediction performance measured by the PCC criterion, so larger values indicate better performance. Note that the MALP has the same PCC as the LSLP for all three parameter specifications. This is an expected result since the PCC is invariant with respect to linear transformations. On the other hand, the prediction performance as measured by the CCC criterion in panel B shows some discrepancy between the MALP and the LSLP. For any parameter specification, the CCC of the MALP is higher than that of the LSLP, implying that the MALP is preferable to the LSLP with respect to the CCC criterion. Furthermore, observe that the median values of the CCC from the MALP are close to the $|\rho|$ values (0.816, 0.5, and 0.3) in the parameter set used for the data generation. Note that theoretically, $|\rho|$, which is $\gamma$ since $p = 1$, is the value of the CCC if the parameters are known, but since we are using the estimated MALP, there is sampling variability in these empirical CCC values, and some of them exceed the value of $|\rho|$. Lastly, panel C shows the prediction performance measured by the MSE criterion, so smaller values indicate better performances. Note that the MSE values vary with respect to the parameter sets, but the LSLP shows better performance than the MALP regardless of the parameter selection. The results indicate that there is no uniformly best predictor with respect to all three criteria. As such, the choice of a predictor should depend on the measure of predictive performance that one utilizes. If one is interested in the CCC criterion, the MALP would be the better choice for the predictor; whereas, if one considers the MSE criterion, the LSLP would be the better predictor.

\subsection{Computer Experiment 3}
\label{EXP_3}

Computing the asymptotic variance of the EMALP under a non-normal joint distribution for $(Y,X)$ may be non-trivial due to the need to calculate the covariance matrix $\Gamma$. For the non-normal model described in equation (\ref{structural_model}), $\Gamma$ could still be obtained straight-forwardly since the conditional distribution of $Y$, given $X = x_0$, is still normal and since $x_0 \in \{0,1\}$. However, we did not obtain the expressions for the elements of $\Gamma$ since we opted instead to examine the consistent estimators of the asymptotic variances of the EMALP and the ELSLP described in subsection \ref{subsec: estimating asymptotic variance}.
In this subsection, we perform a Monte Carlo study for this non-normal model to evaluate agreement of the estimates of the asymptotic variances of the EMALP and the ELSLP with their respective empirical variances, and also to examine the shapes of their sampling distributions.

\begin{figure}[ht!]
\centering
	\includegraphics[height=170pt]{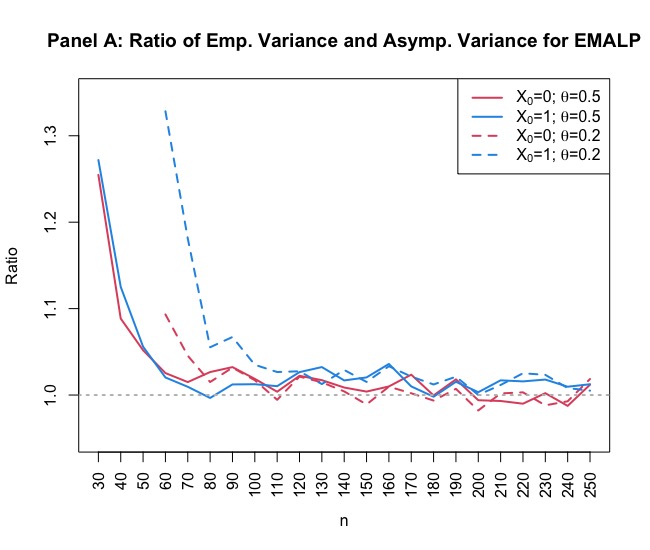}
    \includegraphics[height=170pt]{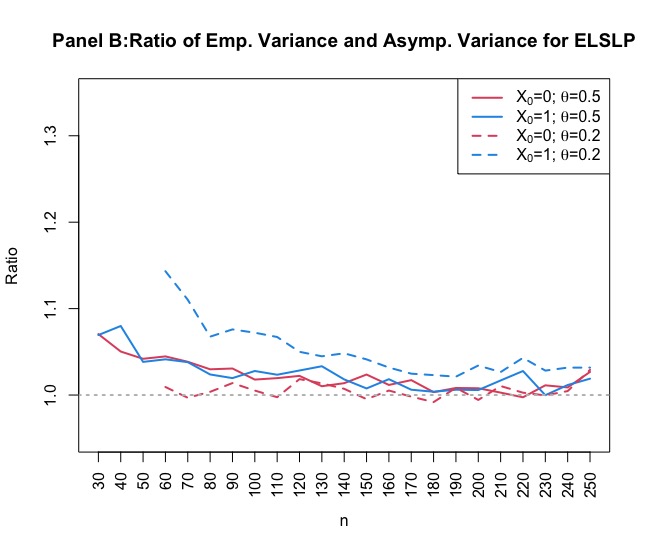}
		\caption{Ratios of empirical variance and mean of the estimates of the asymptotic variance for EMALP and ELSLP at $x_0=0\;\&\;1$ and for $\theta=0.5\;\&\;0.2$ for varying sample size, $n \in \{30, 40, \ldots, 250\}$, based on MReps $=20000$ Monte Carlo replications. }
    \label{Simul_3_Comp}
\end{figure}

In the simulation for the non-normal model in (\ref{structural_model}), the following parameter values were arbitrarily chosen: $\alpha = 5$, $\beta = 2$, and $\sigma = 2$. In addition, $\theta=0.5$ and $\theta=0.2$ were used so we have symmetric and asymmetric situations. The number of simulation replications was MReps $=20000$. For each replication in the simulation, a sample of size $n$ was generated according to the model. For each sample, the EMALP and ELSLP predictions at $x_0 \in \{0,1\}$ were obtained, as well as the estimates of their asymptotic variances. The respective means of the estimates of the asymptotic variances were then obtained for the MReps replications. The (sample) variances of the MReps EMALP and ELSLP predictions were also obtained, and these as the empirical variances.  For the EMALP and the ELSLP, the ratios between the empirical variance and the mean of the estimates of the asymptotic variance, at $x_0=0$ and $x_0 = 1$, were obtained for sample sizes, $n=30,40,\ldots, 250$.  Figure \ref{Simul_3_Comp} pictorially summarizes the results of this simulation study.

As to be expected, when $n$ is small, the estimates of the asymptotic variances of the EMALP and ELSLP may be unsatisfactory to use since they could underestimate, substantially, the true variances of the predictors. This underestimation is more pronounced with the EMALP compared to the ELSLP, especially under the asymmetric situation. As the sample size increases, however, there is more agreement between the empirical variances and the means of the estimates of the asymptotic variances. Thus, when the sample size is large, for the EMALP and ELSLP, the respective estimates of the asymptotic variances provide satisfactory estimates of the true variances. 
\begin{figure}[ht!]
\centering
	\includegraphics[height=170pt]{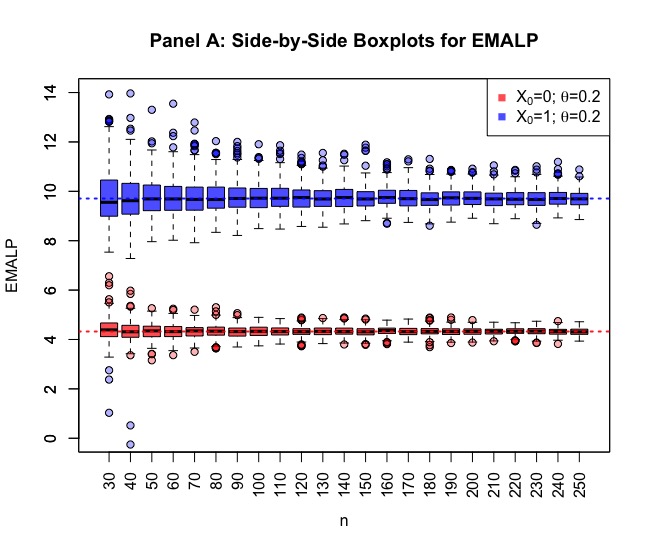}
    \includegraphics[height=170pt]{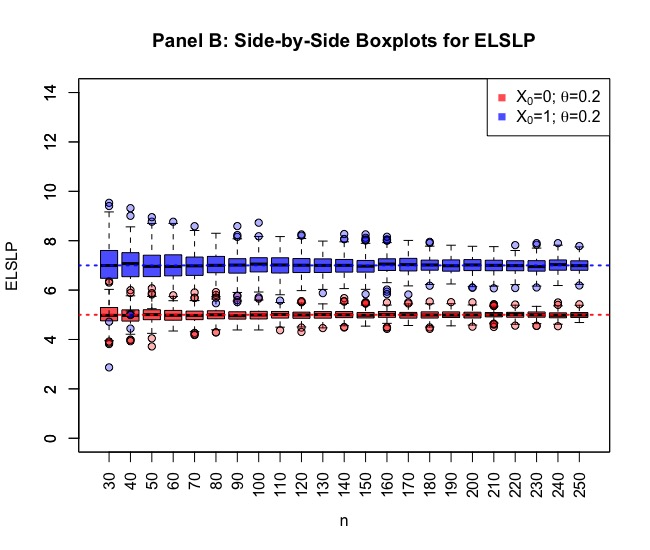}
		\caption{Side-by-side boxplots of EMALP and ELSLP  predictions for $x_0=0\;\&\;1$ and $\theta=0.2$ for varying values of the sample size $n$. Each boxplot is based on a random sample of size 300 from the 20000 predictions in the Monte Carlo study. The dotted lines correspond to the predictor's asymptotic mean.}
    \label{Simul_3_BOX}
\end{figure}
In these simulations we also  examined the shapes of the sampling distributions of the EMALP and ELSLP predictions as $n$ changes. These shapes are depicted in Figure \ref{Simul_3_BOX} using side-by-side boxplots, with each boxplot based on a random sample of size 300 from the 20000 Monte Carlo replications. We used only 300 out of the 20000 values to improve the visibility of the boxplots. Note that, as $n$ increases, these sampling distributions become close to being symmetric, bell-shaped, and centered at the predictors respective asymptotic means, the latter indicated by the dotted lines.  As such, these results from this modest Monte Carlo study also partly demonstrate empirically that the asymptotic results for the EMALP and ELSLP hold for the non-normal setting described in (\ref{structural_model}).

\section{Confidence and Prediction Intervals}
\label{sec: Intervals}

\subsection{Computational Approaches for Standard Errors}
\label{SECTION_COMPU_ASYMP_VAR}
The asymptotic normality is an important property of the estimated MALP that provides an approximate method for uncertainty quantification through the expression of the asymptotic variance $\sigma_\txt{MA:MVN}^2$ given in Corollary \ref{coro: under MVN, p >=1, predictor}, which could be used for constructing confidence or prediction intervals, assuming muitivariate normality. {For general non-normal joint distributions for $(Y,X)$, the asymptotic variance could be estimated by the procedures described in subsection \ref{subsec: estimating asymptotic variance}.} An alternative computer-intensive approach for approximating the variance of the estimated MALP is via resampling approaches such as the jackknife or bootstrap procedures; see \citet[Chapter 10]{Efro:2016}. The resulting approximate variances are reasonable alternatives to the asymptotic variance and may even be better for small-to-moderate sample sizes $n$. 

For the jackknife and bootstrap approaches, suppose we observe a random sample $\{\bs{Y},\bs{X}\}=\{(Y_i,X_i)\}_{i=1}^n$ and the estimated predictor $\hat{y}^\star(x_0)=\hat{y}^\star(x_0;\{\bs{y},\bs{x}\})$ has the form in (\ref{EMALP}). For the jackknife approach, we create a sample without the $j$th component $	\{	\bs{Y}_\txt{(j)},\bs{X}_\txt{(j)}\}  =  \{(Y_1,X_1)^{\trp},\ldots,(Y_\txt{j-1},X_\txt{j-1})^{\trp},(Y_\txt{j+1},X_\txt{j+1})^{\trp}\ldots,(Y_\txt{n},X_\txt{n})^{\trp}\}^{\trp}$.
Then, the jackknife estimated predictor based on the observed sample can be defined $\hat{y}_{(j)}^\star\left(x_0\right)  =  \hat{y}^\star\left(x_0;\left\{	\bs{y}_{(j)},\bs{x}_{(j)}\right\}\right)$.
Lastly, we define the jackknife estimate of variance for $\hat{y}^\star\left(x_0\right)$: 
\begin{equation}\label{JACKKNIFE_VAR}
	\hat{\sigma}^2_\txt{JK}(x_0) = \frac{n-1}{n}\sum_{j=1}^n\left(\hat{y}_{(j)}^\star\left(x_0\right)-\bar{\hat{y}}_\txt{JK}^\star\left(x_0\right)\right)^2\;\text{ with }\;\bar{\hat{y}}_\txt{JK}^\star\left(x_0\right)=\frac{1}{n}\sum_{j=1}^n\hat{y}_{(j)}^\star \left(x_0\right).
\end{equation}

For the bootstrap approach, we draw a bootstrap sample of size $n$ from $\{\bs{Y},\bs{X}\}$ with equal probability and with replacement: $\{	\bs{Y}^*,\bs{X}^*\}  =  \{(Y_1^*,X_1^*)^{\trp},(Y_2^*,X_2^*)^{\trp},\ldots,(Y_n^*,X_n^*)^{\trp}\}^{\trp}$.
Based on this bootstrap sample, the associated individual boostrap estimate of MALP is $\hat{y}^{\star*}\left(x_0\right)  =  \hat{y}^\star\left( x_0; \left\{ 	\bs{y}^*,\bs{x}^*\right\}\right)$.
Repeating this resampling procedure $B$ times, we obtain the bootstrap replicates: $\hat{y}^{\star*}_1(x_0),\;\hat{y}^{\star*}_2(x_0),\;\ldots,\; \hat{y}^{\star*}_B(x_0)$. Lastly, we define the bootstrap estimate of variance for $\hat{y}^\star\left(x_0\right)$:
\begin{equation}\label{BOOTSTRAP_VAR}
			\hat{\sigma}^2_\txt{BS}(x_0) = \frac{1}{B-1}\sum_{j=1}^B\left(\hat{y}_j^{\star*}\left(x_0\right)-\bar{\hat{y}}_\txt{BS}^{\star*}\left(x_0\right)\right)^2\;\text{ with }\;\bar{\hat{y}}_\txt{BS}^{\star*}\left(x_0\right)=\frac{1}{B}\sum_{j=1}^B\hat{y}_{j}^{\star*} \left(x_0\right).
\end{equation}
The resulting $\hat{\sigma}^2_\txt{JK}(x_0)$ and $\hat{\sigma}^2_\txt{BS}(x_0)$ are the jackknife and bootstrap estimates of $\sigma^2_{\txt{MA}}(x_0)/n$.

\subsection{Interval Estimation of MALP}
\label{CONFIDENCE_INTERVALS}

Given the asymptotic normality in Corollary \ref{coro: asy normality 2}, an asymptotic $100(1-\alpha)\%$ confidence interval (CI) for $\tilde{Y}^\star(x_0)$ would be as follows, where $z_\alpha = \Phi^{-1}(1-\alpha)$, the $100(1-\alpha)$th quantile of the standard normal distribution, whose distribution and quantile functions are denoted, respectively, by $\Phi(\cdot)$ and $\Phi^{-1}(\cdot)$:
\[
\Gamma[x_0,(y,x);\alpha]  =  \left[\hat{y}^\star(x_0)\pm z_{\alpha/2}\tfrac{\sigma_\txt{MA}(x_0)}{\sqrt{n}} \right].
\]
Note that the standard error term $\tfrac{\sigma_\txt{MA}(x_0)}{\sqrt{n}}$ is unknown. The first three CIs are obtained when this standard error is estimated by the asymptotic standard error obtainable in Corollary \ref{coro: under MVN, p >=1, predictor} under the multivariate normality assumption, jackknife standard deviation obtained in (\ref{JACKKNIFE_VAR}), and the bootstrap standard deviation obtained in (\ref{BOOTSTRAP_VAR}). We denote these CIs by $\Gamma_1[x_0,(y,x);\alpha]$, $\Gamma_2[x_0,(y,x);\alpha]$, and $\Gamma_3[x_0,(y,x);\alpha]$, respectively. An alternative option for the construction of CIs for MALP is a fully nonparametric approach based on resampling procedures. We adopt some of these approaches for our illustration: the bootstrap-$t$ CI procedure, $\Gamma_4[x_0,(y,x);\alpha]$; the standard bootstrap percentile CI procedure, $\Gamma_5[x_0,(y,x);\alpha]$; and the bias-corrected accelerated percentile CI procedure, $\Gamma_6[x_0,(y,x);\alpha]$. For forms and details of these CIs, see {Section C in the Appendices}. See also \citet[Chapter 5]{DH:1997} and \citet[Chapter 11]{Efro:2016} for the constructions and discussions regarding the resampling-based CI procedures. 

For comparison of these procedures, we provide simulation results for the case of $p=2$ {and under a trivariate normal distribution}. The set of parameters of the simulation coincides with the first experiment with $\gamma = 0.5$ in Section \ref{EXP_1}, so the results could be understood in light of Figures 1 and 2. The point where the prediction is made was randomly chosen to be at $(x_1,x_2)=(3.177,6.457)$ whose squared Mahalanobis distance from the mean is 5.071. For a confidence level $1-\alpha=0.95$, we performed simulations with sample sizes $n\in\{50,100,200\}$ for the CI procedures, and their empirical coverage probabilities and average lengths were computed based on  $10000$ simulation replications. For the bootstrap-based CIs, we utilize $2000$ resamplings and $30$ sub-resamplings. 

\begin{table}[ht!]
\caption{Performance of the different CIs: in panel A, the empirical coverage probabilities for the CIs with the standard errors; and, in panel B, the average standardized (multiplied by $\sqrt{n}$) lengths for the CIs with the standard errors. The CI procedures are constructed for 0.95 nominal level and the performance measures are calculated based on 10000 replications.}
\centering
{\small
\begin{tabular}{c|c|cccccc}
\hline\hline
Panel A & n     & $\Gamma_1$: Asymp.    & $\Gamma_2$: Jack    & $\Gamma_3$: Boot     & $\Gamma_4$: Boot-t    & $\Gamma_5$: Percentile    & $\Gamma_6$: BCa    \\\hline
                        & 50  & 0.943  & 0.948  & 0.947  & 0.937  & 0.941  & 0.932  \\
                                                                          Coverage      &   &(.0023) & (.0022) & (.0022) & (.0024) & (.0024) & (.0025) \\
                                                                           Probability     & 100 & 0.946  & 0.948  & 0.946  & 0.942  & 0.943  & 0.938  \\
                                                                       (Standard          &  & (.0023) &( .0022) & (.0023) & (.0023) & (.0023) & (.0024)\\
                                                                         Error)       & 200 & 0.95   & 0.95   & 0.948  & 0.946  & 0.949  & 0.945  \\
                                                                                &  & (.0022) & (.0022) & (.0022) & (.0023) & (.0022) & (.0023) \\ \hline\hline
Panel B & n     & $\Gamma_1$: Asymp.    & $\Gamma_2$: Jack    & $\Gamma_3$: Boot     & $\Gamma_4$: Boot-t    & $\Gamma_5$: Percentile    & $\Gamma_6$: BCa    \\\hline
                                                                            & 50  & 17.042 & 18.221 & 18.384 & 18.272 & 18.539 & 18.092 \\
                                                                          Average      &  & (.0564) & (.0672) & (.0651) & (.069)  & (.0641) & (.0573)\\
                                                                          Length      & 100 & 16.716 & 17.221 & 17.131 & 17.147 & 17.275 & 17.182 \\
                                                                          (Standard      &  & (.0347) &(.039)  & (.0399) & (.0418) & (.0389) & (.0377) \\ 
                                                                            Error)    & 200 & 16.563 & 16.813 & 16.697 & 16.849 & 16.812 & 16.780  \\
                                                                                & & (.023)  & (.0262) & (.0272) & (.0292) & (.0259) & (.0257)\\ \hline\hline
\end{tabular}}
\label{COMPARISON_CI}
\end{table}

The empirical coverage probabilities and average lengths standardized by the sample size are summarized in Table \ref{COMPARISON_CI} with the corresponding estimated standard errors. 
In terms of coverage probability, $\Gamma_2$ and $\Gamma_3$ show satisfactory results for all three sample sizes. In contrast, $\Gamma_4$, $\Gamma_5$, and $\Gamma_6$ have lower coverage rates relative to what is desired, although these rates improve as the sample size is increased. Lastly, $\Gamma_1$ shows less coverage rate than desired for $n=50$, but becomes satisfactory when $n=100$ and $n=200$ as the empirical coverage probabilities are consistent with the desired level 0.95. With regards to the standardized average length, $\Gamma_1$ consistently shows good performance by having a shorter average length. This becomes more pronounced when $n=100$ and $n=200$ since it also possesses the desired coverage rate. $\Gamma_2$ and $\Gamma_3$ have stable performances, providing moderate expected lengths. On the other hand, those of $\Gamma_5$ and $\Gamma_6$ have low coverage rates when $n$ is small, which contributed to their having moderately shorter average lengths. Finally, $\Gamma_4$ exhibited the worst performance when $n=200$, providing the largest average length; additionally, the relatively moderate lengths observed when $n=50$ and $n=100$ may not be very informative since they may be explained by the unsatisfactory coverage rates.

\subsection{Prediction Intervals for \texorpdfstring{$Y(x_0)$}{Y(x0)}}

While the confidence interval provides an estimate together with a measure of uncertainty for the true MALP, in practical settings we may be more interested in predicting the value of a new observation $Y$ at $X=x_0$ together with a measure of uncertainty regarding the prediction. In this subsection, we therefore consider this problem of constructing a prediction interval (PI) for the new observation $Y(x_0)$ based on the MALP. 

\begin{theo}\label{PREDICTION_INT}
Suppose $(Y,X)$ follows a multivariate normal distribution with non-singular $\Sigma$ and $\Sigma_\txt{YX} \ne 0$. Then, the $100(1-\alpha)\%$ prediction interval for $Y(x_0)$ based on MALP becomes:
\begin{equation}
\label{PI based on MALP}
\left[
\left(\hat{Y}^\star(x_0) + \hat{b}(x_0)\right) \pm
z_{\alpha/2} S_\txt{Y} \sqrt{1 - \hat{\gamma}^2} \sqrt{ 1 + \frac{1}{n} \hat{D}_\txt{MA}^2(x_0)}\right],
\end{equation}
where 
$\hat{b}(x_0) = \left(1-\tfrac{1}{\hat{\gamma}}\right)S_\txt{XX}^{-1}S_\txt{YX}(x_0-\bar{X})$ and
$$\hat{D}_\txt{MA}^2(x_0) = \frac{2}
{1+{\hat{\gamma}}} +  \frac{1}{{\hat{\gamma}}^2} (x_0-\bar{X}) S_\txt{XX}^{-1}(x_0-\bar{X})^{\trp} - \left\{\frac{(1-{\hat{\gamma}}^2)}{S_\txt{Y}^2 {\hat{\gamma}}^4}\right\} \left\{ S_\txt{XX}^{-1} S_\txt{YX}(x_0 - \bar{X})\right\}^2.$$   
\end{theo}

\begin{proof}
{See Section A.2 in the Appendices.}
\end{proof}

Note that $\hat{Y}^\star(x_0) + \hat{b}(x_0) = \hat{Y}^\dagger(x_0)$, and the PI in (\ref{PI based on MALP}) tends to be wider compared to the PI based on the LSLP:
\begin{equation}\label{SHORTEST_PI}
    \left[\hat{Y}^\dagger(x_0)\pm z_{\alpha/2}S_\txt{Y}\sqrt{(1-\hat{\gamma}^2)}\sqrt{1+\tfrac{1}{n}\left\{1+(x_0-\bar{X})S_\txt{XX}^{-1}(x_0-\bar{X})^{\trp}\right\}} \right],
\end{equation}
indicating the larger variability accruing with the MALP relative to the LSLP. For small to moderate $n$, when the asymptotic approximations may not be adequate, the prediction intervals based on the LSLP and the MALP may be improved by estimating the standard error $\sigma_\txt{LS}(x_0)/\sqrt{n}$ and $\sigma_\txt{MA}(x_0)/\sqrt{n}$ and the bias $b(x_0)$ through computational approaches such as the jackknife or the bootstrap methods described in Section \ref{SECTION_COMPU_ASYMP_VAR}.

\section{Illustrative Data Analyses}
\label{sec: Illustrations}

In this section, we demonstrate the prediction procedures using two real data sets. {The {\tt R} programs that implement the  prediction procedures, as well as those used for performing the simulation or computer experiment studies in this paper, are in the publicly accessible {\tt R} package, called the {\tt malp} package (\citet{malp}).} 
The first data set, with $n = 46$, pertains to eye measurements used in \citet{APDS:2011}. This is a simple case in that there is only one predictor variable, so $p=1$. The second data set contains body fat data with $n=252$ and with more than one predictor variable, so $p > 1$. 
\subsection{Eye Data Set}
In ophthalmology, central subfield macular thickness (CSMT) measurements can be obtained by optical coherence tomography (OCT). As introduced in Figure \ref{OCT_DATA}, \citet{APDS:2011} focused on two types of OCT: time-domain Stratus OCT, the most widely used method prior to 2006; and spectral-domain Cirrus OCT, a more advanced method. As Cirrus OCT replaces Stratus OCT, the agreement between the measurements from the two methods is of interest to researchers in the field. For this purpose, \citet{APDS:2011} provided a CCC-based conversion formula from the Cirrus OCT measurement to the Stratus OCT measurement.

In the data set, both OCTs were measured from 46 subjects, i.e., 92 eyes, but only 61\% of these observations were selected based on the reliability of the OCTs (signal strength $\geq$ 6 for  both approaches). 
The computed correlations are $\hat{\rho}=0.783$ and $\hat{\rho}^\txt{C}=0.200$. Based on the data set, two conversion formulas were introduced: i) $x \mapsto x_1=x-60$ was based on a location shift which adjusts for the mean difference between Cirrus and Stratus OCTs, and this leads to $\hat{\rho}^\txt{C}_{\txt{x}_1\txt{y}}=0.756$; ii) $x \mapsto x_2=0.76x-0.51$ was chosen to maximize the CCC, yielding $\hat{\rho}^\txt{C}_{\txt{x}_2\txt{y}}=0.781$; see \citet[Figure 2]{APDS:2011}. 

\begin{figure}[ht!]
\begin{minipage}{0.34\linewidth}
\centering
\vspace{-20pt}
\begin{eqnarray*}
\text{OS: } 	\hat{y}^\star(x_\txt{0})&=&0.95x_0-44.19\\
	\hat{y}^\dagger(x_\txt{0})  &=& 0.82x_0-12.97;\\
  \text{OD: } \hat{y}^\star(x_\txt{0}) &=& 0.68x_0+20.47\\
 	\hat{y}^\dagger(x_\txt{0})  &=&  0.51x_0+63.51.
\end{eqnarray*}
\hfill
\vfill
\end{minipage}
\begin{minipage}{0.65\linewidth}
\centering
	\vspace{-5pt}
	\subfloat{\includegraphics[height=130pt]{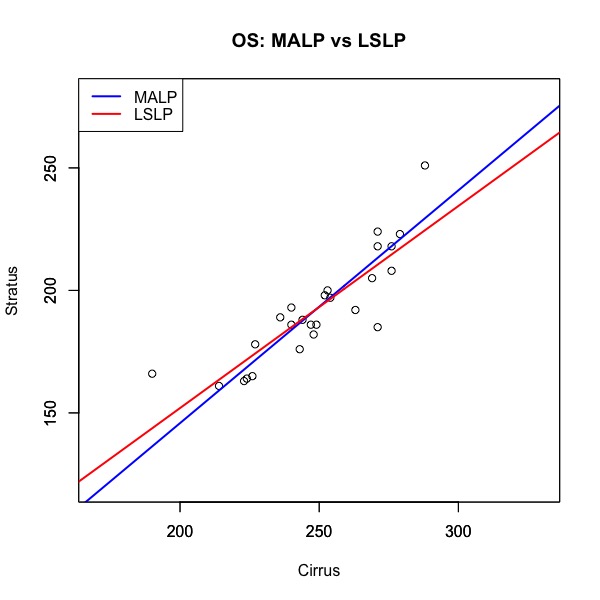}}
	\subfloat{\includegraphics[height=130pt]{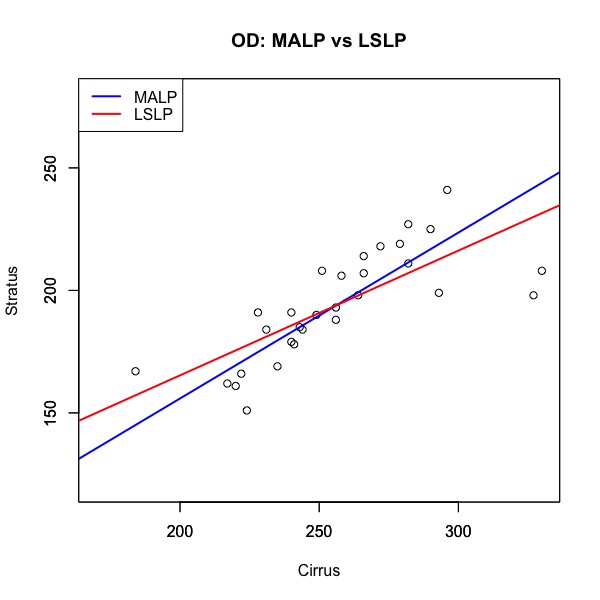}}
\end{minipage}
		\caption{ The comparison between the EMALP and the ELSLP for Stratus OCT based on Cirrus OCT: OS (left eye) and OD (right eye) are presented separately.}
		\label{EYE_ILLUST}
		\vspace{-10pt}
\end{figure}

From a prediction perspective, we follow \citet{APDS:2011}'s idea of obtaining the Stratus OCT using the Cirrus OCT. We start with an \emph{illustrative} case using a plug-in approach: $\hat{y}^\star(x_\txt{0}) = \hat{\mu}_\txt{y}+\text{Sgn\{$\hat{\rho}$\}}(\hat{\sigma}_\txt{Y}/\hat{\sigma}_\txt{X})(x_\txt{0}-\hat{\mu}_\txt{x})$ and $\hat{y}^\dagger(x_\txt{0}) = \hat{\mu}_\txt{y}+\hat{\rho}(\hat{\sigma}_\txt{Y}/\hat{\sigma}_\txt{X})(x_\txt{0}-\hat{\mu}_\txt{x})$.
We consider the OS (left eye) and the OD (right eye) measurements separately in order to mitigate the impact of their correlation ($n_\txt{OS}=26$ and $n_\txt{OD}=30$).  The predicted values from these predictors, together with the observed values, are then plotted in Figure \ref{EYE_ILLUST}.

 Using the observed values and the predicted values, we computed the PCC, CCC, and MSE, and these are summarized in the second main column labeled `Illustrative Case' in Table \ref{EYE_EVALUATION}. Note, however, that we are using the observed data twice: for the construction of the predictors and then for the evaluation of their performance, hence such comparisons could be biased due to data double-dipping.
\begin{table}[ht]
\caption{Prediction performances of MALP and LSLP measured by PCC, CCC, and MSE.}
\label{EYE_EVALUATION}
\centering
{\footnotesize
\begin{tabular}{c|cc|cc|cc|cc}
\hline\hline
Prediction&\multicolumn{4}{c|}{Illustrative Case} & \multicolumn{4}{c}{Realistic Case}\\
\hline
Eye&\multicolumn{2}{c|}{OS (Left Eye)} & \multicolumn{2}{c|}{OD (Right Eye)}&\multicolumn{2}{c|}{OS (Left Eye)} & \multicolumn{2}{c}{OD (Right Eye)}\\
\hline
 Predictor& MALP & LSLP & MALP & LSLP & MALP & LSLP & MALP & LSLP\\
 \hline\hline
 PCC& 0.868  &0.868 & 0.752  &0.752& 0.601&0.601&0.667&0.667 \\
 CCC&  0.868& 0.859 &0.752& 0.722&0.560&0.466&0.614&0.556 \\
 MSE&  122.735& 114.636 &  228.977& 200.573&488.458&406.421&343.476&277.980 \\
 \hline\hline
\end{tabular}
}
\end{table}
To obtain a more appropriate evaluation of the performance and comparison, which circumvents the data double-dipping, of $\hat{y}^\star(x_\txt{0})$ and $\hat{y}^\dagger(x_\txt{0})$ using this eye data, we utilize a resampling approach. We randomly split the data in half into training and testing sets so that the training set is used for constructing the estimated MALP and the estimated LSLP, respectively. The test set is used for evaluating their performance, where the predictions from the predictors using the Cirrus OCT values are compared to the observed Stratus OCT values. This process is repeated $\mbox{\rm Nreps}=2000$ times. For the resulting 2000 values of the PCC, CCC, and MSE, we then obtain their respective means. The results are summarized in the third main column labeled `Realistic Case' in Table \ref{EYE_EVALUATION}.

The results in Table \ref{EYE_EVALUATION} indicate that the MALP performs better than LSLP when the criterion is the CCC, but the situation is reversed when the criterion is the MSE. Thus, as noted earlier in the computer experiments section, it behooves for the researcher to decide first what desirable property the predictor should have prior to its choice: i) if the goal is to maximize agreement, then the MALP is preferable; while  ii) if the goal is to minimize the MSE, then the LSLP is preferable.  

\begin{figure}[ht!]
\begin{minipage}{0.45\linewidth}
\centering
\underline{\textbf{Conversion Formulae}}
\begin{eqnarray*}
	\hat{y}^\star(x_\txt{0})  &=& 0.765x_0-0.341;\\
	\hat{y}_\txt{Abedi}(x_\txt{0})  &=& 0.760x_0-0.510.
\end{eqnarray*}
\hfill
\end{minipage}
\begin{minipage}{0.45\linewidth}
\centering
	\vspace{-5pt}
		\includegraphics[height=150pt]{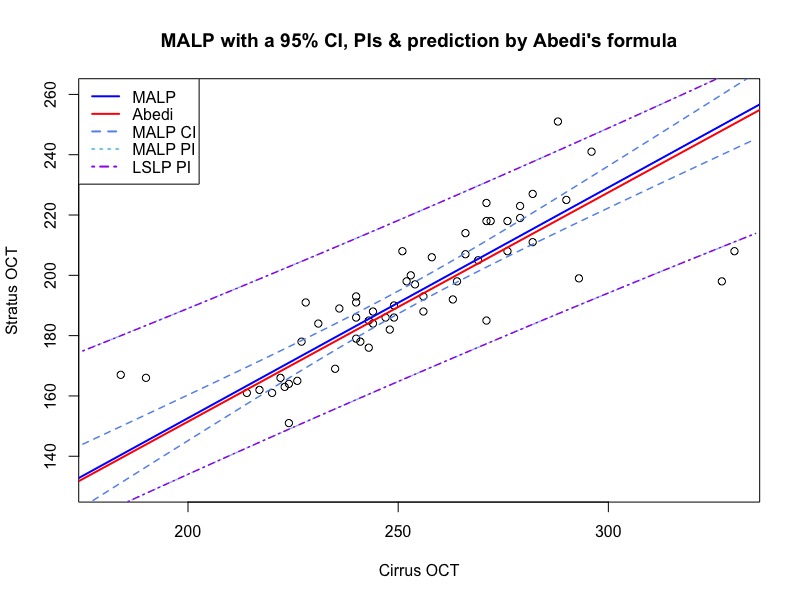}
		\vspace{-15pt}
\end{minipage}
		\caption{Scatter plots of two OCTs superimposed with the estimated prediction lines based on two conversion formulae: the estimated MALP and Abedi's conversion formula. Also included are the 95\% asymptotic CI for the MALP and the 95\% asymptotic prediction intervals based on the LSLP and the MALP. }
		\label{EYE_COMP}
		\vspace{-10pt}
\end{figure}

To obtain a comparable conversion formula with that of \citet{APDS:2011}, we then combined the OD and OS observations, though it should be noted that this is a violation of the independence assumption since it is expected that the left and right eye measurements will be correlated. Using our procedure, we obtained the conversion function given in the first equation in Figure \ref{EYE_COMP}, and we also provide the conversion function of \citet{APDS:2011}, which is the second equation in the same figure. Observe that the two prediction functions (lines) depicted in the plot in Figure \ref{EYE_COMP} are almost identical. 
The slight discrepancy is that the equation of \citet{APDS:2011} was obtained through a grid search, while our procedure was obtained from an analytical derivation, hence it is more accurate. In fact, the CCC of the observed and predicted values from the two equations, which are both about 0.783, differed by less than $10^{-5}$, with the one from our conversion equation just a tad higher. In addition, the 95\% asymptotic normal confidence interval for MALP is presented around the estimated MALP, as well as the prediction intervals based on the formulas in (\ref{PI based on MALP}) and (\ref{SHORTEST_PI}). Observe that the two prediction interval curves are not so different owing to the fact that $\hat{\gamma} =  {0.783}$. In addition, note that the prediction interval curves are not centered on the estimated MALP line, rather they are centered on the estimated LSLP line as previously shown, though this line is not depicted in the plot for simplicity.

\subsection{Body Fat Data Set}
The percent body fat is an important bodily characteristic that serves as a marker for the health status of an individual. Being able to infer its value from easily-measured or determined bodily characteristics, such as age, weight, height, circumference measurements, or skin-fold measurements is therefore of interest, e.g., \citet{BW:1974,KW:1977}. For example, one could obtain the percent body fat from body density via underwater weighting based on Siri's equation $100\times \text{Body fat}= 495/\text{Body Density}-450$ as in \citet{Siri:1956}. Including body density (BD) and percent body fat (PBF), the data set of interest contains 13 additional variables such as age (in years), weight (WGT, in pounds), height (HGT, in inches); several body circumference measurements (in cm): neck (NCK), chest (CST), abdomen (ABD), hip, thigh (TGH), knee (KN), ankle (ANK), biceps (BCP), forearm (FA), and wrist (WRT) for 252 men. The data set was originally in \citet{Pen:1985} and became available to the public courtesy of Dr.\ A.\ Garth Fisher. 

\begin{figure}[ht]
\centering
\includegraphics[height=160pt]{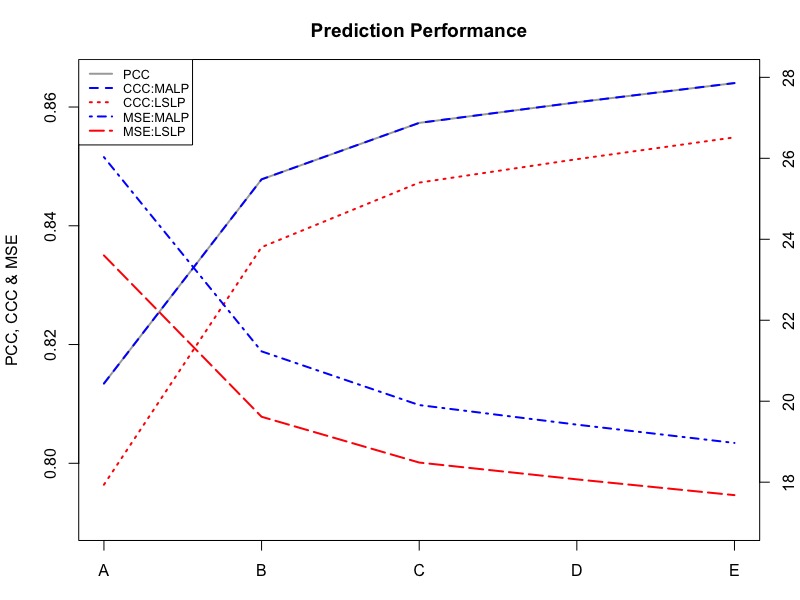}
\caption{Plots of the performance measures (PCC, CCC, and MSE) of the EMALP and ELSLP for the body fat data as the size of subsets of predictors in the model increases. Note that the PCC and CCC of the MALP coincide.}
\label{BODY_FAT_PERF}
\end{figure}

Using the data set, our goal is to provide an illustration of the MALP procedure to predict the percent body fat based on other easily-measured or determined body characteristics. While there are several variables that have high correlations with the percent body fat variable, some of those are highly correlated with each other, e.g., see Table 3 in {Section D of the Appendices} for details. 
Using the exhaustive search method based on LSLP, we selected the subsets of variables that yielded the largest coefficient of determination, $R^2$, for the specified subset sizes of 1, 2, 4, 6, and 8. For each of these subsets of predictor variables, we obtained the estimated MALP and estimated LSLP functions, and calculated the percent body fat predicted values from each predictor for all $n = 252$ subjects. Thus, note immediately that we are double-dipping on the data in that we used it to construct the predictors, and then used the data again to obtain the predicted values. 
\begin{table}[ht]
\caption{Coefficients of the MALP and LSLP with the body fat variable as a dependent variable for different subsets of predictor variables, together with their corresponding performance measures computed on the whole data set.}
\label{BODY_FAT_RESULT}

\centering

{\scriptsize
\begin{tabular}{c|cc|cc|cc|cc|cc} \hline\hline
    & \multicolumn{10}{c}{Coefficients of the Linear Predictors}  \\\hline
   Subset & \multicolumn{2}{c|}{A} & \multicolumn{2}{c|}{B} & \multicolumn{2}{c|}{C} & \multicolumn{2}{c|}{D} & \multicolumn{2}{c}{E} \\ \hline
  Predictor  & MALP & LSLP & MALP& LSLP & MALP & LSLP & MALP & LSLP & MALP  & LSLP    \\\hline
Intercept &-52.68 &-39.28 &-57.64&-45.95&-43.84&-34.85&-47.62&-38.32&-29.24&-22.66\\    
ABD & 0.776    & 0.631 & 1.167 & 0.990 & 1.161 & 0.996 & 1.059 & 0.912 & 1.093 & 0.945 \\
WGT &\multirow{7}{*}{--}   & \multirow{7}{*}{--} &    -0.175 &-0.148 &-0.158 & -0.136&  -0.159     &  -0.136  & -0.104  & -0.090  \\
FA  &      &      & \multirow{6}{*}{--}      & \multirow{6}{*}{--}    & 0.552&0.473  & 0.568      &    0.489 & 0.597  &   0.516  \\
WRT &      &      &      &     & -1.756&-1.506 & -2.067      &    -1.779&-1.778   &     -1.537  \\
AGE &      &      &   &       &  \multirow{4}{*}{--}      &  \multirow{4}{*}{--}     &0.073       &    0.063 & 0.076  &    0.066    \\
TGH &      &      &    &       &       &      & 0.256      &      0.220& 0.350  &    0.302    \\
NCK &      &      &     &      &       &      &  \multirow{2}{*}{--}      &  \multirow{2}{*}{--}     & -0.540&-0.467    \\
HIP &      &      &     &      &      &       &        &       &-0.226  &  -0.195 \\ \hline\hline
PCC & 0.813 & 0.813 &  0.848 & 0.848  & 0.857 & 0.857 & 0.861 & 0.861 & 0.864 & 0.864    \\ \hline
CCC & 0.813 & 0.796 &  0.848 & 0.836  & 0.857 & 0.847 & 0.861 & 0.851 & 0.864 & 0.855    \\ \hline
MSE & 26.029& 23.601& 21.232 & 19.616 & 19.905 & 18.485 & 19.421 & 18.070 & 18.969 & 17.680  \\  \hline \hline            
\end{tabular}}

\end{table}
Table \ref{BODY_FAT_RESULT} contains the coefficients of these linear predictors for each of the subsets, together with their associated performance measures given by the PCC, CCC, and MSE, calculated using the observed and predicted values of the percent body fat variable. 
Figure \ref{BODY_FAT_PERF} presents these performance values of the MALP and LSLP pictorially. 
As expected from classical multiple linear regression theory, the PCC of the LSLP, which is also the PCC of the MALP from an earlier result, equals the multiple correlation coefficient, as could be seen from Table \ref{BODY_FAT_RESULT}.  In addition, observe that as the subset of predictor variables used in the model increases in size, the performance of the predictors also improves, though there is a diminishing rate of improvement as could be seen from Figure \ref{BODY_FAT_PERF}. However, as we point out in the Concluding Remarks section, these measures need to be adjusted for the number of predictor variables used, akin to the adjusted-$R^2$ in classical multiple linear regression, if a proper comparison of performance of prediction functions with a different number of predictor variables is to be done. This issue remains an open research problem. 

Lastly, the panels in Figure 8 in the {Appendices} show the scatter plots of the pairs of observed and predicted values from each of the fitted prediction functions. Looking at the scatter plots, for both the MALP and the LSLP, the point clouds appear to cluster around the 45$^\circ$ line, with those for the MALP tending to be a tad more closely clustered to this line than those for the LSLP, which is to be expected from its development since it is focused towards achieving a higher agreement between predicted and predictand values. But note also that there is more variability in the MALP predictions compared to the LSLP predictions, which is the price one pays by trying to achieve a higher degree of agreement, consistent with the universal adage that {\em there is no free lunch!}

\section{Concluding Remarks} 
\label{sec: Conclusions}

In this study, we investigated a new predictor, called the MALP, which is designed to maximize the agreement between the predictor and predictand as measured by the $\CCC$. For the  case with a single-dimensional covariate or feature, the predictor function coincides with those derived through other approaches, such as the geometric mean regression or the least-triangle regression. However, the development of our MALP allows an extension to the case with multi-dimensional covariates or features, which may not be possible or easy to achieve with the other approaches relying on geometric motivations. In addition, the MALP has a computational profile that is of the same order as the least-squares linear predictor (LSLP) since it is obtainable as a linear transformation of the LSLP with coefficients depending on the multiple correlation coefficient. 

Distributional properties of the estimated MALP were obtained. Asymptotic normality was obtained under general joint distributions using the Multivariate Central Limit Theorem and the Delta Method.  A very neat closed-form expression of the asymptotic variance was derived under the multivariate normal distribution model. Computer experiments showed that the quality of the normal approximation and the approximations of asymptotic mean and variance are quite satisfactory when the multiple correlation coefficient is not too close to zero and the sample size is at least moderate. It was also demonstrated via the simulation studies for a few sets of parameter values under normality that the interval estimators for the MALP are satisfactory. 
The MALP-based prediction procedures were also demonstrated using two real data sets: an eye data set and a body fat data set.

Comparisons between MALP and LSLP were also performed. The MALP and LSLP belong to the class of linear predictors that maximize the $\PCC$ between the predictor and the predictand, with the MALP uniquely maximizing the $\CCC$, whereas the LSLP is uniquely minimizing the $\MSE$. Whether one should prefer the LSLP over the MALP, or the MALP over the LSLP, depends on  the performance measure that is desired for the predictor. Thus, if one is seeking a linear predictor that will have a high CCC with what is being predicted, then the MALP should be preferred; while if the MSE is the desired criterion, then the LSLP is the preferred prediction function.

We conclude by pointing out some open research problems related to the maximum agreement predictor. First, when comparing the performance of MALPs with different sets of predictor variables, simply computing their CCCs could be a misleading comparison measure since they are not adjusted for the number of predictor variables being used, cf., in the illustrative example using the body fat data set. So an important question is how could the CCC be properly adjusted to account for the number of predictor variables, akin to the adjusted-$R^2$ in classical multiple linear regression? Second, there is the problem of developing hypothesis tests and confidence intervals for the coefficients of the MALP.  Are there analogous procedures to the inferential procedures in classical multiple linear regression? Third, when there are many potential predictor variables or features, how could the MALP approach be utilized for variable selection? This may require a resolution of the first problem described above. Another avenue of investigation is to incorporate a regularization component in the MALP derivation, leading to a so-called {\em regularized} MALP.

\section*{Acknowledgements}
We are very grateful to Dr.\ Gelareh Abedi for providing and letting us utilize the eye data set in the illustration section and in the {\tt R} package {\tt malp}. 
%

\section*{Disclaimer}
E.\ Pe\~na was Program Director in the Division of Mathematical Sciences at the National Science Foundation (NSF) from 2020-2023. He received support (NSF Grant 2049691) for research, which included some work in this paper. Any opinions, findings, conclusions, or recommendations expressed in this material are those of the authors and do not necessarily reflect the views of NSF.

\begin{center}
{\large\bf R-Package Information}
\end{center}
The R-package {\tt malp} implements the maximum agreement linear prediction and provides the two data sets which are used in the paper. A vignette that describes how to use the different functions is also included in the package. It is available online on the development page \url{https://github.com/pchausse/malp}. To install the package in R, install and load the package {\tt devtools} and run the command {\tt install\_github("pchausse/malp")}.

\bibliographystyle{imsart-nameyear}
\bibliography{main.bib}

\newpage


\begin{appendices}
This Appendices portion consists of the proof of Theorem 1 in section \ref{Proofs}, additional results of the computer experiments in section \ref{SUPPLE_SEC_EXPERIMENT}, the forms of confidence interval procedures in section \ref{SUPPLE_CI}, and the description of the body fat data set in section \ref{SUPPLE_BF_DATA}. 
\setcounter{figure}{0}   
\setcounter{table}{0}  
\section{Proofs}
\label{Proofs}

In this section we provide proofs of Theorem 1 and Theorem 4.

\subsection{Proof of Theorem 1}

\begin{proof}
Consider the class of linear predictors $\mathcal{H}_\txt{LP}$. We seek an  element  $\tilde{Y}^\star(\cdot) \in \mathcal{H}_\txt{LP}$ such that $\rho^\txt{c}_{\txt{Y}\tilde{\txt{Y}}^\star}:=\rho^\txt{c}(Y,\tilde{Y}^\star(X)) = \sup_{\{\tilde{Y} \in \mathcal{H}_\txt{LP}\}} \rho^\txt{c}(Y,\tilde{Y}(X))$. Under the linearity condition, the moments for the MALP can be expressed in the following concrete forms:
\begin{equation*}
\mu_{\tilde{\txt{Y}}}  \ = \ \alpha+\mu_\txt{X}\beta,\;\; \sigma^2_{\tilde{\txt{Y}}} \ = \ \beta^{\trp} \Sigma_{\txt{XX}}\beta, \text{ and }\sigma_{\txt{Y}\tilde{\txt{Y}}} \ = \ \Sigma_{\txt{YX}}\beta.	
\end{equation*}
Then, the $\CCC$ between the linear predictor and the predictand becomes:
\begin{equation}\label{CCC_LINEAR}
\rho^\txt{c}_{\txt{Y}\tilde{\txt{Y}}} \ = \ \rho^\txt{c}(Y,\tilde{Y}(X)) \ = \ \frac{2\Sigma_\txt{YX}\beta}{\sigma_\txt{Y}^2+\beta^{\trp}\Sigma_\txt{XX}\beta+(\mu_\txt{Y}-\alpha-\mu_\txt{X}\beta)^2}	
\end{equation}
from the definition of CCC. First, we can maximize $\rho^\txt{c}_{\txt{Y}\tilde{\txt{Y}}}$ by taking $\alpha^\star=\mu_\txt{Y}-\mu_\txt{X}\beta$ for any given $\beta$. In such a case, the form of $\CCC$ in (\ref{CCC_LINEAR}) is simplified:
\begin{equation}\label{EQUATION_1}
	\rho^\txt{c}_{\txt{Y}\tilde{\txt{Y}}} \ = \ \frac{2\Sigma_\txt{YX}\beta}{\sigma_\txt{Y}^2+\beta^{\trp}\Sigma_\txt{XX}\beta}.
\end{equation}
By taking the logarithm on both sides, the equation becomes 
\begin{equation*}
\log\rho^\txt{c}_{\txt{Y}\tilde{\txt{Y}}} \ = \ \log(2\Sigma_\txt{YX}\beta)-\log(\sigma_\txt{Y}^2+\beta^{\trp}\Sigma_\txt{XX}\beta).
\end{equation*}
To optimize with respect to $\beta$, we obtain the gradient and equate it to zero:
$$\nabla_\beta\log\rho^\txt{c}_{\txt{Y}\tilde{\txt{Y}}} \ = \ \frac{\Sigma_\txt{XY}}{\Sigma_\txt{YX}\beta}-\frac{2\Sigma_\txt{XX}\beta}{\sigma_\txt{Y}^2+\beta^{\trp}\Sigma_\txt{XX}\beta} \ \stackrel{\text{set}}{=} \ 0.$$
{Multiplying by $\beta^{\trp}$, we obtain
$$1 = \frac{2\beta^{\trp}\Sigma_\txt{XX}\beta}{\sigma_\txt{Y}^2+\beta^{\trp}\Sigma_\txt{XX}\beta}$$}
so that
\begin{equation}\label{CONSTRAINT_1}
	\sigma_\txt{Y}^2 \ = \ \beta^{\trp}\Sigma_\txt{XX}\beta.
\end{equation}
Replacing $\beta^{\trp}\Sigma_\txt{XX}\beta$ by $\sigma_\txt{Y}^2$ in (\ref{EQUATION_1}), 
the maximal $\rho^\txt{c}_{\txt{Y}\tilde{\txt{Y}}}$ is of  form
\begin{equation}\label{EQUATION_2}
	\rho^\txt{c}_{\txt{Y}\tilde{\txt{Y}}} \ = \ \frac{\Sigma_\txt{YX}\beta}{\sigma_\txt{Y}^2}.
\end{equation}
Now, we need to maximize with respect to $\beta$ subject to the condition (\ref{CONSTRAINT_1}). Set the Lagrange equation as follows:
\begin{equation*}
	\mathcal{L}(\beta,\lambda) \ = \ \Sigma_\txt{YX}\beta+\lambda(\beta^{\trp}\Sigma_\txt{XX}\beta-\sigma_\txt{Y}^2).	
\end{equation*}
Obtaining the gradient for $\beta$ and $\lambda$ and equating it to zero
\begin{equation*}
	\nabla_\beta\mathcal{L}(\beta,\lambda) \ = \ \Sigma_\txt{XY}+\lambda2\Sigma_\txt{XX}\beta \ \stackrel{\text{set}}{=} \ 0\;\text{ and }\;
	\nabla_\lambda\mathcal{L}(\beta,\lambda) \ = \ \beta^{\trp}\Sigma_\txt{XX}\beta - \sigma_\txt{Y}^2 \ \stackrel{\text{set}}{=} \ 0,
\end{equation*}
we have the following equation:
\begin{equation}\label{BETA}
	\beta \ = \  -\frac{\Sigma_\txt{XX}^{-1}\Sigma_\txt{XY}}{2\lambda}.
\end{equation}
The constraint in (\ref{CONSTRAINT_1}) then becomes 
\begin{equation*}
	\sigma_\txt{Y}^2 \ = \ \left(-\frac{\Sigma_\txt{YX}\Sigma_\txt{XX}^{-1}}{2\lambda}\right)\Sigma_\txt{XX}\left(-\frac{\Sigma_\txt{XX}^{-1}\Sigma_\txt{XY}}{2\lambda}\right) \ = \ \frac{\Sigma_\txt{YX}\Sigma_\txt{XX}^{-1}\Sigma_\txt{XY}}{(2\lambda)^2}.
\end{equation*}
Therefore, 
\begin{equation}\label{LAMBDA_PM}
	2\lambda \ = \ \pm\sqrt{\frac{\Sigma_\txt{YX}\Sigma_\txt{XX}^{-1}\Sigma_\txt{XY}}{\sigma_\txt{Y}^2}}.
\end{equation}
In order to choose the sign of $\lambda$, note that it must be $\Sigma_\txt{YX}\beta\ge0$ in (\ref{EQUATION_2}) to maximize $\rho^\txt{c}_{\txt{Y}\tilde{\txt{Y}}}$. Then, 
\begin{equation}\label{EQUATION_3}
	\rho^\txt{c}_{\txt{Y}\tilde{\txt{Y}}}\ = \ \frac{\Sigma_\txt{YX}\beta}{\sigma_\txt{Y}^2} \ = \ \frac{\Sigma_\txt{YX}\left(-\frac{\Sigma_\txt{XX}^{-1}\Sigma_\txt{XY}}{2\lambda}\right)}{\sigma_\txt{Y}^2} \ = \ \left(-\frac{1}{2\lambda}\right)\frac{\Sigma_\txt{YX}\Sigma_\txt{XX}^{-1}\Sigma_\txt{XY}}{\sigma_\txt{Y}^2}.
\end{equation}
Thus, it has to be $2\lambda<0$ in (\ref{LAMBDA_PM}), so that the optimal $\alpha^\star$ and $\beta^\star$ can be obtained as desired from (\ref{BETA}): 
\begin{equation*}
		\beta^\star \ = \ \frac{\sigma_\txt{Y}\Sigma_\txt{XX}^{-1}\Sigma_\txt{XY}}{\sqrt{\Sigma_\txt{YX}\Sigma_\txt{XX}^{-1}\Sigma_\txt{XY}}}\;\text{ and }\; \alpha^\star \ = \ \mu_\txt{Y}-\mu_\txt{X}\beta^\star.
\end{equation*}
The resulting MALP is $\tilde{Y}^\star(X) = \alpha^\star+X\beta^\star$ with $\alpha^\star$ and $\beta^\star$ given above. Lastly, the maximum CCC can be obtained from (\ref{EQUATION_3}):
\begin{equation}\label{MAXIMUM_CCC_GAMMA}
	\rho^\txt{c}_{\txt{Y}\tilde{\txt{Y}}^\star} \ = \ \frac{(\Sigma_\txt{YX}\Sigma_\txt{XX}^{-1}\Sigma_\txt{XY})/\sigma_\txt{Y}^2}{\sqrt{(\Sigma_\txt{YX}\Sigma_\txt{XX}^{-1}\Sigma_\txt{XY})/\sigma_\txt{Y}}} \ = \ \frac{\sqrt{\Sigma_\txt{YX}\Sigma_\txt{XX}^{-1}\Sigma_\txt{XY}}}{\sigma_\txt{Y}} =\gamma.
\end{equation}

To show uniqueness of the MALP, suppose we have two MALPs, $\tilde{Y}^\star_1(x) = \alpha^\star + x\beta^\star$ and $\tilde{Y}^\star_2(x) = \alpha^{\star\star} + x\beta^{\star\star}$. Then, we have the following relations from the first part of Remark 1-iii in the paper:
\begin{equation*}
    \rho^\txt{c}_{\txt{Y}\tilde{\txt{Y}}_1^\star}\le \rho_{\txt{Y}\tilde{\txt{Y}}_1^\star}\le\gamma\;\; \text{ and }\;\; \rho^\txt{c}_{\txt{Y}\tilde{\txt{Y}}_2^\star}\le \rho_{\txt{Y}\tilde{\txt{Y}}_2^\star}\le\gamma.
\end{equation*}
In addition, due to the result in (\ref{MAXIMUM_CCC_GAMMA}), the following equalities hold:
\begin{equation*}
        \rho^\txt{c}_{\txt{Y}\tilde{\txt{Y}}_1^\star}=\rho^\txt{c}_{\txt{Y}\tilde{\txt{Y}}_2^\star}= \rho_{\txt{Y}\tilde{\txt{Y}}_1^\star}=\rho_{\txt{Y}\tilde{\txt{Y}}_2^\star}=\gamma.
\end{equation*}
Now, from the second part of Remark 1-iii, the above equalities hold when 
$\E[\tilde{Y}^\star_1(X)]=\E[\tilde{Y}^\star_2(X)]=\E[Y]$ and $\Var[\tilde{Y}^\star_1(X)]=\Var[\tilde{Y}^\star_2(X)]=\Var[Y]$. However, the only possible case in order for these conditions to hold is when $\alpha^\star = \alpha^{\star\star}$ and $\beta^\star = \beta^{\star\star}$, thereby completing the proof of the uniqueness of the MALP.
\end{proof}

\subsection{Proof of Theorem 4}
When the distribution $F$ of $(Y,X)$ follows a multivariate normal with known parameters, then
\begin{equation}\label{DIST_NEW_OBS}
    Y(x_0) | (X = x_0) \sim \mathcal{N}\left(\tilde{Y}^\dagger(x_0) = \mu_\txt{Y}+\Sigma^{-1}_\txt{XX}\Sigma_\txt{XY}(x_0-\mu_\txt{X}),\sigma_\txt{Y}^2(1-\rho^2)\right).
\end{equation}
We present prediction intervals for $Y(x_0)$ under this distributional setting.
We commence with the situation where the parameter values are known. Starting with the pivotal quantity 
\begin{displaymath}
Q_1 = \frac{Y(x_0) - \tilde{Y}^\dagger(x_0)}
{\sigma_\txt{Y} \sqrt{1-\rho^2}}
\end{displaymath}
which, given $X = x_0$, has a standard normal distribution, the resulting $100(1-\alpha)\%$ prediction interval for $Y(x_0)$ is
\begin{equation}\label{PI_KNOWN_PARA}
    \left[\tilde{Y}^\dagger(x_0)\pm z_{\alpha/2}\sigma_\txt{Y}\sqrt{1-\rho^2}\right].
\end{equation}

For the unknown parameter case, we start with LSLP based on the following result:
\begin{equation*}
    \hat{Y}^\dagger(x_0)=\bar{Y}+S_\txt{XX}^{-1}S_\txt{XY}(x_0-\bar{X})\sim \mathcal{AN}\left(\tilde{Y}^\dagger(x_0),\tfrac{1}{n}\sigma_\txt{LS}^2(x_0)\right),
\end{equation*}
such that $\sigma_\txt{LS}^2(x_0)$ is consistently estimated by $$\hat{\sigma}^2_\txt{LS}(x_0)=S_\txt{Y}^2(1-\hat{\gamma}^2)\left[1+(x_0-\bar{X})S_\txt{XX}^{-1}(x_0-\bar{X})^{\trp}\right]$$ with $\hat{\gamma}^2={S_\txt{YX}S_\txt{XX}^{-1}S_\txt{XY}}/{S_\txt{Y}^2}$. {We note, however, that there could be other consistent estimators of $\sigma_\txt{LS}^2(x_0)$ (e.g., unbiased estimators) which could have better performance when the sample size is small to moderate, and they will be provided as an option in the {\tt R} package {\tt malp}.} Note that $\E\left[Y(x_0)-\hat{Y}^\dagger(x_0)\middle|X=x_0\right]=\tilde{Y}^\dagger(x_0)-\tilde{Y}^\dagger(x_0)=0$ and $\Var\left[Y(x_0)-\hat{Y}^\dagger(x_0)\middle|X=x_0\right]=\sigma_\txt{Y}^2(1-\gamma^2)\left\{1+\tfrac{1}{n}\left[1+(x_0-\mu_\txt{X})\Sigma_\txt{XX}^{-1}(x_0-\mu_\txt{X})^{\trp}\right]\right\}$.
Since the conditional variance can be consistently estimated by $$S_\txt{Y}^2(1-\hat{\gamma}^2)\left\{1+\tfrac{1}{n}\left[1+(x_0-\bar{X})S_\txt{XX}^{-1}(x_0-\bar{X})^{\trp}\right]\right\},$$ an asymptotic pivotal quantity  is
\begin{equation}\label{PI_PIVOT}
   Q_2 =  \frac{Y(x_0)-\hat{Y}^\dagger(x_0)}{S_\txt{Y}\sqrt{1-\hat{\gamma}^2}\sqrt{1+\tfrac{1}{n}\left[1+(x_0-\bar{X})S_\txt{XX}^{-1}(x_0-\bar{X})^{\trp}\right]}} \stackrel{d}{\longrightarrow} \mathcal{N}(0,1).
\end{equation}
Using this pivotal quantity, the shortest length asymptotic $100(1-\alpha)\%$ prediction interval for $Y(x_0)$ is
\begin{equation}
   \left[\hat{Y}^\dagger(x_0)\pm z_{\alpha/2}S_\txt{Y}\sqrt{(1-\hat{\gamma}^2)}\sqrt{1+\tfrac{1}{n}\left[1+(x_0-\bar{X})S_\txt{XX}^{-1}(x_0-\bar{X})^{\trp}\right]} \right].
\end{equation}
%
Such a prediction interval provides a quantification of the uncertainty inherent in the prediction of $Y(x_0)$ using $\hat{Y}^\dagger(x_0)$. 

How about if we use the estimated MALP $\hat{Y}^\star(x_0)$ as the predicted value of $Y(x_0)$? How do we quantify the uncertainty involved in such a prediction? We may proceed analogously as in the preceding except now utilize the MALP instead of the LSLP, starting with the difference: $Y(x_0) - \hat{Y}^\star(x_0)$. 
First, recall the asymptotic result, conditional on $X = x_0$, in which we have
\begin{equation*}
    \hat{Y}^\star(x_0)=\bar{Y}+\tfrac{1}{\hat{\gamma}}S_\txt{XX}^{-1}S_\txt{XY}(x_0-\bar{X})\sim\mathcal{AN}\left(\tilde{Y}^\star(x_0),\tfrac{1}{n}\sigma_\txt{MA}^2(x_0)\right),
\end{equation*}
where $\sigma_\txt{MA}^2$ is defined in Theorem 3.1. Thus, we have $\E\left[Y(x_0)-\hat{Y}^\star(x_0)\middle|X=x_0\right]=Y^\dagger(x_0)-Y^\star(x_0)=\left(1-\tfrac{1}{\gamma}\right)\Sigma_\txt{XX}^{-1}\Sigma_\txt{XY}(x_0-\mu_\txt{X}) \equiv b(x_0)$.
Therefore, under multivariate normality, we have that, under $X_0 = x_0$ and for large $n$,
\begin{equation}
\label{pivot Q3}
Q_3 = \frac{Y(x_0) - \hat{Y}^\star(x_0) - b(x_0)}{\sigma_\txt{Y} \sqrt{1-\gamma^2} \sqrt{1 + \frac{1}{n} D_\txt{MA}^2(x_0)}} \stackrel{\cdot}{\sim} N(0,1),
\end{equation}
where $$D_\txt{MA}^2(x_0) = \frac{2}
{1+{\gamma}} +  \frac{1}{{\gamma}^2} (x_0-\mu_\txt{X})\Sigma_\txt{XX}^{-1}(x_0-\mu_\txt{X})^{\trp} - \left[\frac{(1-{\gamma}^2)}{\sigma_\txt{Y}^2 {\gamma}^4}\right] \left[ \Sigma_\txt{XX}^{-1} \Sigma_\txt{XY}(x_0 - \mu_\txt{X})\right]^2.$$
Using this as a pivotal quantity, and assuming first that we know $b(x_0)$ and $\sigma_\txt{MA}^2(x_0)$, an approximate $100(1-\alpha)\%$ prediction interval for $Y(x_0)$ is
\begin{equation*}
\left[
(\hat{Y}^\star(x_0) + b(x_0)) \pm z_{\alpha/2} \sigma_\txt{Y} \sqrt{1-\gamma^2} \sqrt{1 + \frac{1}{n} D_\txt{MA}^2(x_0)}
\right].
\end{equation*}
The bias $b(x_0)$ is estimated by $\hat{b}(x_0) = \left(1-\tfrac{1}{\hat{\gamma}}\right)S_\txt{XX}^{-1}S_\txt{XY}(x_0-\bar{X})$, while $\sigma_\txt{Y}^2 (1-\gamma^2) (1 + \frac{1}{n} D_\txt{MA}^2(x_0))$ is consistently estimated by
$ S_\txt{Y}^2 (1 - \hat{\gamma}^2) \left[1 + \frac{1}{n} \hat{D}_\txt{MA}^2({x}_0)\right]$
with 
\[\hat{D}_\txt{MA}^2(x_0) = \frac{2}
{1+{\hat{\gamma}}} +  \frac{1}{{\hat{\gamma}}^2} (x_0-\bar{X}) S_\txt{XX}^{-1}(x_0-\bar{X})^{\trp} - \left[\frac{(1-{\hat{\gamma}}^2)}{S_\txt{Y}^2 {\hat{\gamma}}^4}\right] \left[ S_\txt{XX}^{-1} S_\txt{XY}(x_0 - \bar{X})\right]^2.\]
Again, there could be other consistent estimators of $\sigma_\txt{MA}^2(x_0)$ which could have better performances when the sample size is small to moderate.
As such, an approximate prediction interval for $Y(x_0)$, when using the estimated MALP as the predictor, is given by 
\begin{equation}
\left[
(\hat{Y}^\star(x_0) + \hat{b}(x_0)) \pm
z_{\alpha/2} S_\txt{Y} \sqrt{1 - \hat{\gamma}^2} \sqrt{ 1 + \frac{1}{n} \hat{D}_\txt{MA}^2(x_0)}\right].
\end{equation}

\section{Additional Computer Experiments}
\label{SUPPLE_SEC_EXPERIMENT}

This section complements the computational experiments reported in Section 4 of the paper. The first and second subsections provide additional results of the first computational experiment: the $p=2$ case and the LSLP case, respectively. The third subsection investigates relationships between predictor and predictand for the estimated MALP (EMALP) and the estimated LSLP (ELSLP), respectively. The last subsection ascertains the asymptotic normality results of the EMALP and the ELSLP with respect to predicting the $Y(x_0)$-values at fixed $x_0$-values under the three different parameter sets in Table 1.

\subsection{Computer Experiment 1: 
EMALP with $p=2$}
\label{SUPPLE_SUBSEC_EXPERIMENT_1}
\begin{figure}[ht]
    \centering
    \includegraphics[height=230pt]{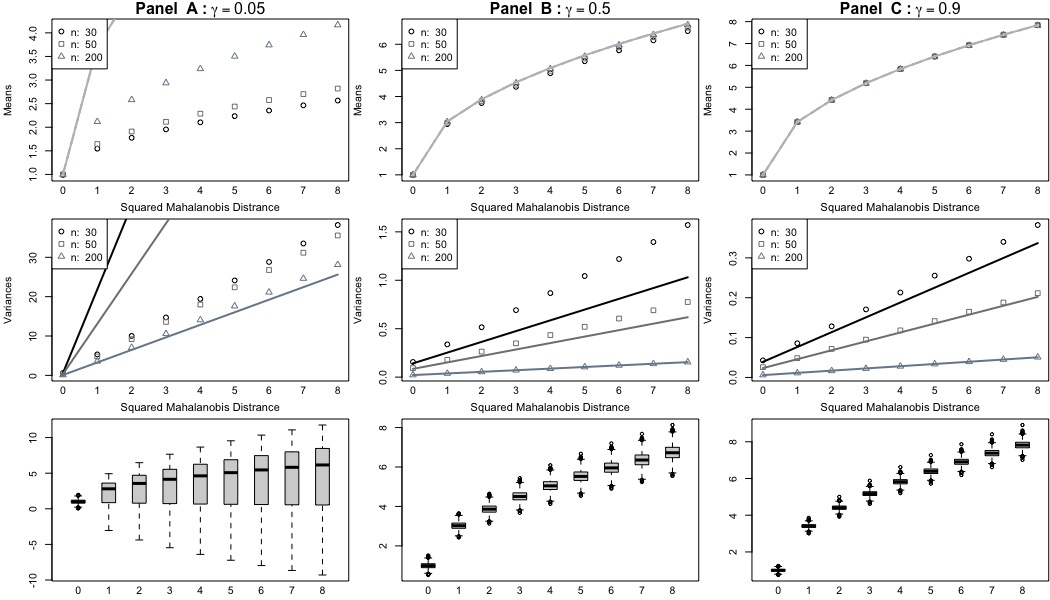}
    \caption{{\small The approximation quality for the EMALP for different $(\gamma,n)$ and $p=2$: the empirical means and variances (dotted) of $\hat{Y}_0^\star$s with their asymptotic approximations (solid curves) under the parameter sets. Also depicted are the boxplots for the case where $n=200$.}}
    \label{Simul_11_MAP_P}
\end{figure}

\begin{figure}[ht]
    \centering
    \includegraphics[height=135pt]{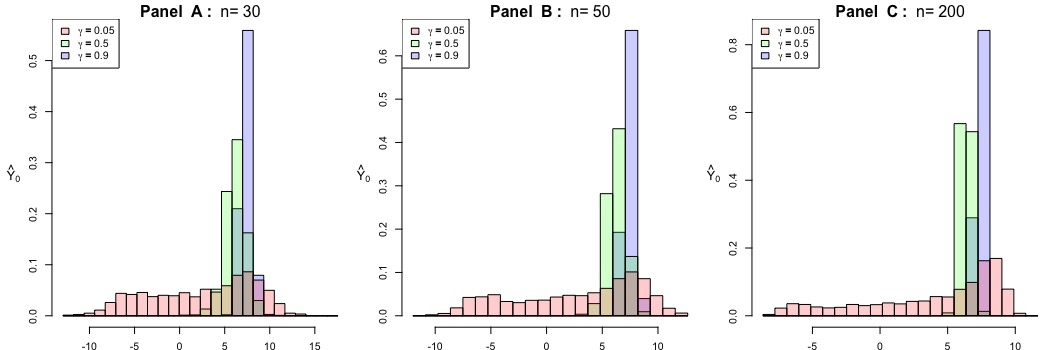}
    \caption{{\small Empirical histograms of the EMALP $\hat{Y}_0^\star$s when the squared Mahalanobis distance is 7 for different values of $(\gamma,n)$ and $p=2$ for the parameter sets based on 2000 replications.}}
    \label{Simul_12_MAP_P}
\end{figure}

We set $\mu = (2,3,1)$, and three covariance matrices were selected for $\Sigma$ in order that their respective $\gamma$ values are approximately 0.05, 0.5, and 0.9. The prediction points are chosen to be the intersections in the first quadrant between the contours where the squared Mahalanobis distances are $0,\;1,\;2,\;\ldots,\; 7$ and a line crossing through $\mu_\txt{X}$ with a {randomly chosen positive slope equal to $u_1/u_2$, where $u_1$ and $u_2$ are realizations of two independent uniform(0,1) random variables. }The resulting plots are in Figures \ref{Simul_11_MAP_P} and \ref{Simul_12_MAP_P}.

In general, the asymptotic approximation quality when $p=2$ is similar to that when $p=1$, with respect to $n$ and $\gamma$. When $\gamma$ is small as in Panel A, it shows unsatisfactory approximations for the means and variances for small $n$. However, the approximations improve as $n$ increases. Observe also from the boxplots that when $\gamma$ is small, there are more outliers on one side, which is indicative of a highly left-skewed distribution. The approximation quality improves as $\gamma$ increases as seen in Panels B and C. For the means, the empirical means lie on the curve depicting the asymptotic mean function; whereas, for the variances, there are still discrepancies when the sample size is small. We surmise, and some of our (unreported) simulation results provide support to this statement, that, as $p$ increases, larger $n$ will be needed for the asymptotic approximations to be acceptable. In this simulation, the asymptotic variance is linear with respect to the squared Mahalanobis distance due to the choice of the set of $x_0$-values. However, in general, the asymptotic variance is not linear with respect to the squared Mahalanobis distance when $p >1$.


\subsection{Computer Experiment 1: ELSLP with $p=1$}
\label{SUPPLE_SUBSEC_EXPERIMENT_1_LSLP}
\begin{figure}[ht]
\centering
    \includegraphics[height=230pt]{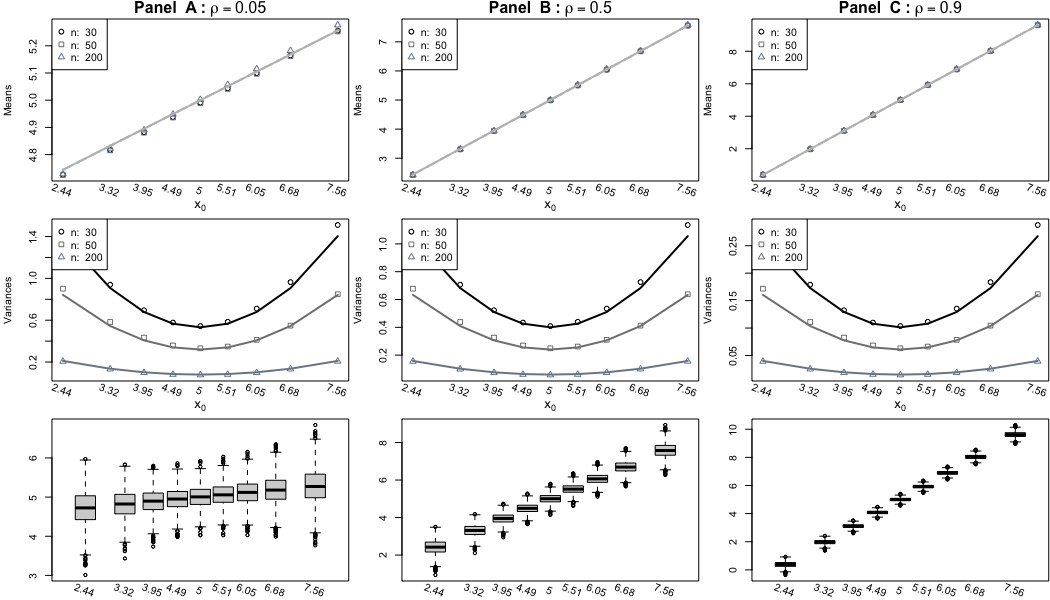}
    \caption{{\small The approximation quality for the ELSLP for different $(\rho,n)$: the empirical means and variances (dotted) of $\hat{Y}_0^\dagger$s with their asymptotic approximations (solid curves). Also depicted are the boxplots for the case when $n=200$. }}
    \label{Simul_11_LSP}
\end{figure}

\begin{figure}[ht!]
\centering
    \includegraphics[height=135pt]{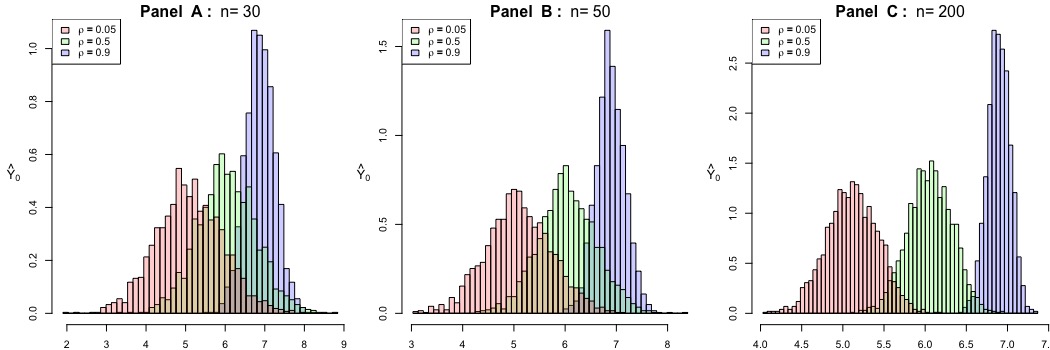}
    \caption{{\small Empirical histograms of the ELSLP $\hat{Y}_0^\dagger$s at $x_0=6.05$ for different values of $(\rho,n)$ under parameter set 1 based on 2000 replications.}}
    \label{Simul_12_LSP}
\end{figure}

The experimental results in Figure \ref{Simul_11_LSP} show that the quality of normal approximation for the ELSLP is somewhat better than that of the MALP. In particular, the LSLP consistently provides a bell-shaped distribution for $\rho=0.05$, whereas the MALP showed unsatisfactory approximation results for the same $\rho$ value for the sample sizes considered in the study.
This can also be observed in the histograms in Figure \ref{Simul_12_LSP}, showing consistently bell-shaped distributions, compared to the bimodal distribution observed in the MALP case with $\rho=0.05$.

\subsection{Computer Experiment 3}
\label{SUPPLE_SUBSEC_EXPERIMENT_2}
\begin{figure}[ht]
    \centering
    \includegraphics[height=230pt]{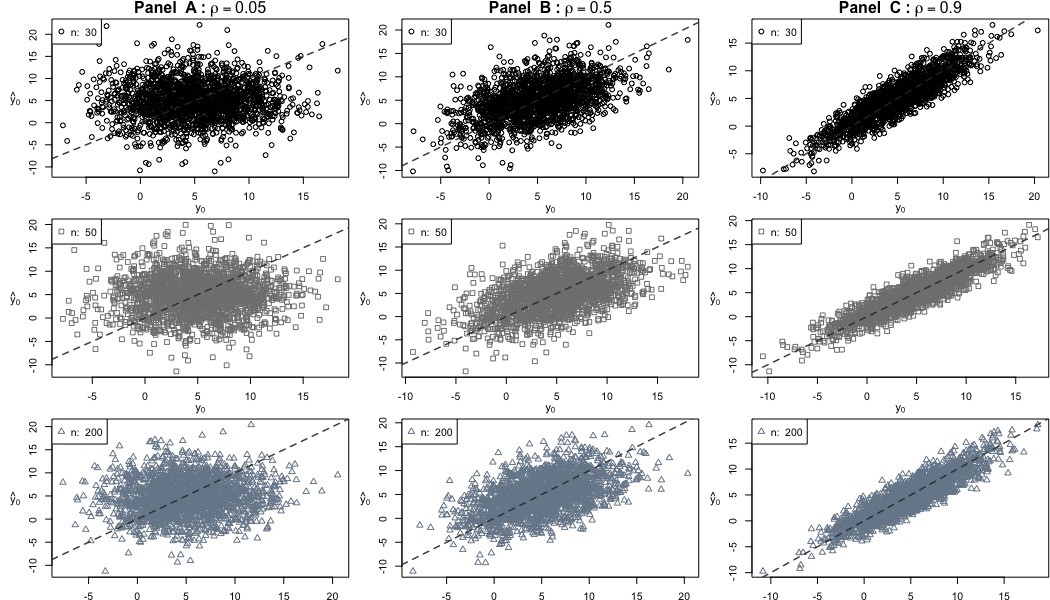}
    \caption{{\small Scatter plots of $Y_0$ and $\hat{Y}_0^\star$, together with the $45^\circ$ line, for different ($\rho,n$) combinations.}}
    \label{Simul_2_MAP}
\end{figure}

In the third set of computer experiments, again for the specified parameter set 1 in Table 1 with $\rho$ replaced by a value in $\{0.05, 0.5,0.9\}$ and sample size $n$, for the $l$th replication with $l = 1,\ldots,\MReps$, a random sample $(X_l,Y_l) =  \{(X_{li},Y_{li}), i=1,\ldots,n\}$, as well as a single realization $(X_{0l},Y_{0l})$, is generated from the bivariate normal distribution with parameters $(\mu_\txt{X},\mu_\txt{Y},\sigma_\txt{X},\sigma_\txt{Y},\rho)$. Using this sample, the MALP and LSLP predictions of $Y_{0l}$, given $X_{0l}$, are obtained, denoted by $\hat{Y}_{0l}^\star$ {and $\hat{Y}_{0l}^\dagger$, respectively}. Figure \ref{Simul_2_MAP} consists of scatterplots of the pairs $\{(Y_{0l},\hat{Y}_{0l}^\star), l=1,\ldots,\MReps\}$ for different $\rho$ and $n$ values. Observe that as $\rho$ becomes larger, the major axis of these scatterplots is the $45^\circ$ line, which is to be expected from the MALP approach.
For the pairs of $Y_0$ and $\hat{Y}_0^\star$, the empirical versions of $\CCC$, $\PCC$, and $\MSE$ (more precisely, the $\MSPE$) are utilized, and their values associated with the $Y_0$ and $\hat{Y}_0^\star$ are summarized in Table \ref{Scatter_MAP}. 

\begin{figure}[ht]
\centering
    \includegraphics[height=230pt]{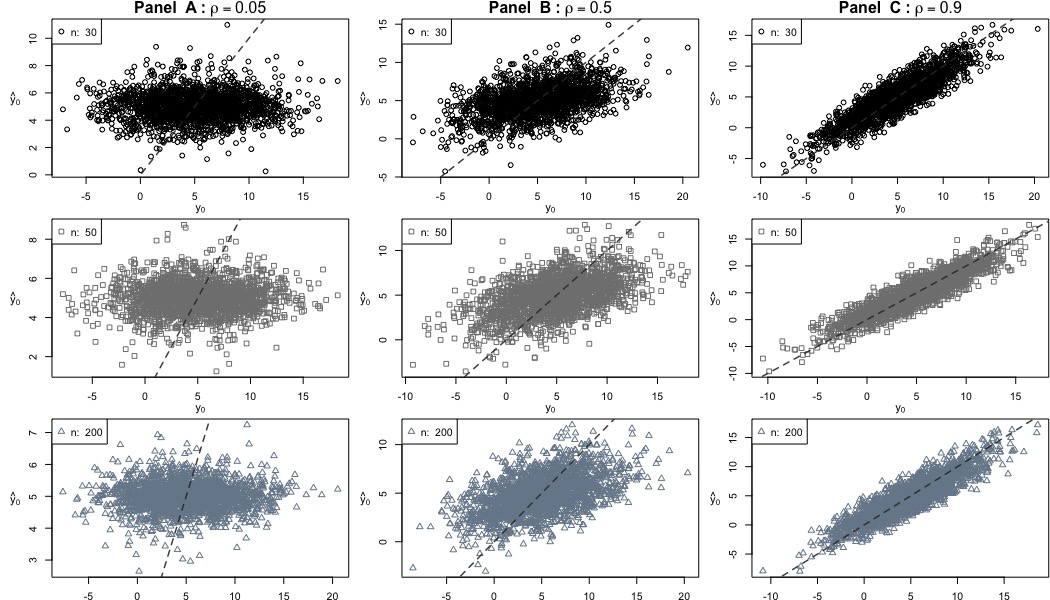}
        \caption{{\small Scatter plots of the pairs $(Y_0, \hat{Y}_0^\dagger)$ superimposed by $45^\circ$ line for different ($\rho,n$).}}
    \label{Simul_2_LSP}
\end{figure}

Note that MALP is designed to maximize the $\CCC$ between $Y_0$ and $\hat{Y}_0$. Therefore, the scatter plots in Figure \ref{Simul_2_MAP} are clustered around the $45^\circ$ line; whereas, the scatter plots associated with LSLP in Figure \ref{Simul_2_LSP} are not clustered on the $45^\circ$ line, especially when $|\rho|$ is small. Recall that the $\CCC$ could not exceed the $\PCC$ in absolute value, and if the parameters are known, the CCC and PCC of $Y_0$ and $\hat{Y}_0^\star$ are both equal to $\gamma =  |\rho|$. Since we are estimating MALP based on the sample data, the CCC and PCC of the $Y_0$ and the predictions based on the estimated MALP need not anymore equal $|\rho|$. These two phenomena are both reflected in Table \ref{Scatter_MAP}. In addition, compared to $\CCC$ and $\PCC$, the $\MSE$ becomes smaller as $|\rho|$ becomes larger since clearly, the $X$-values will contain more information about the $Y$-values as $|\rho|$ increases, hence a higher predictive content. 
\begin{table}[ht]
	\caption{{\small The performance of EMALP measured by empirical $\PCC$, $\CCC$, and $\MSE$ from 2000 $Y_0$ and $\hat{Y}_0^\star$ for different ($\rho$, $n$) combinations.}}
\label{Scatter_MAP}
\centering
{\small
\begin{tabular}{c|ccc|ccc|ccc}
\hline\hline
 & \multicolumn{3}{|c|}{$n=30$} & \multicolumn{3}{|c|}{$n=50$} & \multicolumn{3}{|c}{$n=200$} \\ \hline
$\rho$  &  0.05     &  0.5     & 0.9     & 0.05      &  0.5     &   0.9   &  0.05     & 0.5      & 0.9     \\\hline
$\PCC$  &   0.014               &     0.452             &   0.897                  &    0.001               &    0.486              &  0.900                &     0.063               &   0.511               &  0.903               \\
 $\CCC$  &   0.014               &     0.451             &   0.897            &     0.001               &    0.486              &  0.900            & 0.062               &   0.510               &  0.903              \\
$\MSE$ &      33.680              &     17.849            &   3.372                &      32.349              &    16.836             &  3.153          &   31.105              &   15.859              &  3.294               \\
\hline\hline
\end{tabular}}
\end{table}

Compared to the scatterplots of the MALP in Figure \ref{Simul_2_MAP}, the scatterplots of the LSLP in Figure \ref{Simul_2_LSP} are tilted from the $45^\circ$ line. This tendency is more pronounced for small $\rho$, but the scatterplot becomes closer to the $45^\circ$ line when $\rho=0.9$. 

\begin{table}[ht]
	\caption{{\small The performance of the ELSLP measured by empirical $\PCC$, $\CCC$, and $\MSE$ from 2000 $Y_0$ and $\hat{Y}_0^\dagger$ for different ($\rho$, $n$) combinations. }}
\label{Scatter_LSP}
\centering
{\small
\begin{tabular}{c|ccc|ccc|ccc}
\hline\hline
 & \multicolumn{3}{|c|}{$n=30$} & \multicolumn{3}{|c|}{$n=50$} & \multicolumn{3}{|c}{$n=200$} \\ \hline
$\rho$  &  0.05     &  0.5     & 0.9     & 0.05      &  0.5     &   0.9   &  0.05     & 0.5      & 0.9     \\\hline
$\PCC$  &   0.047               &     0.472             &   0.896                         &   0.024               &    0.467              &  0.899                              &     0.014               &   0.474               &  0.905                   \\
 $\CCC$  &   0.023               &     0.391             &   0.891                     &     0.010               &    0.380              &  0.893                     & 0.003               &   0.377               &  0.900                   \\
$\MSE$ &      16.541              &     13.435            &   3.149                         &     16.627              &    12.768             &  3.086                      &   15.592              &   12.867              &  2.979                       \\
\hline\hline
\end{tabular}}
\end{table}

\subsection{Computer Experiment 4}
\label{SUPPLE_SUBSEC_EXPERIMENT_4}
\begin{figure}[ht]
\centering
	\includegraphics[height=135pt]{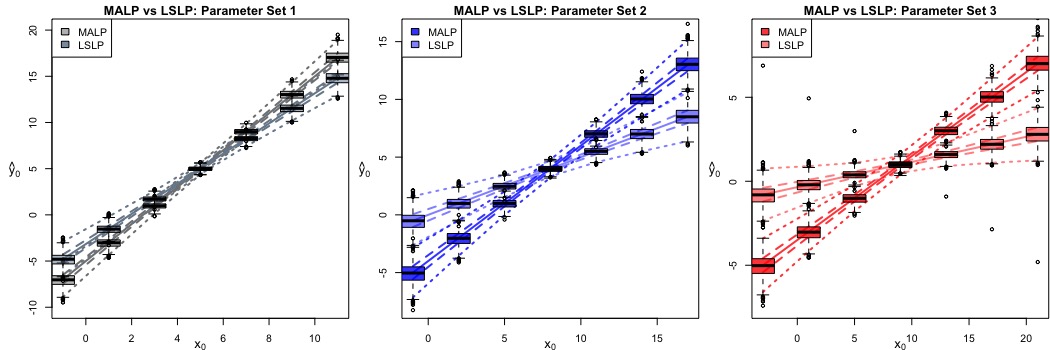}
\caption{{\small The comparisons between the EMALP and the ELSLP: the functional forms are compared at fixed points $x_0$'s for the three different parameter sets. The approximations to the normal distributions are also evaluated through the boxplots at each location, superimposed by the corresponding theoretical quantiles. For the three parameter sets, the associated $\rho$-values are  0.816, 0.5, and 0.3, from left-to-right panels.}}
\label{Figure_Asym_Normal}
\end{figure}

In this subsection, we provide the results of another experiment to compare the functional forms of MALP and LSLP, along with the quality of the normal approximation at fixed $x_0$-values or locations. The comparison was made for the three parameter sets in Table 1. We first set up the locations of $x_0$'s to be 0, $\pm1$, $\pm2$, and $\pm3$ standard deviations from $\mu_\txt{X}$ for each parameter set. 
Then, we construct $\hat{Y}^\star(x_0)$ and $\hat{Y}^\dagger(x_0)$ with $n=100$ sample sizes for each $x_0$ point. By repeating the process, we construct the side-by-side boxplots in Figure \ref{Figure_Asym_Normal} based on $\MReps=1000$ MALPs and LSLPs, superimposed with the 0.005, 0.25, 0.5, 0.75, 0.995 theoretical quantiles from the normal distribution, with parameters specified in Theorem 3.1 of the main paper. 

While the MALP and LSLP functions intersect at the point $x_0=\mu_\txt{X}$, the slopes of MALP are always steeper than those of LSLP, and the difference becomes larger as $|\rho|$ becomes smaller from the left plots to the right plots. 
The normal approximations provided by the asymptotic normality results appear good for this particular set of parameters and sample size, as could be discerned from the boxplots of the predictions from the estimated MALP and LSLP fitting well the corresponding quantiles of the approximating normal distributions. In particular, note that the sizes of boxes match well with the theoretical interquartile ranges, represented by dashed lines, from the approximating normal distributions. While the asymptotic standard errors of the EMALP are greater than those of the ELSLP, these differences are not so noticeable because the standard errors became quite small due to the scaling by the square root of the sample size $n=100$.

\section{Forms of Confidence Intervals}
\label{SUPPLE_CI}
We presented three approaches for estimating the standard error of $\hat{Y}^\star(x_0)$, which are through the asymptotic theory, the jackknife, and the bootstrap. Approximate confidence intervals (CIs) for MALP $\tilde{Y}^\star(x_0)$ could then be constructed using the approximate normality and the estimated standard errors from the three approaches mentioned. These CIs for $\tilde{Y}^\star(x_0)$ are provided below, where $z_{\alpha/2} = \Phi^{-1}(1-\alpha/2)$ is the $(1-\alpha/2)$th quantile of the standard normal distribution, and we utilize the estimated variances provided by the asymptotic normality result, the jackknife, and the bootstrap procedures. 
\begin{eqnarray}
	\Gamma_1[x_0,(x,y);\alpha]  &=&  \left[\hat{y}^\star(x_0)\pm z_{\alpha/2}\tfrac{\hat{\sigma}_\txt{MA}(x_0)}{\sqrt{n}} \right];\label{ASYMPTOTIC_NORMAL_CI}\\
	\Gamma_2[x_0,(x,y);\alpha]  &=&  \left[\hat{y}^\star(x_0)\pm z_{\alpha/2}\hat{\sigma}_\txt{JK}(x_0) \right];\label{JACKKNIFE_CI}\\ 
	\Gamma_3[x_0,(x,y);\alpha]  &=& \left[\hat{y}^\star(x_0)\pm z_{\alpha/2}\hat{\sigma}_\txt{BS}(x_0) \right]\label{PARAMETRIC_BOOT_CI}. 
\end{eqnarray}

 For the bootstrap-$t$ procedure, we emulate the pivot of the $t$-based confidence interval $$T(x_0)=\frac{\hat{y}^\star(x_0)-\tilde{y}^\star(x_0)}{\hat{\sigma}_\txt{MA}(x_0)/\sqrt{n}}$$ using the $B$ bootstrap samples as well as the $B'$ bootstrap sub-samples such that $$T_b^*(x_0)=\frac{\hat{y}^{\star *}_b(x_0)-\hat{y}^\star(x_0)}{\hat{\sigma}_\txt{BS}^*(x_0)},$$ where $\hat{\sigma}_\txt{BS}^*(x_0)$ is obtained by the formula in the bootstrap procedure but with the $B'$ sub-resamples. 
The estimated sampling distribution of $T(x_0)$ becomes $\hat{G}_t(z)=\frac{1}{B}\sum_{b=1}^B I\{T_b^*(x_0) \le z\}$, called a bootstrap-$t$ distribution.
The resulting $(1-\alpha)100\%$ bootstrap-$t$ confidence interval is
\begin{equation}
	\Gamma_4[x_0,(x,y);\alpha]  =  \left[\hat{y}^\star(x_0)-\hat{G}_t^{-1}(1-\alpha/2)\hat{\sigma}_\txt{BS}(x_0), \hat{y}^\star(x_0)-\hat{G}_t^{-1}(\alpha/2)\hat{\sigma}_\txt{BS}(x_0) \right],
\end{equation}
where $\hat{G}^{-1}_t(\alpha)$ is the $\alpha$th-quantile of the bootstrap-$t$ distribution and $\hat{\sigma}_\txt{BS}(x_0)$ is from the bootstrap procedure.
 For the percentile approach, we utilize $B$ bootstrap samples to obtain the same number of bootstrap replications $\hat{y}^{\star*}(x_0)$. The resulting bootstrap distribution is $\hat{G}_{p}(z)=\frac{1}{B}\sum_{b=1}^B I\{\hat{y}^{\star*}_b(x_0)\le z\}$.
Then, we define the $(1-\alpha)100\%$ bootstrap percentile confidence interval as follows:
\begin{equation}
	\Gamma_5[x_0,(x,y);\alpha]  =  \left[\hat{G}_p^{-1}(\alpha/2), \hat{G}_p^{-1}(1-\alpha/2)   \right],
\end{equation}
where $\hat{G}^{-1}_p(\alpha)$ is the $\alpha$th-quantile of the bootstrap distribution. Note that this approach is purely nonparametric without assuming any distributions compared to the parametric bootstrap procedure in (\ref{PARAMETRIC_BOOT_CI}). While the bootstrap sample size of $B=200$ was enough to estimate the standard error, bootstrap CI procedures generally require larger $B$ such as 1000 or 2000.
To introduce a more refined approach, we first estimate the bias-correction parameter as follows: $z_0  =  \Phi^{-1}(\hat{G}(\hat{y}^\star(x_0)))$.
In words, $z_0$ is the theoretical quantile for the proportion of the bootstrap replications that are less than the observed $\hat{y}^\star(x_0)$. Moreover, the acceleration parameter, $a$, which measures the skewness of the bootstrap distribution and can be estimated by using the jackknife replications, is
$$\hat{a}  =  \sum_{j=1}^n\frac{\left(\hat{y}^\star_{(j)}(x_0)-\hat{y}^\star_{(\bullet)}(x_0)\right)^3 }{6\left\{\sum_{j=1}^n\left(\hat{y}^\star_{(j)}(x_0)-\hat{y}^\star_{(\bullet)}(x_0)\right)^2\right\}^{3/2}}.$$
Based on these two estimated parameters, we define the bias-corrected and accelerated (BCa) bootstrap percentile CI as follows:
\begin{eqnarray*}
	\Gamma_6[x_0,(x,y);\alpha]  &=&  \left[\hat{G}^{-1}\left\{\Phi\left(z_0+\tfrac{z_0-z_{\alpha/2}}{1-\hat{a}(z_0-z_{\alpha/2})}\right) \right\}, \hat{G}^{-1}\left\{\Phi\left(z_0+\tfrac{z_0+z_{\alpha/2}}{1-\hat{a}(z_0+z_{\alpha/2})} \right)\right\} \right].
	\end{eqnarray*}
While the BCa procedure generally provides better estimation results, its actual performance needs to be evaluated in our problem setting. In particular, the shape of the bootstrap distribution would be an important factor for comparison with the previously established asymptotic normality.

\section{Body Fat Data Set Description}
\label{SUPPLE_BF_DATA}

This section provides additional information about the body fat data set \citep{Pen:1985}. 
The study leading to this dataset was motivated by the difficulty of obtaining the percentage of body fat, which is usually obtained through underwater weighing, so the aim is to predict the percent body fat using easily-measured characteristics. The dataset consists of 15 variables and 252 observations. The variables are percent body fat (PBF), body density (BD), age (in years), weight (WGT, in pounds), height (HGT, in inches), and various body circumference measurements (in cm): neck (NCK), chest (CST), abdomen (ABD), hip (Hip), thigh (TGH), knee (KN), ankle (ANK), biceps (BCP), forearm (FA), and wrist (WRT). Note that the PBF variable was calculated from BD based on the equation in \cite{Siri:1956}:
$\text{PBF}= 495/\text{BD}-450$. Table~\ref{BODY_FAT_CORR} provides the correlation matrix for the 15 variables in the data set. 

\begin{table}[ht]
\caption{{\small Correlation between variables from the body fat data}}
\label{BODY_FAT_CORR}
{\tiny{
\centering
\begin{tabular}{c|rrrrrrrrrrrrrrr}

\hline\hline
&BD & PBF & Age &WGT& HGT & NCK &CST& ABD & Hip& TGH & KN &ANK& BCP& FA& WRT    \\
\hline
BD         & 1.00              & -0.99 &-0.28        & -0.59     & 0.10 &-0.47 &-0.68                & -0.80 &-0.61& -0.55& -0.50& -0.26 & -0.49 & -0.35& -0.33      \\
PBF         & -0.99             & 1.00               & 0.29      & 0.61                            & -0.09                         & 0.49  & 0.70        & 0.81             & 0.63       & 0.56  & 0.51  & 0.27 & 0.49 & 0.36 & 0.35 \\
Age             & -0.28             & 0.29               & 1.00      & -0.01                           & -0.17                         & 0.11  & 0.18        & 0.23& -0.05& -0.20 & 0.02& -0.11 & -0.04 & -0.09 & 0.21     \\
WGT          & -0.59             & 0.61& -0.01         & 1.00      & 0.31                            & 0.83                          & 0.89  & 0.89        & 0.94             & 0.87       & 0.85  & 0.61  & 0.80 & 0.63 & 0.73      \\
HGT         & 0.10              & -0.09& -0.17        & 0.31      & 1.00                            & 0.25                          & 0.13  & 0.09        & 0.17             & 0.15       & 0.29  & 0.26  & 0.21 & 0.23 & 0.32       \\
NCK            & -0.47             & 0.49               & 0.11      & 0.83                            & 0.25                          & 1.00  & 0.78        & 0.75             & 0.73       & 0.70  & 0.67  & 0.48 & 0.73 & 0.62 & 0.74 \\
CST           & -0.68             & 0.70               & 0.18      & 0.89                            & 0.13                          & 0.78  & 1.00        & 0.92             & 0.83       & 0.73  & 0.72  & 0.48 & 0.73 & 0.58 & 0.66 \\
ABD         & -0.80             & 0.81               & 0.23      & 0.89                            & 0.09                          & 0.75  & 0.92        & 1.00             & 0.87       & 0.77  & 0.74  & 0.45 & 0.68 & 0.50 & 0.62 \\
Hip             & -0.61             & 0.63 &-0.05         & 0.94      & 0.17                            & 0.73                          & 0.83  & 0.87        & 1.00             & 0.90       & 0.82  & 0.56  & 0.74 & 0.55 & 0.63       \\
TGH           & -0.55             & 0.56 &-0.20         & 0.87      & 0.15                            & 0.70                          & 0.73  & 0.77        & 0.90             & 1.00       & 0.80  & 0.54  & 0.76 & 0.57 & 0.56      \\
KN            & -0.50             & 0.51               & 0.02      & 0.85                            & 0.29                          & 0.67  & 0.72        & 0.74             & 0.82       & 0.80  & 1.00  & 0.61 & 0.68 & 0.56 & 0.66 \\
ANK           & -0.26             & 0.27& -0.11         & 0.61      & 0.26                            & 0.48                          & 0.48  & 0.45        & 0.56             & 0.54       & 0.61  & 1.00  & 0.48 & 0.42 & 0.57     \\
BCP          & -0.49             & 0.49 &-0.04         & 0.80      & 0.21                            & 0.73                          & 0.73  & 0.68        & 0.74             & 0.76       & 0.68  & 0.48  & 1.00 & 0.68 & 0.63       \\
FA         & -0.35             & 0.36& -0.09         & 0.63      & 0.23                            & 0.62                          & 0.58  & 0.50        & 0.55             & 0.57       & 0.56  & 0.42  & 0.68 & 1.00 & 0.59       \\
WRT          & -0.33             & 0.35               & 0.21      & 0.73                            & 0.32                          & 0.74  & 0.66        & 0.62             & 0.63       & 0.56  & 0.66  & 0.57 & 0.63 & 0.59 & 1.00\\
\hline\hline
\end{tabular}}}
\end{table}

\begin{figure}[!htpp]
\centering
\subfloat{\includegraphics[scale=0.21]{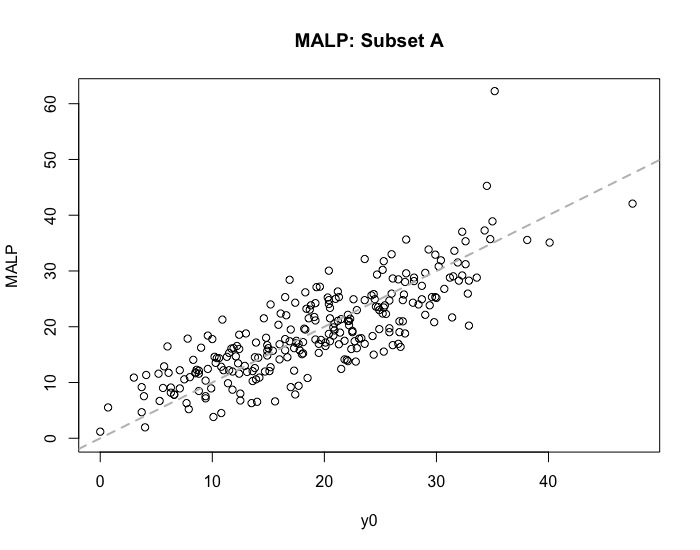}}
\subfloat{\includegraphics[scale=0.21]{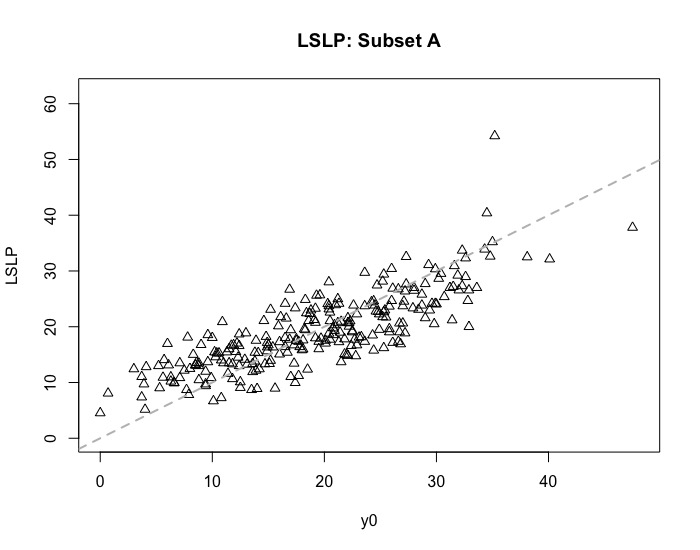}}\\\vspace{-15pt}
\subfloat{\includegraphics[scale=0.21]{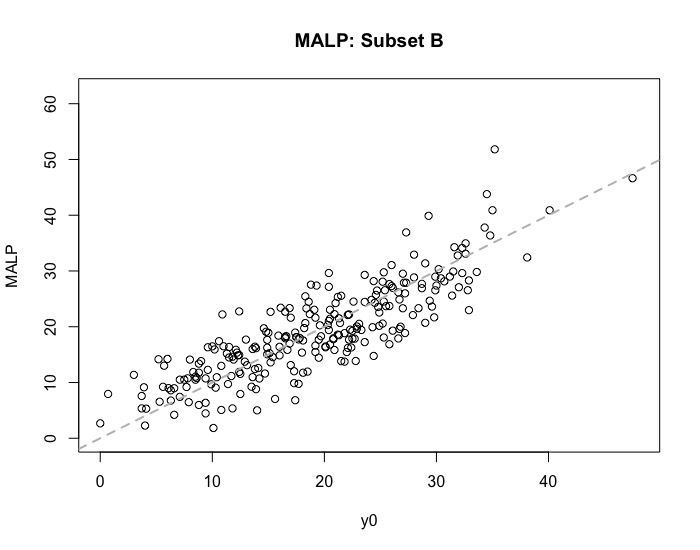}}
\subfloat{\includegraphics[scale=0.21]{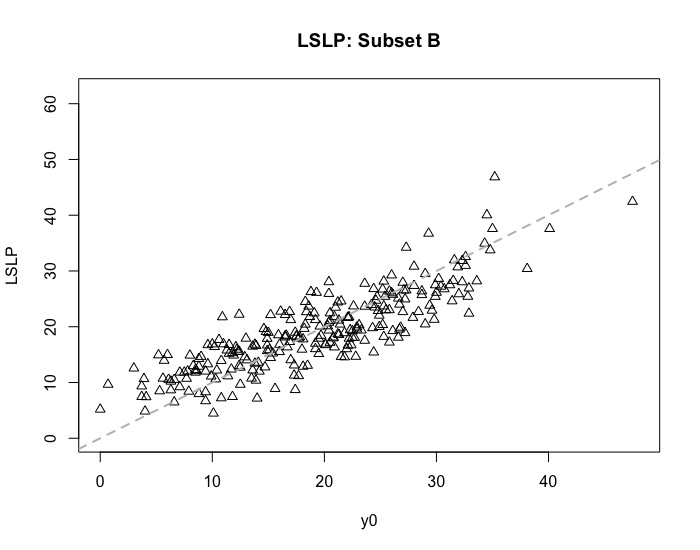}}\\\vspace{-15pt}
\subfloat{\includegraphics[scale=0.21]{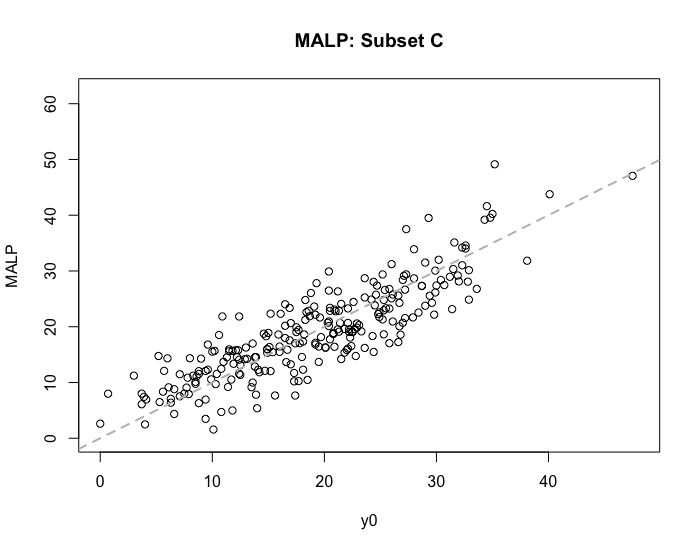}}
\subfloat{\includegraphics[scale=0.21]{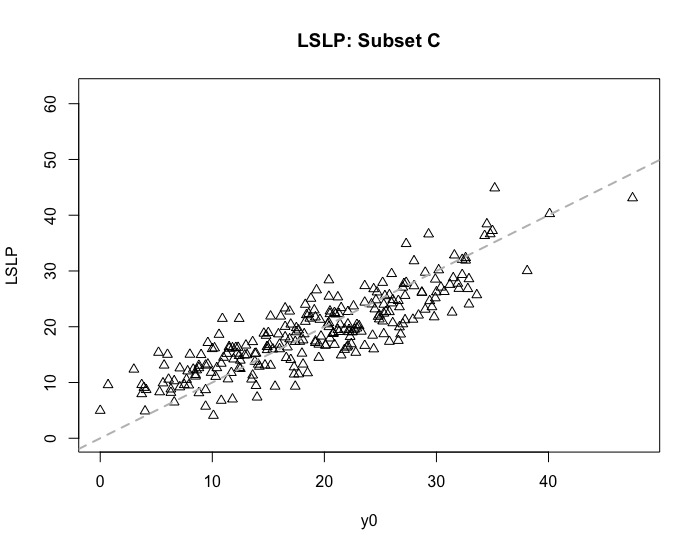}}\\\vspace{-15pt}
\subfloat{\includegraphics[scale=0.21]{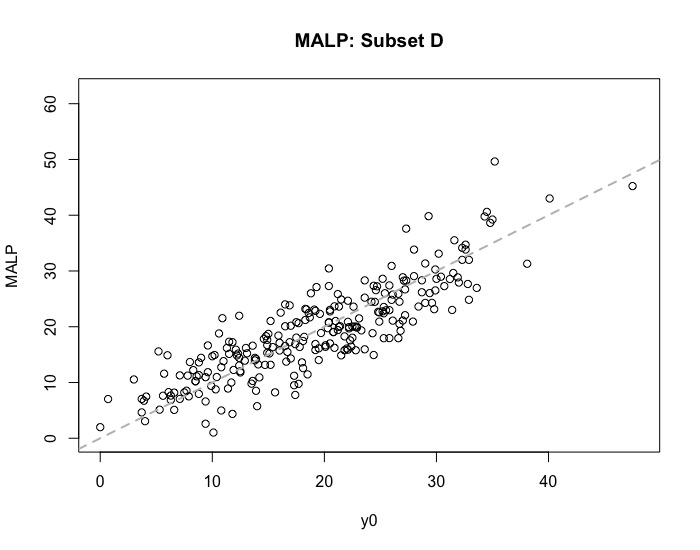}}
\subfloat{\includegraphics[scale=0.21]{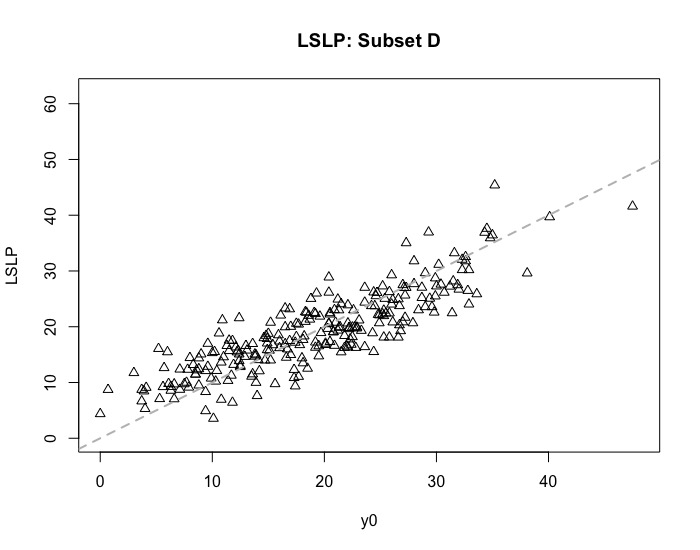}}\\\vspace{-15pt}
\subfloat{\includegraphics[scale=0.21]{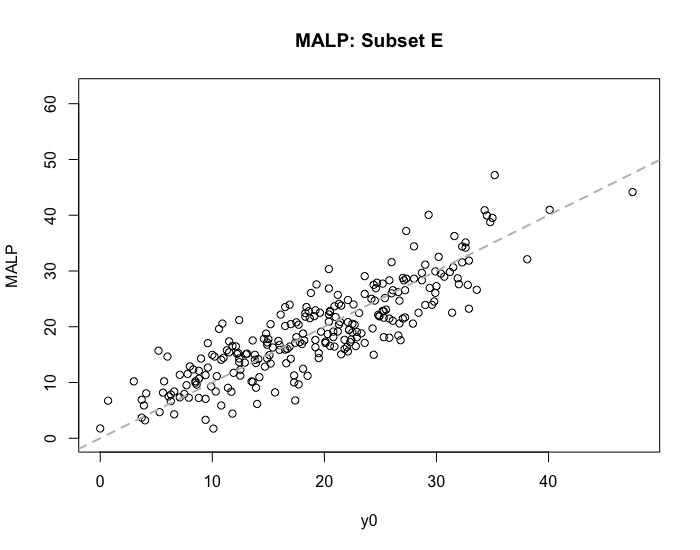}}
\subfloat{\includegraphics[scale=0.21]{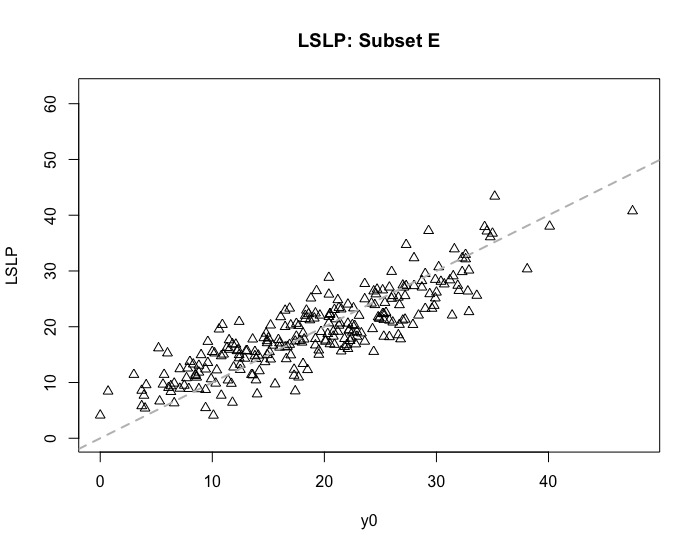}}
\caption{{\small Scatter plots between $Y_0$ and $\hat{Y}_0$, with the predicted values obtained via the MALP and LSLP, for different (increasing) subsets of predictor variables, with the dashed line being the $45^\circ$ line.}}
\label{BODY_FAT_Scatter}
\end{figure}

In the paper, the EMALP and ELSLP were obtained for increasing sizes of subsets containing the set of the best predictive variables with respect to the coefficient of determination. Subsets A, B, C, D, and E correspond, respectively, to the best subsets with 1, 2, 4, 6, and 8 variable(s). Figure \ref{BODY_FAT_Scatter} presents the scatter plots between predictor and predictand with respect to these subsets. Observe that the shape of the point cloud associated with the MALP predictions is more aligned with 45$^\circ$ line than the shape of the point cloud associated with the LSLP predictions.

\end{appendices}

\end{document}